\documentclass{aa}  
\newcommand{\lsim}{\lower 2pt \hbox{$\, \buildrel {\scriptstyle
<}\over {\scriptstyle \sim}\,$}}  \newcommand{\gsim}{\lower 2pt
\hbox{$\, \buildrel {\scriptstyle >}\over {\scriptstyle \sim}\,$}}
\usepackage{appendix}
\usepackage{graphicx}
\usepackage{txfonts}
\usepackage{natbib}
\usepackage{rotating}
\defcitealias{Roth05}{Roth et al. 2005 }
\defcitealias{Bonnet04}{Bonnet et al. 2004 }
\defcitealias{Arribas98}{Arribas et al. 1998 }

\bibpunct{(}{)}{;}{a}{}{,} % to follow the A&A style.

\begin{document}

\title{VLT-VIMOS integral field spectroscopy of luminous and ultraluminous infrared galaxies}

\subtitle{III: the atlas of the stellar and ionized gas distribution.}

\author{
J. Rodr\'iguez-Zaur\'in\inst{1}, S. Arribas\inst{1}, A. Monreal-Ibero\inst{2,4,5}, L. Colina\inst{1}, A. Alonso-Herrero\inst{1} and J. Alfonso-Garz\'on\inst{3}}
\offprints{Javier Rodr\'iguez:\\ {\tt jrz@damir.iem.csic.es} \smallskip}

\institute{
Instituto de Estructura de la Materia (CSIC), C/Serrano 121, 28006, Madrid, Spain.\\\email{jrz@damir.iem.csic.es}
\and
European Organization for Astronomical Research in the Southern Hemisphere (ESO); Karl-Schwarzschild-Strasse 2 D-85748, Garching bei M\"unchen, Germany.
\and
Laboratorio de astrof\'isica espacial y f\'isica fundamental (LAEFF),Apartado 78, E-28691, Villanueva de la Ca\~nada, Madrid, Spain. 
\and
Astrophysikalisches Institut Postdam, An der Sternwarte 16, D-14482 Potsdam, Germany
\and
Instituto de Astrof\'isica de Andaluc\'ia. CSIC. Glorieta de la Astronom\'ia, s/n, 18008,Granada, Spain.
}

\date{  }

\abstract {Luminous and ultraluminous infrared galaxies (LIRGs and
ULIRGs) are much more numerous at higher redshifts than locally,
dominating the star-formation rate density at redshifts $\sim$1 --
2. Therefore, they are important objects in order to understand how
galaxies form and evolve through cosmic time. Local samples provide a
unique opportunity to study these objects in detail.} {We aim to
characterize the morphologies of the stellar continuum and the ionized
gas (H${\alpha}$) emissions from local sources, and investigate how
they relate with the dynamical status and IR-luminosity of the
sources.}  {We use optical (5250 -- 7450~\AA) integral field
spectroscopic (IFS) data for a representative sample of 38 sources (31
LIRGs and 7 ULIRGs), taken with the VIMOS instrument on the VLT.}{We
present an atlas of IFS images of continuum emission, H${\alpha}$
emission, and H${\alpha}$ equivalent widths for the sample. The
morphologies of the H${\alpha}$ emission are substantially different
from those of the stellar continuum. The H${\alpha}$ images frequently
reveal extended structures that are not visible in the continuum, such
as HII regions in spiral arms, tidal tails, rings, bridges, of up to
few kpc from the nuclear regions. The morphologies of the continuum
and H${\alpha}$ images are studied on the basis of the C$_{2kpc}$
parameter, which measures the concentration of the emission within the
central 2 kpc. The C$_{2kpc}$ values found for the H${\alpha}$ images
are higher than those of the continuum for the majority (85\%) of the
objects in our sample. On the other hand, most of the objects in our
sample ($\sim$62\%) have more than half of their H${\alpha}$ emission
outside the central 2 kpc. No clear trends are found between the
values of C$_{2kpc}$ and the IR-luminosity of the sources. On the
other hand, our results suggest that the star formation in advance
mergers and early-stage interactions is more concentrated than in
isolated objects. Finally, we compared the H${\alpha}$ and infrared
emissions as tracers of the star-formation activity. We find that the
star-formation rates derived using the H${\alpha}$ luminosities
generally underpredict those derived using the IR luminosities, even
after accounting for reddening effects.}{}


\authorrunning{J. Rodr\'iguez Zaur\'in et al.:}
\titlerunning{VLT-VIMOS integral field spectroscopy of luminous and ultraluminous infrared galaxies. III.}
\maketitle 
%
%________________________________________________________________

\section{Introduction}

The advent of the new infrared (IR), sub-millimeter (sub-mm) and
millimeter (mm) facilities such as for example the {\it Spitzer} Space
Telescope, the Submillimeter (sub-mm) Common User Bolometer Array
(SCUBA), and the Max-Planck Millimeter Bolometer (MAMBO) array, have
made it possible to extend deep cosmological surveys from the
UV/optical to the IR, sub-mm and mm wavelengths
\citep[e.g.][]{Hughes98,Perezgonzalez05,
Coppin06,Austermann09,Lonsdale09}. These surveys revealed the presence
of high-z galaxies with luminosities, morphologies and sizes
consistent with those of the local luminous (LIRGs, 10$^{11} <
$L$_{\rm IR} <$ 10$^{12}$ L$_{\odot}$, L$_{\rm IR}$ = 8 - 1000 $\mu$m)
and ultraluminous (ULIRGs, L$_{\rm IR} >$ 10$^{12}$L$_{\odot}$)
infrared galaxies \citep[e.g.][]{Conselice05}. Follow-up studies of
these high-z (U)LIRGs showed that a large fraction of these objects
were actively forming stars at redshifts 1 -- 4. Furthermore, they
dominate the star-formation rate (SFR) density at redshifts z $>$ 1,
and form a large fraction of the newborn stars at redshifts z $\sim$
1.5 \citep{Perezgonzalez05}.

A great effort has been made during the last years to study in detail
the properties of LIRGs and ULIRGs in the local universe
\citep[e.g.][]{Tacconi02,Dasyra06a,Alonso-Herrero06,Armus07,Nardini09,
Rodriguez-Zaurin10,Clements10}, and at moderate redshifts (z $\sim$ 1
-- 2) \citep[e.g.][]{Farrah09,Fiolet09}. These studies have provided
us with some detailed information about the physical processes taking
place in these objects. However, to have a comprehensive picture of how
galaxies form and evolve from the distant to the local universe we
need detailed studies of (U)LIRGs kinematics, internal structure,
stellar populations and excitation conditions. These detailed studies
require high S/N two-dimensional spectroscopic information with both
high angular and spectral resolutions.

Up to date, optical and near-IR IFS studies of local (U)LIRGs have
usually concentrated on individual objects
\citep[e.g.][]{Colina99,Arribas01,Garcia-Marin06,Bastian06,Bedregal09,
Lipari09a,Lipari09b} or relatively small samples
\citep{Colina05,Monreal-Ibero06,Garcia-Marin09a}. To remedy this
situation we are carrying out a program with the aim of studying the
internal structure and kinematics of a large ($\sim$ 70),
representative sample of LIRGs and ULIRGs using several optical and
near-IR integral field spectroscopic facilities.

This is the third of a series of studies based on VLT-VIMOS
observations. \cite{Arribas08}, hereafter Paper I, presented the
sample, data reduction and analysis techniques, as well as preliminary
results obtained for two individual sources, IRAS F06076-2139 and IRAS
F12115-4656. A detailed study of the ionization in the extra-nuclear
extended regions can be found in \cite{Monreal-Ibero10},
hereafter Paper II.

In this paper we present an atlas of reconstructed maps of continuum,
H${\alpha}$ line emission flux and H${\alpha}$ equivalent width
(H${\alpha}$-EW) tracing the stellar component, the ongoing
star-formation activity and the presence of ionizing shocks or an
active galactic nucleus (AGN). We also perform a basic structural
analysis of these images and look for trends and correlations between
the morphological properties of the objects and other properties of
LIRGs/ULIRGs. Finally, we compare the H${\alpha}$ and IR luminosities
as tracers of the star-formation activity.

Throughout the paper we will consider H$_{0}$ = 70
kms$^{-1}$Mpc$^{-1}$, $\Omega_{\rm \Lambda}$ = 0.7, $\Omega_{\rm M}$ = 0.3.

\section{The VIMOS sample: observations, data reduction and line fitting}

\subsection{The sample}

The IFS (U)LIRG survey is a large program that started with the aim of
studying the 2D-internal structure and kinematics of low-z LIRGS and
ULIRGs. The survey was carried out using integral field spectroscopic
facilities in both the northern (INTEGRAL, \citealt{Arribas98}; PMAS,
\citealt{Roth05}) and the southern (VIMOS, \citealt{LeFevre03};
SINFONI, \citealt{Bonnet04}) hemispheres, and includes $\sim$ 70
sources. The VIMOS sample discussed in this paper contains a total of
38 galaxies, which are listed in Table 1. Thirty one of these galaxies
are classified as LIRGs. The LIRG subsample is drawn from the IRAS
Revised Bright Galaxy Sample \citep[RBGS,][]{Sanders03}, and has a
mean redshift of 0.024. The other seven objects in the sample are
classified as ULIRGs, and were selected from the IRAS 1 Jy sample of
ULIRGs \citep{Kim98a}, the RBGS, and from the HST/WFPC2 snapshot
sample (ID 6346 PI: K.Borne). The ULIRG subsample has a mean redshift
of 0.069. Possible biases owing to the higher redshifts of the subsample
of ULIRGs are discussed later in the paper.

The VIMOS sample is not complete either in luminosity or
distance. However, one of the aims of our project was to investigate
how the properties of (U)LIRGs correlate with the different
morphologies of the objects. Therefore, since the VIMOS sample is
certainly representative of the different morphologies within the
(U)LIRG phenomenon, it is adequate for the purposes of this work.

\begin{table*}
{\tiny \centering
%\begin{minipage}{140mm}
\resizebox{1.0\textwidth}{!}{
\begin{tabular}{ccccccccccccc}
\hline
\hline
ID1  & ID2   &$\alpha$& $\delta$& $z$ & References  &  D      &scale       & $\log L_{\mathrm{IR}}$ & Class & Spectral   &  References\\
IRAS & Other &(J2000) & (J2000) &     &             &  (Mpc)  &(pc/arcsec) & (L$_\odot$)            &       & Classification&\\
  (1)&   (2) &   (3)  &   (4)   &  (5)&    (6)      &  (7)    &(8)         &    (9)                 & (10)  &    (11)    & (12) \\
\hline
F01159$-$4443 &   ESO~244$-$G012 &01:18:08.1 & -44:27:40 &  0.022903 &1 & 99.8  &462 & 11.48 & 1   & H(N),H/S(S)& 4,5\\
F01341-3735 & ESO-297-G011 / G012&01:36:24.0 & -37:19:14 &  0.017305 &2 & 75.1  &352 & 11.18 & 1   & H(both) & 4,5\\ 
F04315$-$0840 &         NGC 1614 &04:34:00.0 & -08:34:46 &  0.015938 &3 & 69.1  &325 & 11.69 & 2   & H & 1,5\\
F05189$-$2524 &                  &05:21:01.4 & -25:21:46 &  0.042563 &4 & 188.2 &839 & 12.19 & 2   & S& 3\\
F06035$-$7102 &                  &06:02:54.5 & -71:03:08 &  0.079465 &5 & 360.7 &1501& 12.26 & 1   & H & 2\\
F06076$-$2139 &                  &06:09:45.1 & -21:40:22 &  0.037446 &5 & 165.0 &743 & 11.67 & 1   & - & -\\
F06206$-$6315$^a$ &              &06:21:00.9 & -63:17:23 &  0.092441 &5 & 423.3 &1720& 12.27 & 1   & S & 2\\
F06259$-$4708 & ESO~255$-$IG 007 &06:27:21.1 & -47:10:38 &  0.038790 &6 & 171.1 &769 & 11.91 & 1   & H(N) & 4\\
F06295$-$1735 &  ESO 557$-$G002  &06:31:46.3 & -17:37:15 &  0.021298 &7 & 92.7  &431 & 11.27 & 0   & H & 5\\
F06592$-$6313 &                  &06:59:40.3 & -63:17:53 &  0.022956 &5 & 100.0 &464 & 11.22 & 0   & H & 5\\
F07027$-$6011& AM~0702$-$601     &07:03:27.5 & -60:16:05 &  0.031322 &5 & 137.4 &626 & 11.64 & 0   & S(N) & 4\\
F07160$-$6215& NGC~2369          &07:16:37.7 & -62:20:37 &  0.010807 &2 & 46.7  &221 & 11.16 & 0   & ...&...\\
08355$-$4944 &                   &08:37:02.3 & -49:54:32 &  0.025898 &8 & 113.1 &521 & 11.60 & 2   & ...&...\\
08424$-$3130 & ESO~432$-$IG006   &08:44:27.6 & -31:41:41 &  0.016165 &5 & 70.1  &329 & 11.04 & 1   & ...&...\\
F08520$-$6850 & ESO~60$-$IG016   &08:52:31.2 & -69:01:59 &  0.046315 &9 & 205.4 &909 & 11.83 & 1   & ...&...\\
09022$-$3615 &                   &09:04:12.8 & -36:27:02 &  0.059641 &5 & 267.0 &1153& 12.32 & 2   & ...&...\\
F09437+0317   & IC-563/ 564      &09:46:20.3 & +03:03:22 &  0.020467 &10& 89.0  &415 & 11.21 & 1/0 & ...&...\\
F10015$-$0614 & NGC-3110         &10:04:02.7 & -06:28:35 &  0.016858 &2 & 73.1  &343 & 11.31 & 0   & H & 5\\
F10038$-$3338 & IC2545           &10:06:04.2 & -33:53:04 &  0.034100 &5 & 149.9 &679 & 11.77 & 2   & ...&... \\
F10257$-$4339$^b$ & NGC~3256     &10:27:52.4 & -43:54:25 &  0.009354 &2 & 40.4  &192 & 11.69 & 2   & H & 7\\
F10409$-$4556 & ESO~264$-$G036   &10:43:07.0 & -46:12:43 &  0.021011 &11& 91.4  &425 & 11.26 & 0   & H/L & 5\\
F10567$-$4310 & ESO~264$-$G057   &10:59:02.4 & -43:26:33 &  0.017199 &5 & 74.6  &350 & 11.07 & 0   & H & 5\\
F11255$-$4120$^b$ &ESO~319$-$G022&11:27:56.1 & -41:37:06 &  0.016351 &5 & 70.9  &333 & 11.04 & 0   & H & 5\\
F11506$-$3851 & ESO~320$-$G030   &11:53:12.0 & -39:07:54 &  0.010781 &3 & 46.6  &221 & 11.30 & 0   & H &  6\\
F12043$-$3140$^b$&ESO~440$-$IG 058&12:06:53.0& -31:57:05 &  0.023203 &12& 101.1 &468 & 11.37 & 1   & L/HII(N)/HII(S)& 5,6\\
F12115$-$4656 & ESO~267$-$G030   &12:14:12.6 & -47:13:37 &  0.018489 &5 & 80.3  &375 & 11.11 & 0   & H & 5\\
12116$-$5615 &                   &12:14:21.4 & -56:32:32 &  0.027102 &8 & 118.5 &545 & 11.61 & 2/0 & ...&... \\
F12596$-$1529$^a$&MCG$-$02$-$33$-$098& 13:02:20.5&-15:46:05&0.015921 &13& 69.0  &324 & 11.07 & 1   & H(both) & 1,5\\
F13001$-$2339 & ESO~507$-$G070   &13:02:51.3 & -23:55:09 &  0.021702 &14& 94.5  &439 & 11.48 & 2/0/1& L & 5\\
F13229$-$2934 & NGC 5135         &13:25:43.0 & -29:49:54 &  0.013693 &2 & 59.3  &280 & 11.29 & 0   & S & 4,5\\
F14544$-$4255 & IC 4518          &14:57:43.1 & -43:08:01 &  0.015728 &15 & 68.2  &320 & 11.11 & 1  & S(W) & 5\\
F17138$-$1017 &                  &17:16:36.3 & -10:20:40 &  0.017335 &16& 75.2  &352 & 11.41 & 2/0 & H & 5\\
F18093$-$5744 & IC 4687/4686     &18:13:38.6 & -57:43:36 &  0.017345 &17& 75.3  &353 & 11.57 & 1   & H(both)& 4,5\\
F21130$-$4446$^c$ &              &21:16:19.0 & -44:33:32 &  0.092554 &5 & 423.9 &1722& 12.09 & 2   & H & 2\\
F21453$-$3511 & NGC 7130         &21:48:19.6 & -34:57:05 &  0.016151 &10& 70.0  &329 & 11.41 & 2   & L/S & 1,5\\
F22132$-$3705 & IC 5179          &22:16:10.0 & -36:50:36 &  0.011415 &18& 49.3  &234 & 11.22 & 0   & H & 1\\
F22491$-$1808$^c$ &              &22:51:49.0 & -17:52:28 &  0.077760 &5 & 352.5 &1471& 12.17 & 1   & H & 3\\
F23128$-$5919$^c$ & AM 2312-591  &23:15:46.6 & -59:03:14 &  0.044601 &19& 197.5 &878 & 12.06 & 1   & H/L/S & 2,4\\
\hline
\hline
\end{tabular}}
\caption{General properties of the (U)LIRGs in the VIMOS
  sample. Column (1): object designation in the IRAS Faint source
  catalogue (FSC). For the four sources that are not in the FSC, the
  identification in the IRAS Point source catalogue (PSC) is given,
  which has no prefix 'F'.  Column (2): other name. Columns (3) and
  (4): right ascension (hours, minutes and seconds) and declination
  (degrees, arcminutes and arcseconds) from the IRAS FSC. The
  exceptions are IRAS 08355-4944, IRAS F08424-3130, IRAS 09022-3615
  and IRAS 12116-5615, for which the positions are taken from the IRAS
  PSC. Column (5): redshift of the IRAS sources from the NASA
  Extragalactic Database (NED). The references for the redshift values
  are given in Column (6). REFERENCES.-- 1: \cite{daCosta91}. 2: The
  HI Parkes All Sky Survey Catalogue (HIPAS). 3:
  \cite{deVaucouleurs91}. 4: \cite{Huchra83}. 5: \cite{Strauss92}. 6:
  \cite{Lauberts79}. 7: \cite{Chamaraux99}. 8: \cite{Sanders95}. 9:
  \cite{West81}. 10: \cite{deVaucouleurs76}. 11: \cite{Jones09}. 12:
  \cite{Kaldare03}. 13: \cite{Huchra92}. 14: \cite{Pimbblet06}. 15:
  \cite{Visvanathan96}. 16: \cite{Shier98}. 17: \cite{Martin78}. 18:
  \cite{Mathewson92}. 19: \cite{Hwang07}. Column (7): luminosity
  distances assuming a $\Lambda$CDM cosmology with H$_{0}$ = 70
  kms$^{-1}$Mpc$^{-1}$, $\Omega_{\rm M}$ = 0.7, $\Omega_{\rm M}$ = 0.3
  and using the Edward L. Wright Cosmology calculator, which is based
  on the prescription given by \cite{Wright06}. Column (8):
  scales. Column (9): infrared luminosity ($L_{\rm IR}$) =$L$(8 --
  1000 $\mu$m), in units of solar bolometric luminosity, calculated
  using the fluxes in the four IRAS bands as given in \cite{Sanders03}
  when available. Otherwise, the standard prescription in
  \cite{Sanders96} with the values in the IRAS Point and Faint source
  catalogues \citep{Moshir90} was used. Column (10): Morphology
  class. For those objects for which the morphological classification
  is controversial, the various possible classes are shown in the
  table (see text for details). Column (11): nuclear optical
  spectroscopic classification. H: HII galaxy, L: LINER, S: Seyfert 2
  and Column (12): References for the spectroscopic data.
  REFERENCES.-- 1: \cite{Veilleux95}. 2: \cite{Duc97}. 3:
  \cite{Veilleux99}. 4: \cite{Kewley01}. 5: \cite{Corbett03}. 6:
  \cite{VandenBroek91}. 7: \cite{Lipari00}. \newline $^a$ The
  morphological classification of these objects has been modified with
  respect to that of Paper I. \newline $^b$ In Paper I these four
  objects were named using the designation in the IRAS Point source
  catalogue. For the work presented here we decided to use the
  designation in the IRAS Faint source catalogue. This includes the
  prefix 'F' and a slightly different sequence of numbers. \newline
  $^c$ These objects, although not presented in Table 1 in Paper I,
  were observed and eventually included in our VIMOS sample. }}
\label{Sample}
%\end{minipage}
\end{table*}

\subsection{Morphological classification}

We use here the simple morphological scheme defined in Paper I. In
that paper a preliminary classification was presented for all the
objects discussed here. Here, we revisit this classification and
describe in detail the criteria and the datasets used (see also Paper
II).

The \cite{Arribas08} scheme is a simplified version of that proposed
by \cite{Veilleux02} for ULIRGs, but with only three morphological
classes instead of the five classes (plus four subclasses) presented
in Veilleux et al. (2002). In particular, the three different
morphological classes considered are

\begin{itemize} 

\item Class 0: objects that appear to be single
isolated objects, with a relatively symmetric morphology and without
evidence for strong past or ongoing interaction.

\item Class 1: objects in a pre-coalescence phase with two well
differentiated nuclei separated a projected distance of D $>$ 1.5
kpc. For these objects, it is still possible to identify the
individual merging systems and their corresponding tidal structures
due to the interaction. The limit of 1.5 kpc was considered taking
into account that theoretical models predict a fast coalescence phase
after the nuclei become closer than that distance
\citep[e.g.][]{Mihos96,Bendo00,Naab06}.

\item Class 2: objects with two nuclei separated a projected distance
of D $\leq$ 1.5 kpc or single nucleus with a relatively asymmetric
morphology suggesting a post-coalescence merging phase. For objects
classified as Class 2, it is not possible to individually identify the
interacting systems.
\end{itemize}

Table 1 shows the morphological classification for all the objects
discussed in this paper. For this classification we have used the
Digitized Sky Survey\footnote{http://archive.stsci.edu/dss (DSS)
images, which are available for {\it all} the sources in our sample,
along with the HST images in the archive for the 21 of the sources for
which these images are available}. At this stage it is worth
mentioning that any morphological classification is to some extent a
matter of personal choice. With that in mind the objects in our sample
were classified independently by three of us. The level of agreement
was substantially high (we agreed except for three objects). Finally,
note that some objects in Table 1 have more than one morphological
classification assigned. These objects were particularly hard to
classify. The preferred morphological classification is indicated in
the first place.

\subsection{Observation data reduction and line fitting}

A detailed description of the observations, data reduction and line
fitting techniques can be found in Paper I. To summarize, the
observations were carried out in service mode during periods 76, 78
and 81 using the Integral Field Unit of the VIMOS instrument
\citep{LeFevre03}, on the Very Large Telescope (VLT), with the
high-resolution mode ``HR-Orange'' (grating GG435). The field of view
(FOV) and the spatial scale in this mode are 27 arcsec $\times$ 27
arcsec and 0.67 arcsec per fiber respectively (i.e., 40 $\times$ 40
fibers, 1600 spectra). A square four-pointing dithering pattern was
used, with a relative offset of 2.7 arcsec (i.e. four spaxels). The
exposure time per pointing was in the range 720 -- 850 seconds and
therefore, the total integration time per galaxy is 2880 -- 3400
seconds.

The data were reduced with a combinations of the pipeline recipe {\it
Esorex} (versions 3.5.1 and 3.6.5) included in the pipeline provided
by ESO, and a series of IDL and IRAF customized scripts. {\it Esorex}
was initially used to perform the basic data reduction (bias
subtraction, flat field correction, spectra tracing and extraction,
correction of fiber and pixel transmission and relative flux
calibration). Then, the four quadrants per pointing were reduced
individually and combined into a single data-cube associated to each
pointing. The final ``super-cube'' per object was generated combining
the four independent dithered pointings, containing a total of 1936
spectra. 

During the reduction process, we observed vertical patterns over the
entire FOV in the cases of IRAS F10567-4310 and IRAS F17138-1017,
which affected only the regions of the spectra with relatively low
S/N. These patterns were still present after the flat field
reduction. With the aim of correcting for this effect we tried to
perform the flat-field correction with flat-field exposures taken on
different observing nights. Unfortunately, this attempt failed and the
vertical patterns are visible in the final data cubes of these two
objects. On the other hand, the data cube of IRAS F12596-1529 showed
``zig-zag'' vertical patterns within a region to the east of the VIMOS
FOV. In this case these vertical patterns are caused by an incorrect
fiber tracing during the reduction process. With the aim of correcting
for this effect we first used the approach of changing some of the
input parameters of the recipe {\it vmifucalib} during the reduction,
such as, for example, the ``MaxTraceRejection'' parameter, which sets
the maximum percentage of rejected positions in fiber spectra
tracing. In addition, we also ran the recipe {\it vmifucalib} with a
fiber identification file and with the ``blind'' fiber identification
method (i.e. without a fiber identification file). Finally we tried to
create the final ``super-cube'' using three of the four independent
dithered pointings, leaving out the pointing for which the presence of
the patterns was more important. Unfortunately, none of these attempts
was entirely successful and the best resulting reconstructed maps for
this galaxy are shown in Figure 1.

The wavelength calibration and the fiber-to-fiber transmission
correction were checked using the [O~{\small I}]$\lambda$6300.3 $\AA$
sky line. Because we are going to concentrate on a wavelength range
around the H${\alpha}$ and the [N~{\small
II}]$\lambda$$\lambda$6548.1,6583.4 $\AA$ emission lines, the
[O~{\small I}]$\lambda$6300.3 $\AA$ sky line is suitable because of
its proximity to these lines. In order to give an estimate of the
absolute wavelength calibration accuracy and the spectral resolution
for the sample as a whole, we first fitted the [O~{\small
I}]$\lambda$6300.3 $\AA$ sky line to a single Gaussian profile for all
spectra of each individual source, obtaining a central wavelength and
a {\it FWHM} value of the sky line for each object. Then, we
calculated the mean of these values, obtaining representative values
for the whole sample of 6300.29 $\pm$ 0.07~\AA\footnote{This is in
good agreement with its actual value \citep[6300.304
$\AA$][]{Osterbrock96}} and 1.80 $\pm$ 0.07~\AA. It is worth
mentioning that the spectral resolution is fairly uniform over the
entire FOV for all the objects in the sample. In addition, [O~{\small
I}]$\lambda$6300.3 $\AA$ flux values were also obtained from the
fit. These values were used to derived and correct flat-field
residuals affecting the fiber-to-fiber flux calibration.

   \addtocounter{figure}{+1}
   \begin{figure*}
   \centering
%   \vskip -3cm
%   {\bf IRAS 13229-2934}\\
  \hspace{-1.25cm}\includegraphics[width=16.0cm]{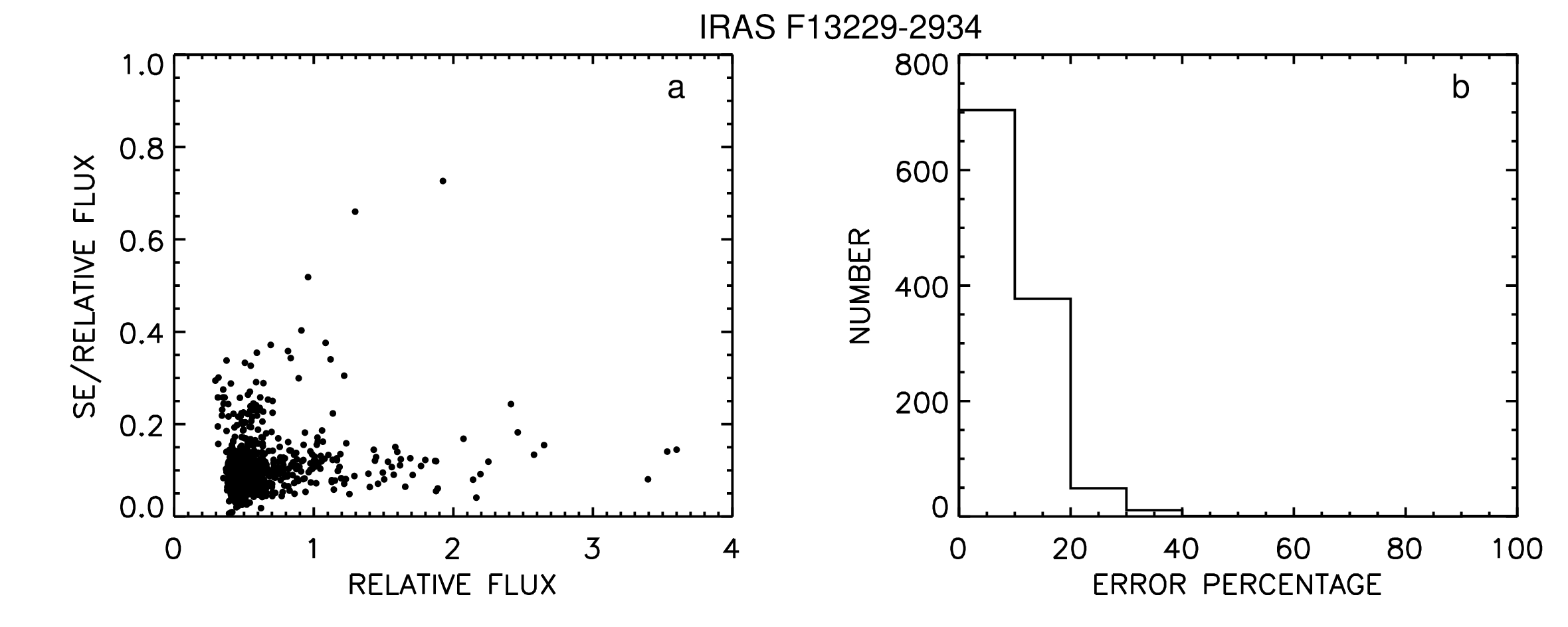}
   \caption{Example of the test performed to estimate the error
   percentage associated to the relative flux for the particular case
   of IRAS 13229-2934. Left: standard error of the mean (SE) plotted
   against the mean value of the flux per spaxel. Right: percentage
   error distribution. We find that the typical error associated to
   the flux in this case is 9.8\%.}
   \label{flux_error}
   \end{figure*}

Because this study is focused on the ionized gas and stellar
structure, it is important to assess the uncertainty associated to the
relative flux after the calibration. With that in mind, we decided to
perform the following test for all the galaxies in the sample. We
first measured for each spaxel and individual pointing the median of
the flux within a ``clean'' (i.e. with no emission/absorption lines)
region of the continuum. We then calculated the mean and the standard
error of the mean (SE) for the values associated to the four dither
pointings. The SE was calculated assuming a normal distribution,
i.e. SE = $\sigma$/$\sqrt{N}$, where $\sigma$ and N are the standard
deviation and the sample size respectively. In this case N = 4,
corresponding to the four flux values considered. At this stage, we
were able to associate a flux percentage error to each point observed
in the selected area. We define as the typical percentage error
associated to the flux for a certain object as the sum of all the flux
percentage errors obtained for each spaxel in the frame divided by the
total number of spaxels considered for the analysis (i.e. leaving out
the {\it bad} spaxels).

Column 2 in Table 2 presents the typical percentage errors for all
galaxies in our sample. We find values ranging from of 3.7\% in the
case of IRAS F10015$-$0614, to 21\% for IRAS F10038$-$3338, with a
mean and a median value for the whole sample of 12\% and 11\%
respectively. Figure 2 shows the outputs of the test for the
particular case of IRAS 13229-2934 (NGC 5135). Figure
\ref{flux_error}a shows the ratio between the standard error of the
mean (SE) and the mean value of the flux per spaxel plotted against
the mean value of the flux per spaxel. Figure \ref{flux_error}b is an
histogram showing the flux percentage error distribution. Is obvious
from this figure that the majority of the spaxels have a typical
percentage error lower than 20\%. We find that the typical error
associated to the flux in this case is 9.8\%.

The emission lines from each galaxy were analyzed by fitting them to
Gaussian profiles with the MPFITEXPR code, which was implemented by
C.B. Markwardt in the IDL environment\footnote {available at
http://purl.com/net/mpfit}. This algorithm allows us to fix wavelength
differences and line intensity ratios according to atomic parameters
when adjusting multiple emission lines (e.g. the
H$\alpha$-\textsc{[N\,ii]} complex). As a first approach, we fitted
automatically all lines to single Gaussian profiles, which produced
adequate fits in most cases. However, for certain regions of the
galaxy, a multi-component fit with two or even three components per
line was required in order to adequately fit the data. This was the
case for 13 of the 38 sources included in this study. These multiple
components are frequently concentrated in the nuclear regions of the
galaxies and only extend few spaxels. The exceptions are IRAS
F04315-0840, IRAS F14544-4255(E), and IRAS F23128-5919, where double
components are visible extending over 12 kpc through the body of the
galaxy. It is worth mentioning that the spatial identification of the
different components can be done unambiguously because they usually
have rather different kinematic properties. A detailed discussion of
the properties of these secondary components is beyond the scope of
this paper and will be addressed in future publications. For those
cases with several kinematic distinct components, the H${\alpha}$ and
the EW maps in Fig. 1 refer to the systemic component, whichh is
usually extended over the entire body of the system. By default the
same line width was considered for all lines of a spectrum. For each
emission line we ended up with the following information: central
wavelength, FWHM, and flux intensity. Finally, we used these magnitudes
and the spatial position of each spaxel to generate an image (a {\it
map}) that can be treated as a standard image of a galaxy.

\section{The atlas of VIMOS continuum, H${\alpha}$ 
and H${\alpha}$-EW maps}

Figure 1 shows DSS or HST images (for the galaxies with HST images
available in the archive), as well as the VIMOS images for all
galaxies in our sample. The second panel shows the continuum images
created by simulating a filter covering the spectral range 6390 --
6490 \AA~(rest frame). When generating these images, we selected a
2${\sigma}$ lower cut as a threshold to distinguish galaxy from
background, where ${\sigma}$ was the root mean square (rms) of the
flux in a region of our FOV free from galaxy emission. The 2${\sigma}$
lower cut was selected based on a detailed comparison between the
VIMOS continuum images and the corresponding DSS and/or HST images of
the galaxies (i.e., after applying a 2${\sigma}$ lower cut to our
VIMOS continuum images these adequately traced the structure observed
in the SDSS and/or HST images).

The third panel in the figure shows the ionized gas emission from the
galaxies, as traced by H${\alpha}$. The lower limit for the
H${\alpha}$ images was selected on the basis of the fit to the
emission line. Only those regions were the S/N ratio was sufficiently
high to perform the fit were used for the figure. The regions with
negligible or no signal and some {\it bad} spaxels have been cleaned
using either customized IDL routines or the routine IMEDIT in {\it
IRAF}. In addition, the forth panel in Fig. 1 shows the H${\alpha}$
equivalent width (EW) in units of Angstroms.

As mentioned in Sect. 2.3, the data cubes of IRAS F10567-4310 and IRAS
F17138-1017 showed vertical patterns over the entire FOV, which are
only important if the S/N is low. Therefore, although they affect the
morphology of the continuum (and the corresponding H${\alpha}$-EW)
images, they have no effect on the H${\alpha}$ emission maps shown in
Fig. 1. On the other hand, the vertical patterns observed in the case
of IRAS F12596-1529 are caused by an incorrect fiber tracing during
the reduction process (see Sect. 2.3 for details), and affect both the
continuum and the H${\alpha}$ images.

The continuum images mainly trace the stellar light from the galaxy,
while the H${\alpha}$ images trace the star-formation activity, or the
presence of ionizing shocks, or an AGN. In this context, Paper II
shows that some of the objects in our sample have line ratios
consistent with ionization by shocks, mainly concentrated in the
extended regions (see the paper for details). However, the study
presented in Paper II excludes the nuclear regions of the sources,
which can be potentially contaminated by AGN emission. In order to
give an idea of the importance of the AGN emission among the objects
in our sample, Col.  9 in Table 1 shows the nuclear spectroscopic
classification (when available) for the objects in our sample. As
shown in the table, only $\sim$ 17\% of the objects in our sample with
nuclear, optical spectroscopic classification (5 of 29) are classified
as Seyfert galaxies, i.e. the AGN contributes significantly at optical
wavelengths. In addition, one object is classified as LINER and five as
ambiguous (HII/Sy, HII/LINER or HII/LINER/Sy). For these objects, the
AGN contribution to the optical emission is less certain. Overall,
$\sim$ 66\% of the objects in our sample with nuclear, spectroscopic
classification in the optical (19 of 29) are classified as HII-like
galaxies. Therefore, although AGN emission might still contribute to
the optical light from these sources, their continuum and nuclear
H${\alpha}$ emissions are dominated by stellar light and
recent/ongoing star-formation activity respectively.

\section{Morphology of the stellar and ionized gas emissions 
and EW(H${\alpha}$) maps.}

The continuum and the H${\alpha}$ emission images show morphologies
that are substantially different for the overwhelming majority of the
sources. The H${\alpha}$ images usually reveal clumpy, extended
structures that are not visible in the corresponding continuum
images. For example, although ring structures are not observed in the
continuum images, they are clearly visible in the H${\alpha}$ images
for some LIRGs in our sample, such as IRAS F01159-4443S, IRAS
F06076-2139, IRAS F11506-3851, IRAS F12043-3140S and IRAS
F12115-4656. The presence of such rings for some of these LIRGs has
already been reported in the past by other authors
\citep[e.g.][]{Alonso-Herrero02,Hattori04,Rampazzo05,Alonso-Herrero06}.
These structures are usually relatively symmetric and centered on the
nucleus of the systems. In most cases they are located in the
circumnuclear region, although in the case of IRAS F11255-4120 the
ring extends up to $\sim$4 kpc away from the nucleus of the galaxy. In
addition, the H${\alpha}$ images of some sources also show other
extended tidal structures such as bridges (e.g. IRAS F01159-4443),
tidal tails (e.g. IRAS F10409-4556) or spiral arms (e.g. F01341-3735N,
IRAS F10567-4310, IRAS F21453-3511 or IRAS 07027-6011N), that extend
up to few kpc ($\sim$3 -- 4 kpc) from the nuclear region of the
galaxy. Is also interesting the case of IRAS F01341-3735S
(ESO-297-G012), where the H${\alpha}$ emission extends along the
galaxy minor axis. As suggested by \cite{Dopita02}, it is possible
that in this galaxy the nuclear starburst is blowing out gas in the
polar direction, similar to the case of M82. Overall, the different
morphologies between the stellar and ionized gas emission are
explained in terms of bright, extranuclear star-formation activity
along the tidal tails or the spiral arms and/or ionizing shocks in the
extra-nuclear extended regions (Colina et al. 2005, Paper II, and
references therein).

It is also interesting to study the morphology of the EW images shown
in Fig. 2. In some cases, these images help to trace the extended
structures seen in H${\alpha}$ emission. For example, a ring of star
formation is inferred from the H${\alpha}$ image in the case of IRAS
12115-4656, but clearly emerges in the corresponding EW image of the
galaxy. In general, these images often show regions with high (EW $>$
100 $\AA$) values that are associated to the large-scale structures
seen in the H${\alpha}$ images (e.g. IRAS F06035-7102, IRAS
F11255-4120 or IRAS F21453-3511). In Paper II we carried out a study
of the different ionization mechanisms for the LIRGs within our
sample. We found log {[N$_{\rm II}$]$\lambda$6583/H$\alpha$} values
substantially lower than $-$~0.2 in the majority of the cases,
consistent with photoionzation by stars (see Paper II for
details). Using the Leitherer et al. 1999 models for solar metalicity,
instantaneous starburst and Salpeter 1955 IMF, these large EW values
correspond to stellar ages of t $\lsim$ 6 Myr.

\subsection{Implications for long-slit spectroscopic studies}

Up to date, most of the spectroscopic studies of LIRGs and ULIRGs have
used long slit observations
\citep[e.g.][]{Veilleux99,Kewley01,Corbett03,Rodriguez-Zaurin09}. In
these studies, the position of the slit is usually selected to cover
the brightest regions of the galaxies observed in continuum
emission. However, as discussed before, the H${\alpha}$ morphologies
are substantially different than those of the continuum for the
majority of the objects in our sample.

\begin{figure}
\hspace{-0cm}\includegraphics[width=8.0cm]{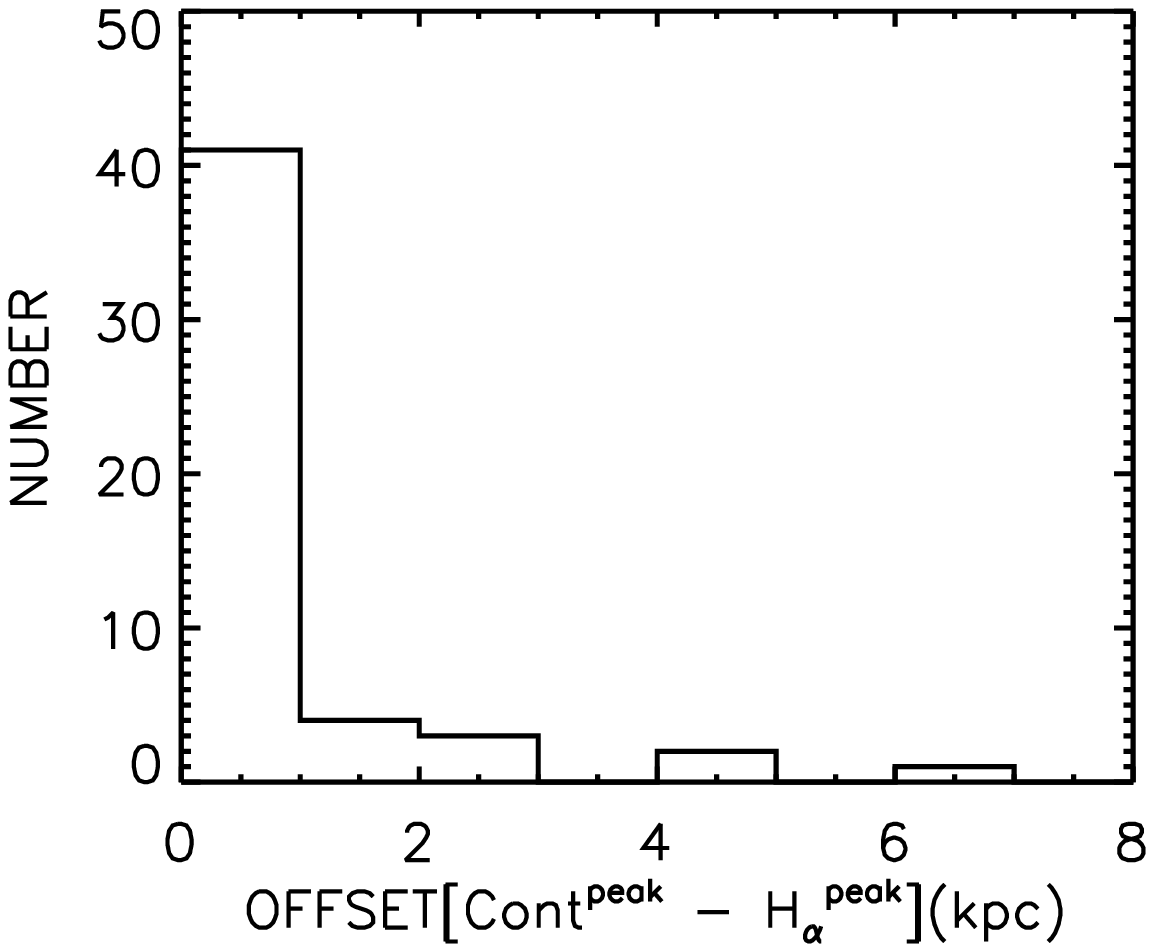}
\caption{Histograms showing the distribution of the offsets, in kpc,
between the high surface brightness peak of continuum and the
H${\alpha}$ emission for the objects shown in Fig. 1.}
\label{offsets_hist}
\end{figure} 

Our IFS dataset allows us to investigate some caveats associated to
long slit spectroscopic studies. For example, positional uncertainties
of the slit may lead to a misclassification of the optical spectrum of
the sources \citep[e.g.][]{Arribas00}. With that in mind, we first
compare the location of the surface brightness peaks in the continuum
and the H${\alpha}$ maps. The surface brightness peaks are determined
using the {\it IRAF} routine {\it PHOT} with a centroid algorithm,
that determines the location of the peaks computing the intensity
weighted means of the marginal profiles in x and y. Column 3 in Table
2 shows the offsets, in kpc, between the peaks of the continuum and
the H${\alpha}$ emission (offset$^{cont}_{H\alpha}$), and Figure
\ref{offsets_hist} is an histogram showing the distribution of such
offsets. For the purpose of the figure the individual nuclei in double
or triple systems have been considered separately. The figure shows
that for the majority of the objects (41 of the 51 individual sources
considered) the peaks of the stellar and the ionized gas emissions are
separated by less than 1 kpc. The median offset$^{cont}_{H\alpha}$
value for the whole sample is 0.2 kpc. Note that 10 sources ($\sim$
20\%) show peak separations higher than 1 kpc. Some extreme cases are
the LIRG IC564 and the ULIRG IRAS 21130-4446. The former has the peak
of the H${\alpha}$ emission $\sim$6 kpc to the east from peak of the
continuum emission. In the case of IRAS 21130-4446, the peak of the
continuum is located to the north of the system, while the maximum
H${\alpha}$ emission is shifted $\sim$4 kpc towards the south,
coinciding with the central region of the system.

In addition, we can also estimate the amount of flux that falls
outside the slit for the typical slit-width values used in previous
long-slit studies. With that in mind we used our H$\alpha$ images and
simulated a slit of a width of 2 arcsec at parallactic angles (PA)
0$^{\circ}$ and 90$^{\circ}$, crossing the galaxy through its center,
defined as the peak of the continuum emission. The fraction of the
observed H${\alpha}$ flux within the slit is shown for each galaxy in
Col. (3) in Table 3. For those objects with double nucleus structure,
for which is not possible to study the individual sources separately
(e.g. IRAS F06035-7102), the slit is centered on the brightest nucleus
in the continuum. If we concentrate in the slit PA that includes the
larger fraction of the observed H${\alpha}$ emission for each galaxy,
we find that the percentage of the flux outside the slit is in the
range 17-90\%, with mean and median values of 59\% and 61\%.

This result is important when addressing the long-standing issue of
whether or not optical observations can penetrate to the main power
source of the IR luminosity. If we aim to compare, for example the
star-formation rates (SFRs) obtained using the H$\alpha$ and IRAS IR
luminosities, our results suggest that using the H$\alpha$
luminosities derived from long-slit studies would substantially
underestimate the value of the optical SFRs. We will come back to this
in the following section.

\begin{table*}[!h]
%{\tiny
\centering
%\begin{minipage}{140mm}
%\resizebox{0.75\textwidth}{!}{
\begin{tabular}{lc|c|cc|cc}
\hline
\hline
IRAS & Percentage & offset$^{cont}_{H\alpha}$ &{\it C$^{cont}_{2kpc}$} (6 kpc)& {\it C$^{H\alpha}_{2kpc}$} (6 kpc) & {\it C$^{cont}_{2kpc}$} (all FOV)& {\it C$^{H\alpha}_{2kpc}$} (all FOV)\\
&error & (kpc)&&&&\\
(1) & (2) & (3) & (4) & (5) & (6) & (7)\\
\hline
F01159$-$4443N	&5.8  &0.0 & -            &0.72$\pm$0.06 & -             & 0.61$\pm$0.05 \\   
F01159$-$4443S  &5.8  &0.1 & -            &              & -             &  -   \\
F01341$-$3735N	&12.7 &0.1 &0.27$\pm$0.05 &0.52$\pm$0.09 & 0.16$\pm$0.03 & 0.39$\pm$0.07 \\  
F01341$-$3735S	&8.8  &0.2 &0.42$\pm$0.05 &0.81$\pm$0.10 & 0.41$\pm$0.05 & 0.81$\pm$0.10 \\  
F04315$-$0840	&15.1 &0.1 &0.37$\pm$0.08 &0.61$\pm$0.13 & 0.31$\pm$0.07 & 0.62$\pm$0.13 \\  
F05189$-$2524   &10.1 &0.1 &0.41$\pm$0.06 &0.56$\pm$0.08 & 0.29$\pm$0.04 & 0.56$\pm$0.08 \\  
F06035$-$7102E  &10.4 &1.0 &0.17$\pm$0.03 &0.24$\pm$0.03 & 0.04$\pm$0.01 & 0.06$\pm$0.01 \\  
F06035$-$7102W  &10.4 &0.4 & -            &              & -             &  -   \\
F06076$-$2139N	&6.9  &0.0 &0.24$\pm$0.02 &0.75$\pm$0.07 & 0.10$\pm$0.01 & 0.28$\pm$0.03 \\  
F06076$-$2139S  &6.9  &0.0 & -            &  -           &  -            &  -   \\
F06206$-$6135   &7.6  &0.8 &0.11$\pm$0.01 &0.21$\pm$0.02 & 0.06$\pm$0.01 & 0.10$\pm$0.01 \\  
F06259$-$4708N  &11.6 &0.2 &0.25$\pm$0.04 &0.48$\pm$0.08 & 0.17$\pm$0.03 & 0.27$\pm$0.04 \\  
F06259$-$4708C  &11.6 &0.7 &0.28$\pm$0.05 &0.29$\pm$0.05 & 0.22$\pm$0.04 & 0.46$\pm$0.08 \\  
F06259$-$4708S  &11.6 &0.8 &0.27$\pm$0.04 &0.33$\pm$0.05 & 0.17$\pm$0.03 & 0.22$\pm$0.04 \\  
F06295$-$1735	&9.6  &1.1 & -            &0.30$\pm$0.04 & -             & 0.16$\pm$0.02 \\  
F06592$-$6313	&18.0 &0.0 &0.40$\pm$0.10 &0.72$\pm$0.14 & 0.37$\pm$0.10 & 0.70$\pm$0.17 \\  
F07027$-$6011N	&9.6  &0.0 &0.42$\pm$0.06 &0.28$\pm$0.15 & 0.33$\pm$0.04 & 0.23$\pm$0.03 \\  
F07027$-$6011S	&14.5 &0.2 &0.37$\pm$0.08 &0.55$\pm$0.04 & 0.33$\pm$0.07 & 0.55$\pm$0.11 \\  
F07160$-$6215	&17.7 &0.2 &0.29$\pm$0.07 &0.60$\pm$0.11 & 0.29$\pm$0.07 & 0.60$\pm$0.15 \\  
08355$-$4944	&14.0 &0.1 &0.36$\pm$0.07 &0.58$\pm$0.15 & 0.29$\pm$0.06 & 0.55$\pm$0.11 \\  
08424$-$3130NE	&12.4 &0.1 & -            &  -           & -             &  -   \\  
08424$-$3130SW  &12.4 &0.1 &-             &  -           & -             &  -   \\
F08520$-$6850   &12.0 &1.1 &0.17$\pm$0.03 &0.35$\pm$0.06 & 0.06$\pm$0.01 & 0.26$\pm$0.04 \\  
09022$-$3615    &14.4 &0.1 &0.22$\pm$0.04 &0.26$\pm$0.05 & 0.14$\pm$0.03 & 0.23$\pm$0.05 \\  
F09437$+$0317N	&17.4/8.7& 6.3 &  -            & -            & -             &   -  \\  
F09437$+$0317S	&19.0 &4.1 &0.17$\pm$0.05 &0.11$\pm$0.03 & 0.11$\pm$0.03 & 0.08$\pm$0.02 \\ 
F10015$-$0614	&3.7  &0.1 & 0.29$\pm$0.02 &0.26$\pm$0.01 & 0.23$\pm$0.01 & 0.16$\pm$0.01 \\  
F10038$-$3338	&21.0 &0.1 &0.19$\pm$0.06 &0.52$\pm$0.15 & 0.15$\pm$0.04 & 0.52$\pm$0.15 \\  
F10257$-$4339	&6.3  &0.0 &0.26$\pm$0.02 &0.30$\pm$0.03 & 0.26$\pm$0.02 & 0.30$\pm$0.03 \\  
F10409$-$4556	&15.8 &0.2 &0.26$\pm$0.06 &0.21$\pm$0.05 & 0.26$\pm$0.06 & 0.12$\pm$0.03 \\  
F10567$-$4310	&9.4  & -  & -            &0.53$\pm$0.07 & -             & 0.25$\pm$0.03 \\  
F11255$-$4120	&8.0  &0.0 &0.25$\pm$0.03 &0.37$\pm$0.04 & 0.16$\pm$0.02 & 0.24$\pm$0.03 \\  
F11506$-$3851	&18.6 &0.8 &0.54$\pm$0.14 &0.48$\pm$0.12 & 0.54$\pm$0.14 & 0.48$\pm$0.12 \\  
F12043$-$3140N	&8.2  &0.1 &0.39$\pm$0.04 &0.75$\pm$0.08 & 0.35$\pm$0.04 & 0.75$\pm$0.08 \\  
F12043$-$3140S	&8.2  &2.2 &0.19$\pm$0.02 &0.20$\pm$0.02 & 0.18$\pm$0.02 & 0.20$\pm$0.02 \\  
12115$-$4656	&7.1  &0.6 &0.28$\pm$0.03 &0.29$\pm$0.03 & 0.23$\pm$0.02 & 0.29$\pm$0.03 \\  
12116$-$5615	&18.8 &0.2 &0.32$\pm$0.08 &0.75$\pm$0.18 & 0.29$\pm$0.07 & 0.75$\pm$0.18 \\  
F12596$-$1529E	&9.6  &0.5 &  -           & -            & -             & -    \\  
F12596$-$1529W  &9.6  &0.0 &  -           & -            & -             & -    \\
F13001$-$2339	&9.3  &0.2 &0.20$\pm$0.03 &0.51$\pm$0.06 & 0.13$\pm$0.02 & 0.48$\pm$0.06 \\  
F13229$-$4255	&9.8  &0.1 &0.45$\pm$0.06 &0.69$\pm$0.10 & 0.45$\pm$0.06 & 0.69$\pm$0.10 \\  
F14544$-$4255E	&17.2 &2.4 &0.33$\pm$0.08 &0.51$\pm$0.12 & 0.28$\pm$0.07 & 0.18$\pm$0.04 \\  
F14544$-$4255W	&10.8 &0.3 &0.28$\pm$0.04 &0.70$\pm$0.11 & 0.19$\pm$0.03 & 0.70$\pm$0.11 \\  
F17138$-$1017	&16.1 & -  &  -           &0.64$\pm$0.14 & -             & 0.64$\pm$0.14 \\  
F18093$-$5744N	&10.5 &0.5 &0.27$\pm$0.04 &0.32$\pm$0.05 & 0.24$\pm$0.03 & 0.32$\pm$0.05 \\  
F18093$-$5744C	&15.1 &1.5 &  -           &0.80$\pm$0.17 & -             & 0.79$\pm$0.17 \\  
F18093$-$5744S	&9.4  &0.1 &0.28$\pm$0.04 &0.39$\pm$0.05 & 0.27$\pm$0.03 & 0.39$\pm$0.05 \\  
F21130$-$4446   &10.6 &4.2 &0.15$\pm$0.02 &0.26$\pm$0.04 & 0.06$\pm$0.01 & 0.05$\pm$0.01 \\  
F21453$-$3511	&11.2 &0.0 &0.21$\pm$0.03 &0.27$\pm$0.04 & 0.21$\pm$0.03 & 0.27$\pm$0.04 \\  
F22132$-$3705	&5.0  &2.0 &0.17$\pm$0.01 &0.16$\pm$0.01 & 0.11$\pm$0.01 & 0.15$\pm$0.01 \\  
F22491$-$1808   &20.4 &0.4 &0.17$\pm$0.05 &0.17$\pm$0.05 & 0.06$\pm$0.02 & 0.12$\pm$0.03 \\ 
F23128$-$5919N  &8.7  &0.4 &0.18$\pm$0.02 &0.30$\pm$0.04 & 0.08$\pm$0.01 & 0.17$\pm$0.02 \\
F23128$-$5919S  &8.7  &0.1 &  -           &  -           & -             & -    \\
\hline 
\hline
\end{tabular}
\caption{Typical flux-percentage error, offset$^{cont}_{H\alpha}$ and
C$_{2kpc}$ values for the (U)LIRGs in the VIMOS sample. Column (1):
IRAS name. Column (2): Typical flux percentage error for all the
galaxies in our sample. Note that for some multiple systems two or
even three VIMOS pointings were used to cover the emission from each
individual galaxy of the system (see Sect. 3 and Fig. 1 for
details). For these objects, typical flux percentage errors were
estimated for each individual pointing (e.g. IRAS F18093$-$5744). In
the case of IRAS F09437+0317 two pointings were used to cover most of
the emission from the northern source. The typical percentage error
for each pointing is indicated in the table for this source. Column
(3): the offsets in kpc between the peaks of the continuum and the
H${\alpha}$ emissions. Column (4): the concentration {\it C$_{2kpc}$}
of the continuum emission for the galaxies in our sample defining as
``total flux'' the flux within the same physical region (6 kpc
$\times$ 6 kpc) for all galaxies in our sample. Column (5) : same as
Col. (1) but for the H${\alpha}$ images. Column (3): same as Col. (1)
but using the entire FOV of our VIMOS images. Column (4): same as
Col. (2) but using the entire FOV of our VIMOS images.\newline The
formal error associated to the {\it C$_{2kpc}$} values in the table
has been calculated using the typical percentage errors given in
Col. (2).\newline For IRAS F08424-3130, IRAS F12596-1529 and IRAS
F09437$+$0317 (IC564) it is not possible to adequately measure {\it
C$_{2kpc}$} (see text for details). In addition, the low S/N of the
continuum images in the cases of IRAS F01159-4443 and IRAS
F06295-1735, important contamination by a bright star in the field for
the central system of IRAS F18093-5744 and the presence of vertical
patterns in the cases of IRAS F10567-4310 and IRAS F17138-1017,
prevents any attempt to measure {\it C$_{2kpc}$} for the continuum
images of these five sources.}%}
\label{C2kpc}
\end{table*}

\subsection{The concentration of the continuum and the H${\alpha}$ emissions}

With the aim of further investigating the morphologies of the images
shown in Fig. 2 we defined the parameter {\it C$_{2kpc}$} as the
ratio of the flux contained within an aperture of 2 kpc diameter
centered on the nucleus of the object, and the total flux from the
galaxy within the same physical region for all the galaxies in our
sample (6 kpc $\times$ 6 kpc):

\begin{equation}
C_{2kpc}({\rm 6kpc})=  \frac{f_{2kpc}}{f_{6kpc}}.  
\label{C_2kpc}
\end{equation}

The objects in our sample expand over a wide range of redshifts, and
therefore, the physical scale covered by the 27 arcsec $\times$ 27
arcsec VIMOS FOV can be substantially different from one object to
another. In order to avoid possible biases in our results caused by
such an effect, we measured C$_{2kpc}$ within the same physical region
for all the objects in our sample used for this study. We selected a
region of 6 kpc $\times$ 6 kpc, which is the FOV of IRAS F10257-4339
and the minimum FOV covered among the objects in our sample. The {\it
C$_{2kpc}$} values for the continuum ({\it C$^{cont}_{2kpc}$}) and the
H${\alpha}$ ({\it C$^{H\alpha}_{2kpc}$}) images using a physical
region of 6 kpc $\times$ 6 kpc are shown in Col. 4 and 5 in Table 3
and used for Fig. 4 and 5. However, for some of the more distant
and extended galaxies, a 6 kpc $\times$ 6 kpc region covers a small
fraction of the total extension from the galaxy within our VIMOS
FOV. Therefore, with the aim of using the full information within our
dataset we also measured C$_{2kpc}$ using the total,integrated flux
within the entire VIMOS FOV:

\begin{equation}
C_{2kpc} (all~FOV) = \frac{f_{2kpc}}{f (all~FOV)}.
\label{C_2kpc}
\end{equation}

The results are shown in Col. 6 and 7 in Table 2. As seen in Fig.  1,
the extended emission for an important fraction of our objects
($\sim$65\%) is not entirely covered by the entire VIMOS
FOV. Therefore, even the {\it C$_{2kpc}$} values obtained using the
VIMOS FOV are, in general, upper limits. On the other hand, because we
have not corrected for reddening effects, if the extinction is much
higher on the nuclear regions than in the external regions, {\it
C$_{2kpc}$} would be lower limits.

The two nuclei of IRAS 08424-3130 and the brighter, western source of
IRAS~F12596-1529 fall close to the edges of the VIMOS images and it is
not possible to measure the flux within an aperture of 2 kpc in
diameter. In addition, in the case of IC 564 (the northern galaxy of
the double system IRAS F09437$+$0317), two pointings were used during
the observations of this galaxy. Therefore, it is not possible to
adequately estimate {\it C$_{2kpc}$}. No values for the continuum and
the H${\alpha}$ images are presented for these three objects in the
table. On the other hand, the low S/N of the IRAS F01159-443 and IRAS
F06295-1735 continuum images, the vertical patterns in the cases of
IRAS F10567-4310 and IRAS F17138-1017 and important contamination by
stars in the field in the case of IRAS F18093-5744 prevents any
attempt to measure {\it C$^{cont}_{2kpc}$} for these five sources.

Note that for some objects that are morphologically classified as 1
(interacting galaxies), it is not possible to infer {\it C$_{2kpc}$}
for the individual sources separately. Then the {\it C$_{2kpc}$}
values presented in the table are obtained centering the 2 kpc
aperture on the brightest nucleus in the continuum. These sources are
the LIRGs IRAS F01159-4443, IRAS F06076-2139 and IRAS F08520-6850 and
the ULIRGs 06035-7102, IRAS F06206-2139, IRAS F22491-1808 and IRAS
F23128-5919. Although the 2 kpc aperture is centered on the brightest
nucleus, a substantial fraction of the emission from these systems
still falls outside the aperture, more so for the ULIRGs. Possible
biases in the general results owing to this effect will be discussed
below.

\begin{figure}[-h]
\hspace*{-0.5cm}\includegraphics[width=9.0cm]{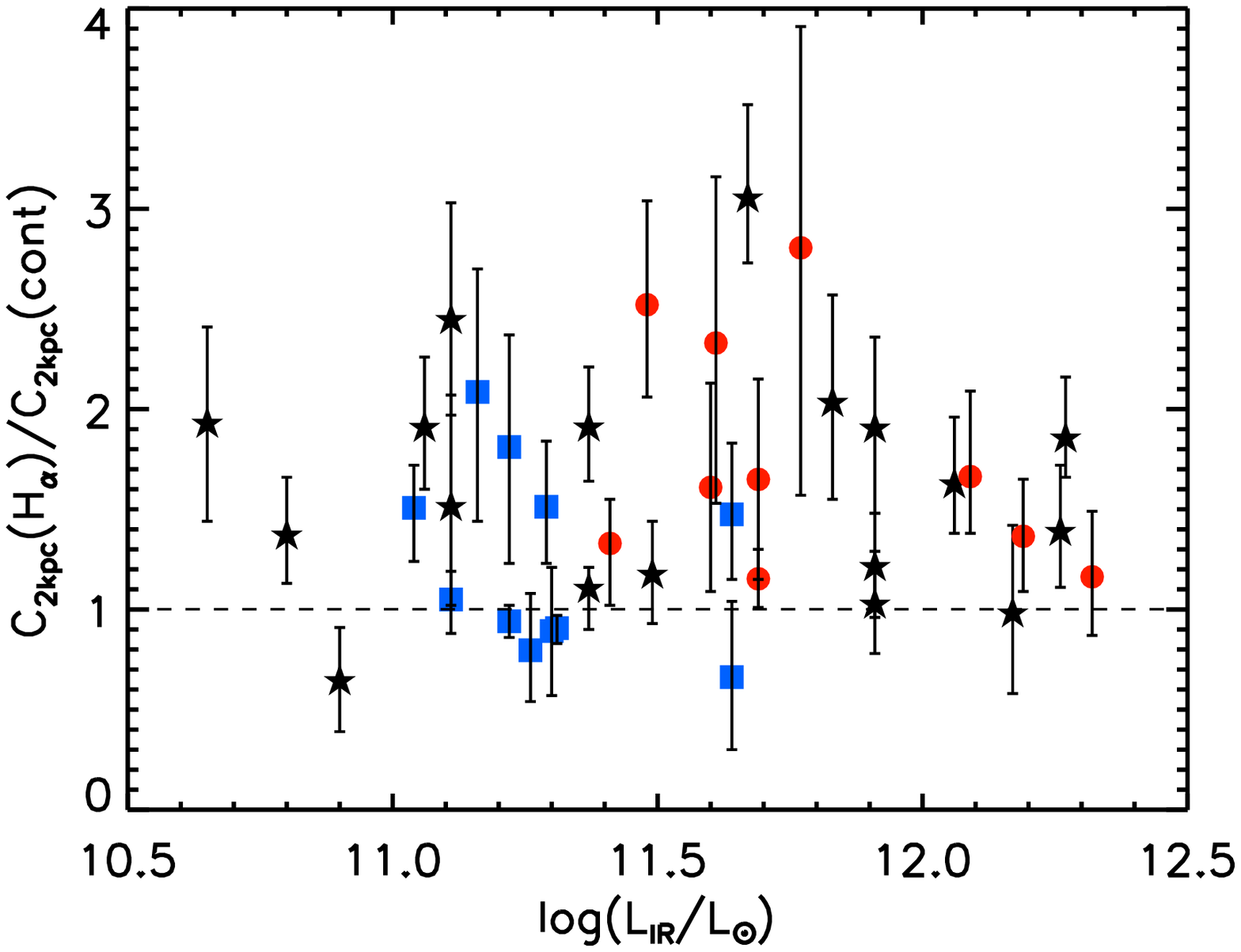}
\caption{Ratio between the concentration {\it C$_{2kpc}$} for the
H${\alpha}$ and the continuum emission plotted against the log of the
infrared luminosity L$_{\rm IR}$. Note that there are three points
with L$_{\rm IR} <$ 10$^{11}$ L$_{\odot}$. These are IRAS F01341-3735N
(ESO-297-G011), IRAS F09437-0317S (IC563) and IRAS F18093-5744S
(IC4689). For these multiple systems at least one of the individual
galaxies falls outside the LIRG luminosity range. The points are
labeled on the basis of our morphological classification. Type 0: blue
squares. Type 1: black stars. Type 2: red circles.}
\label{Havscont}
\end{figure} 

In the first place, we compared the values obtained for the continuum
and the H${\alpha}$ images. Figure \ref{Havscont} shows the ratio
between {\it C$^{H\alpha}_{2kpc}$} and {\it C$^{cont}_{2kpc}$}. For
IRAS F01341-3735, IRAS F09437-0317 and IRAS F18093-5744, the entire
system has a L$_{\rm IR} >$ 10$^{11}$ L$_{\odot}$. However, the
results of \cite{Surace04} show that at least one of the individual
galaxies of these multiple nuclei systems falls outside the LIRG
luminosity range, and therefore, there are points in the figure below
the LIRG threshold (log (L$_{\rm IR}$/L$_{\odot}$) = 11). Figure
\ref{Havscont} shows that the fraction of the H$\alpha$ emission from
the central 2 kpc is higher than that of the continuum for 85\% of the
sources used for the figure (33 of 39). The {\it C$^{cont}_{2kpc}$}
values are in the range of 0.11 - 0.54 with a median value of 0.27,
while values in the range of 0.11 - 0.81 are found for {\it
C$^{H\alpha}_{2kpc}$}, with a median value of 0.39. The six objects
with {\it C$^{H\alpha}_{2kpc}$}/{\it C$^{cont}_{2kpc}$} values lower
than 1.0 are IRAS F07027-6011N, IRAS F09437-0317S, IRAS F10015-0614,
IRAS F10409-4556, IRAS F11506-3851 and IRAS
F22132-3705. Interestingly, the majority of these sources show
H${\alpha}$ morphologies that are substantially different from those
of the continuum. For example, the H$\alpha$ image of IRAS
F07027-6011N shows a chain of knots embedded within the main body of
the galaxy that is separated $\sim$3~kpc from the nucleus and extends
towards the southwest of the galaxy. Comments on this and the other
sources can be found in appendix A.

\begin{figure}
\hspace*{-0.5cm}\includegraphics[width=9.0cm]{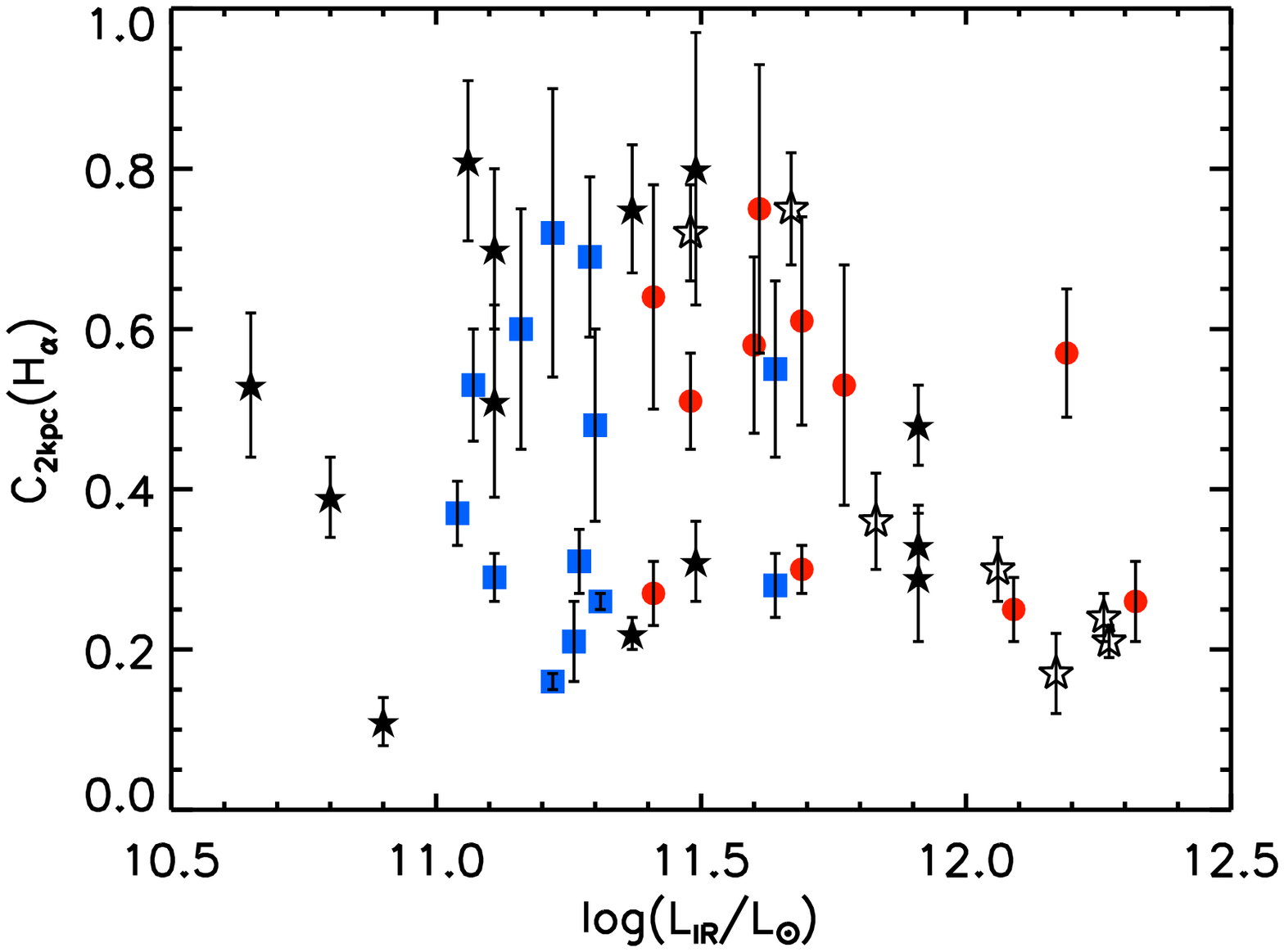}\\
\caption{Concentration {\it C$^{H\alpha}_{2kpc}$} plotted against the
log of the IR luminosity (L$_{\rm IR}$).  Symbols are the same as in
Fig. \ref{Havscont}. The error bars correspond to the formal errors
shown in Table 2. There are seven open stars that correspond to those
objects classified as 1, for which is not possible to infer {\it
C$^{H\alpha}_{2kpc}$} for the individual sources separately. See the
text for further details.}
\label{C_plots}
\end{figure} 

On the other hand, note that $\sim$52\% of the individual sources used
for this study (23 of 37) have more than half of their H${\alpha}$
emission outside the central 2 kpc (this becomes 62\% when using the
entire VIMOS FOV). In principle, this result emphasizes the importance
of the extended star-formation activity in (U)LIRGs \citep[see
also][]{Alonso-Herrero06,Garcia-Marin09a,Garcia-Marin09b}. However, it
is worth mentioning that we have not corrected for reddening effects,
which are usually more important towards the nuclear regions of these
objects. Because reddening effects are expected to be important for
objects such as LIRGs and ULIRGs, it is likely that the fraction of
the H${\alpha}$ emission within the central 2 kpc of the sources is,
once corrected for extinction, larger than the values shown in Table
2. Indeed, ground-based mid-IR studies, which are less affected by
extinction, indicate that at least in ULIRGs without evidence for an
AGN, the star formation activity is concentrated within the central
kpc \citep{Soifer00,Soifer01}, while in a large fraction of LIRGs, the
MIR emission appears to be extended over a few kpc
\citep{Diaz-Santos08}

\subsection{{\it C$^{H\alpha}_{2kpc}$} vs the infrared luminosity and the morphological class}

We investigate in this section whether the more luminous objects have
the star-formation activity more concentrated towards the nuclear
regions. Figure \ref{C_plots} shows {\it C$^{H\alpha}_{2kpc}$} plotted
against the log of infrared luminosity of the sources (L$_{\rm
IR}$). The open stars correspond to those objects morphologically
classified as 1, for which is not possible to infer {\it
C$^{H\alpha}_{2kpc}$} separately for the individual sources. Note that
the four ULIRGs in our sample classified as 1, located in the bottom
right of the figure, fall within this group. As described before, in
these cases we center the 2 kpc aperture on the brightest nucleus, but
most of the emission still falls outside the aperture. If we
concentrate in the figure on the individual sources, no clear evidence
for the presence of correlations between {\it C$^{H\alpha}_{2kpc}$}
and L$_{\rm IR}$ is found. The same conclusion is reached when using
the {\it C$^{H\alpha}_{2kpc}$} obtained using the entire VIMOS FOV.

\begin{figure}
\hspace*{-0.5cm}\includegraphics[width=9.0cm]{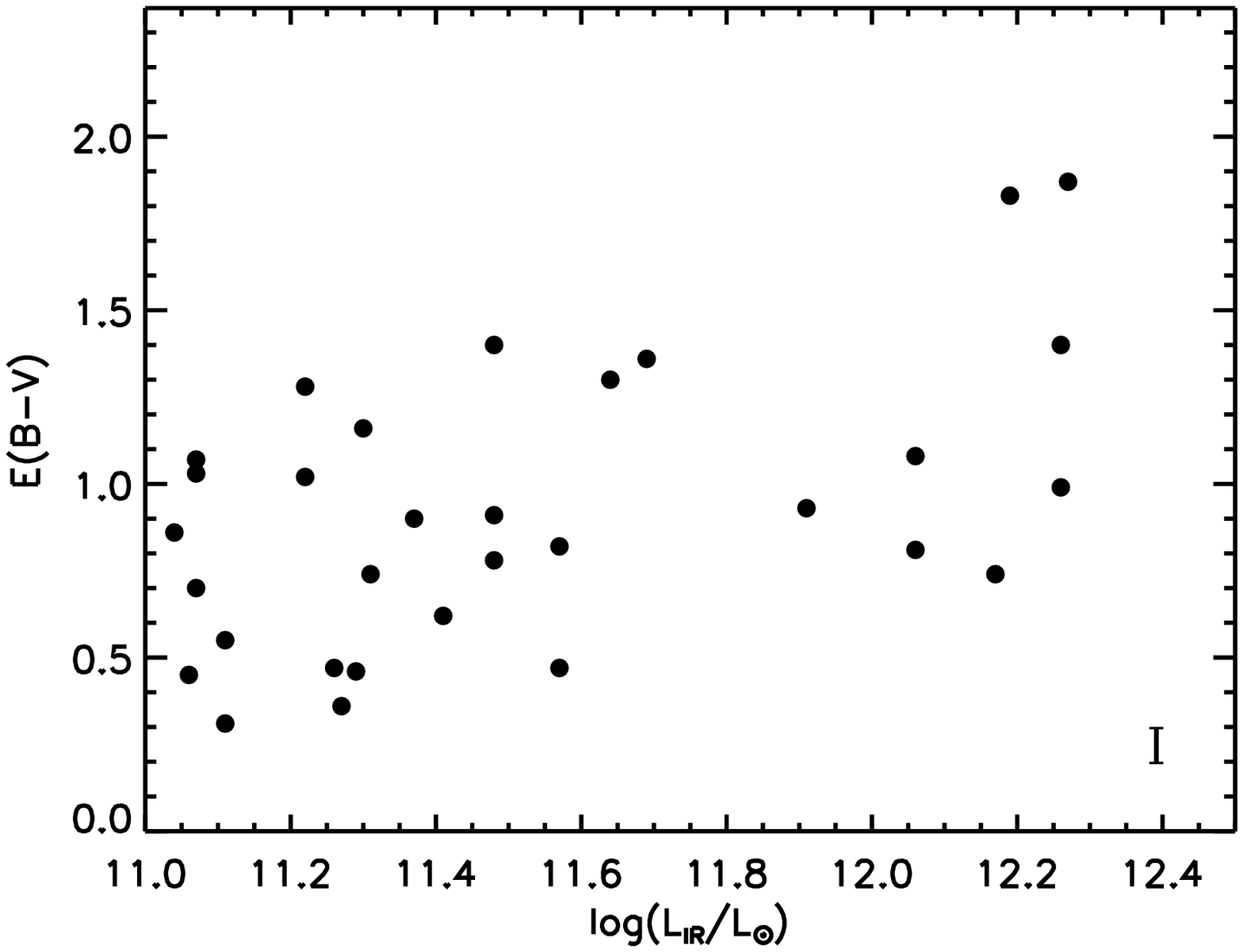}\\
\caption{Reddening coefficient $E(B - V)$ for those objects with
nuclear spectroscopic data available in the literature, plotted
against the IR luminosities of the sources. See Sect. 5 for details on
how the $E(B - V)$ were calculated. The bar in the bottom-right of the
plot represents the error associated with the $E(B - V)$ values.}
\label{reddening}
\end{figure}

A possible explanation for this apparent lack of correlations are
extinction effects. It is possible that for objects with higher
L$_{\rm IR}$ the star-formation activity is indeed more concentrated
in the nuclear regions. However, it is likely that they also have
higher amounts of dust at these locations and therefore, the fraction
of the observed H${\alpha}$ flux within the central 2 kpc is not
significantly higher than that of objects with lower infrared
luminosities. In this context, Fig. \ref{reddening} shows the $E(B -
V)$ values for those objects for which nuclear spectroscopic data are
available in the literature (see Sect. 5 for details), plotted against
the IR~luminosities of the sources. Although with some scatter, the
results shown in the figure suggest that objects with higher
luminosities tend to have higher nuclear $E(B - V)$ values. The mean
and median $E(B - V)$ values for the LIRGs in our sample are 0.83 and
0.86, while these values are 1.24 and 1.08 for the ULIRGs.

To further look for the presence of trends, we calculated the {\it
C$^{H\alpha}_{2kpc}$} median values for the different morphological
classes. We find values of 0.37, 0.50 and 0.53 for objects classified
as 0, 1, and 2 respectively (these values are 0.25, 0.39 and 0.51 when
using the {\it C$^{H\alpha}_{2kpc}$} values, in Col. 6 and 7 in Table
2). The seven objects for which it is not possible to estimate {\it
C$^{H\alpha}_{2kpc}$} for the individual sources separately, and those
with a controversial morphological classification have not been used
when calculating these numbers. These results suggest that the
H$_{\alpha}$ emission from objects classified as 1 and 2 is more
compact than for objects classified as 0. However, as seen in Fig.
\ref{C_plots}, where the points are labeled based on the different
morphological classes, there is a large scatter associated with the
values of {\it C$^{H\alpha}_{2kpc}$}. Samples with a larger number of
objects of each different morphological class are required to confirm
this result.

Consistent with previous studies we find substantial, extended
star-formation activity in objects without evidence of strong past or
ongoing interaction
\citep[e.g.][]{Hattori04,Alonso-Herrero06,Alonso-Herrero09}. Indeed,
\cite{Hattori04} found that the majority of the objects in their
sample with most of the H${\alpha}$ emission in the extended regions
were single objects. It is possible that these objects have undergone
some kind of perturbance in the past (e.g. minor merger). We are
currently carrying out a 2D kinematic study of the galaxies in our
sample that will help to investigate this possibility.

\begin{table*}[!h]
%{\tiny
\centering
%\begin{minipage}{140mm}
\begin{tabular}{lccccccc}
\hline
\hline
IRAS    & Flux H${\alpha}$     &{\it f}$^{\rm slit}$          & $E(B - V)$ & Flux$_{\rm corr}$ H${\alpha}$     & SFR$_{\rm H\alpha}$ & SFR$_{L{\rm IR}}$\\
name    &10$^{-13}$ erg cm$^{-2}$ s$^{-1}$   & PA 0$^{\circ}$/90$^{\circ}$&            &10$^{-13}$ erg cm$^{-2}$ s$^{-1}$   & M$_{\odot}$yr$^{-1}$ & M$_{\odot}$yr$^{-1}$\\
(1)     &   (2)                              &    (3)     &   (4)                              &      (5)             &        (6)     & (7)    \\
\hline
F01159$-$4443$^c$     & 3.88  &0.47/0.55 & 1.41(N),0.91(S)&31.30$\pm$8.44   & 30.14$\pm$8.13    & 51.98   \\
F01341$-$3735N        & 5.97  &0.30/0.24 & 0.45	        &  9.14$\pm$1.65    & 4.96$\pm$0.89     & 6.27    \\
F01341$-$3735S        & 1.98  &0.60/0.67 & 0.24	        &  2.94$\pm$0.67    & 1.60$\pm$0.36	& 19.78   \\
F04315$-$0840         & 14.75 &0.46/0.50 & 1.36	        &  168.7$\pm$45.90  & 77.07$\pm$21.04   & 84.31   \\
F05189$-$2524         & 0.15  &0.67/0.47 & 1.83	        &  6.40$\pm$1.81    &   -	        &  -      \\
F06035$-$7102$^{a,c}$ &	1.40  &0.18/0.26 & 1.40(W),0.99(E)&6.43$\pm$1.55    & 85.36$\pm$20.63   & 313.26  \\
F06076$-$2139         &	0.63  &0.36/0.33 & -	        &-	            &  -                & -       \\
F06206$-$6315         &	0.16  &0.44/0.31 & 1.87	        &  4.96$\pm$1.39    & 91.79$\pm$25.73   & 320.55  \\
F06259$-$4708C$^b$    & 2.82  &0.54/0.36 & 0.93	        &  13.82$\pm$3.58   & 39.73$\pm$10.3    & 139.93  \\
F06259$-$4708N        & 2.28  &0.53/0.66 & 0.13	        &  2.80$\pm$0.60    & 8.04$\pm$1.72     &  -      \\
F06259$-$4708S        & 0.75  &0.58/0.26 & -	        &-	            & -                 &	-      \\
F06295$-$1735         &	3.81  &0.58/0.65 & 0.36	        &  5.06$\pm$0.85    & 4.20$\pm$0.71     & 32.05   \\
F06592$-$6313         &	0.52  &0.26/0.14 & 1.28	        &  6.33$\pm$1.75    & 6.12$\pm$1.7      & 28.57   \\
F07027$-$6011N$^{b}$  & 1.86  &0.30/0.33 & 1.30	        &  12.92$\pm$3.34   & 23.78$\pm$6.15    & 75.14   \\
F07027$-$6011S        & 2.87  &0.55/0.58 & -	        &-	            &-	                & 	-      \\
F07160$-$6215         &	5.95  &0.39/0.21 & -	        &-	            &-	                & 24.97   \\
08355$-$4944          &	25.72 &0.53/0.58 & -	        &-	            &-	                & 68.77   \\
08424$-$3130N         &	0.71  &0.59/0.83 &  -	        &-	            &-	                & 18.94   \\
08424$-$3130S         & 1.36  &0.44/0.44 &  -           &-                  &-                  &-        \\       
F08520$-$6850         &	2.00  &0.30/0.53 & -	        &-	            &-	                & 116.79  \\
09022$-$3615          &	5.59  &0.52/0.57 & -	        &-	            &- 	                & 360.93  \\
F09437$+$0317N$^a$    &	4.39  &0.10/0.11 & -	        &-	            &- 	                & 28.01   \\
F09437$+$0317S        &	3.50  &0.10/0.14 & -	        &-	            &-	                & 	-      \\
F10015$-$0614         &	8.18  &0.15/0.15 & 0.74	        &  13.51$\pm$2.33   & 6.94$\pm$1.20     & 35.14   \\
F10038$-$3338$^a$     &	0.50  &0.50/0.60 & -	        &-	            &-	                & 	101.72 \\
F10257$-$4339         &	80.06 &0.13/0.19 & -	        &-	            &-	                & 	84.61  \\
F10409$-$4556         &	3.72  &0.11/0.19 & 0.47	        &  5.06$\pm$0.85    & 4.08$\pm$0.68     & 31.32   \\
F10567$-$4310         &	2.90  &0.21/0.24 & 1.03	        &  9.38$\pm$2.10    & 5.02$\pm$1.12     & 20.22   \\
F11255$-$4120         &	1.34  &0.27/0.24 & 0.86	        &  3.52$\pm$0.75    & 2.00$\pm$0.36     & 18.88   \\
F11506$-$3851         &	12.90 &0.21/0.26 & 1.16	        &  56.00$\pm$13.4   & 11.62$\pm$2.78    & 34.35   \\
F12043$-$3140S$^b$    & 0.81  &0.20/0.22 & 0.93	        &  1.54$\pm$0.33    & 1.60$\pm$ 0.33    & 40.35   \\
F12043$-$3140N        & 1.54  &0.62/0.73 & 0.00	        &  2.10$\pm$0.44    & 2.07$\pm$0.43     &  -      \\
F12115$-$4656         &	3.91  &0.23/0.26 & 0.55	        &  6.43$\pm$1.16    & 4.00$\pm$0.72     & 22.17   \\
12116$-$5615          &	1.55  &0.63/0.70 & -	        &-	            &-	                & 70.37   \\
F12596$-$1529$^c$     &	3.11  &0.38/0.38 & 1.07(W),0.70(E)& 10.61$\pm$2.51  & 4.85$\pm$1.15     & 20.22   \\
F13001$-$2339         &	0.66  &0.33/0.48 &0.78	        &  2.20$\pm$0.53    & 2.00$\pm$0.46     & 51.98   \\
F13229$-$2934         &	4.30  &0.37/0.39 &0.46	        &  7.38$\pm$1.46    & 2.50$\pm$0.49     & 33.56   \\
F14544$-$4255W$^b$    & 2.75  &0.45/0.32 &0.31	        &  4.01$\pm$0.77    & 1.80$\pm$0.35     & 22.17   \\
F14544$-$4255E        & 0.56  &0.04/0.14 &	-	&	-           & -                 &  -      \\
F17138$-$1017         &	2.12  &0.50/0.14 &	-	&	-           & -                 & 44.25   \\
F18093$-$5744N$^b$    & 15.60 &0.21/0.19 & 0.47	        &  21.83$\pm$3.67   & 12.00$\pm$2.00    & 63.96   \\
F18093$-$5744C        & 4.43  &0.54/0.57 & 0.82	        &  18.09$\pm$4.64   & 9.86$\pm$2.53     &	-      \\
F18093$-$5744S        & 2.73  &0.29/0.29 & -	        &-	            &-	                & 	-      \\
F21130$-$4446         &	0.83  &0.44/0.18 & 0.44         &  1.45$\pm$0.30    & 27.00$\pm$5.52    & 212.53  \\
F21453$-$3511         &	9.48  &0.31/0.24 & 0.62	        &  18.52$\pm$3.66   & 8.71$\pm$1.72     & 44.25   \\
F22132$-$3705         &	11.36 &0.10/0.10 & 1.05	        &  22.49$\pm$4.05   & 5.23$\pm$0.94     & 28.57   \\
F22491$-$1808         &	0.45  &0.37/0.59 & 0.74	        &  1.60$\pm$0.41    & 20.31$\pm$5.15    & 254.62  \\
F23128$-$5919$^c$     &	2.29  &0.46/0.26 & 1.08(N),0.81(S) & 10.10$\pm$2.54 & 38.90$\pm$9.79    & 197.65  \\   
\hline
\hline
\end{tabular}%}
\caption{H$\alpha$ flux emission and derived quantities for the (U)LIRGs
in the VIMOS sample. Column (1): IRAS name. Column (2): H${\alpha}$ fluxes
without reddening correction. An error of 20\% is assumed for these
values. Column (3): fraction of the observed H${\alpha}$ emission within
a slit of width 2 arsec at PAs 0 and 90$^{\circ}$ (see text for
details). Column (4): $E(B - V)$ for those objects with nuclear
H${\alpha}$/H${\beta}$ measurements in the literature. The references
for the nuclear data are those shown in Col. (10) in Table 1. For those
objects with multiple references we have used the more recent
one. Assuming a typical 10\% for the H${\alpha}$/H${\beta}$ ratio
\citep[e.g.][]{Veilleux99} we find that the formal error for the $E(B
- V)$ values in the table is 0.09. Column (5): The H${\alpha}$ flux
values corrected from reddening effects. These values have been
calculated using the $E(B - V)$ values in Col. (4) to deredden the
fraction of the flux within the slit and assuming no reddening outside
the slit. The slit PA including the larger fraction of the H${\alpha}$
emission was used during the process (see text for details). Columns (6)
and (7): the SFRs derived using the H${\alpha}$ luminosity
(L$_{H\alpha}$) corrected from reddening effects, and the IR
luminosity (L$_{\rm IR}$). \newline$^{a}$ These objects were not
observed under photometric conditions. \newline $^{b}$ The SFR$_{\rm
IR}$ values in the table refer to the whole system with the exception
of IRAS F18093-5744, for which the SFR$_{\rm IR}$ refer to the
northern pair (IC4687/IC4686). In addition, there are no
H${\alpha}$/H${\beta}$ nuclear measurements available for the southern
galaxies of IRAS F06295-1735, IRAS F07027-6011 and IRAS F18093-5744,
and the eastern galaxy in IRAS F14544-4255. These galaxies were not
used when calculating SFR$_{\rm H\alpha}$. \newline $^{c}$ For these
four galaxies, it is not possible to estimate precisely the
contribution to the total H${\alpha}$ emission from each individual
nuclei or galaxy. In these cases we used the average reddening value
to correct from reddening effects.}
\label{Ha_flux}
%\end{minipage}
\end{table*}

\section{The H${\alpha}$ emission as a tracer of star formation}

In this section we compare the ionized gas and the IR emissions as
indicators of the star-formation activity. Col. (2) in Table 3 shows
the values of the H${\alpha}$ flux obtained for all galaxies in our
sample. All objects except for IRAS F06035-7102, IRAS
F09437+0317(IC563), and IRAS F10038-3338 were observed under
photometric conditions, as observed in the Paranal-LOSSAM (Line of
Sight Sky Absorption Monitor), which is used to determine the sky
conditions, i.e. photometric, cloudy, or
overcast\footnote{http://archive.eso.org/asm/ambient-server}. For
completeness, the values of the H${\alpha}$ flux for these objects are
shown in the table, although they are not used for the analysis
presented in this section. The values presented in the table were
corrected for Galactic reddening using the \cite{Howarth83} extension
to optical wavelengths of the \cite{Seaton79} reddening law, along
with the $E(B - V)$ values derived from the far-IR based maps of
extinction by \cite{Schlegel98}. The estimated uncertainty for the
absolute fluxes presented in the table is $\lsim$20\%. This value was
estimated by comparing the response curves of the standard stars
observed during each of the three observing periods. Then we calculated
the ``mean response curve'' and the standard deviation from this curve
for each individual standard star. The response curves are within an
uncertainty of 20\% of the mean through the entire useful wavelength
range for all the standard stars observed.

\begin{figure}
\centering
\includegraphics[width=8.5cm]{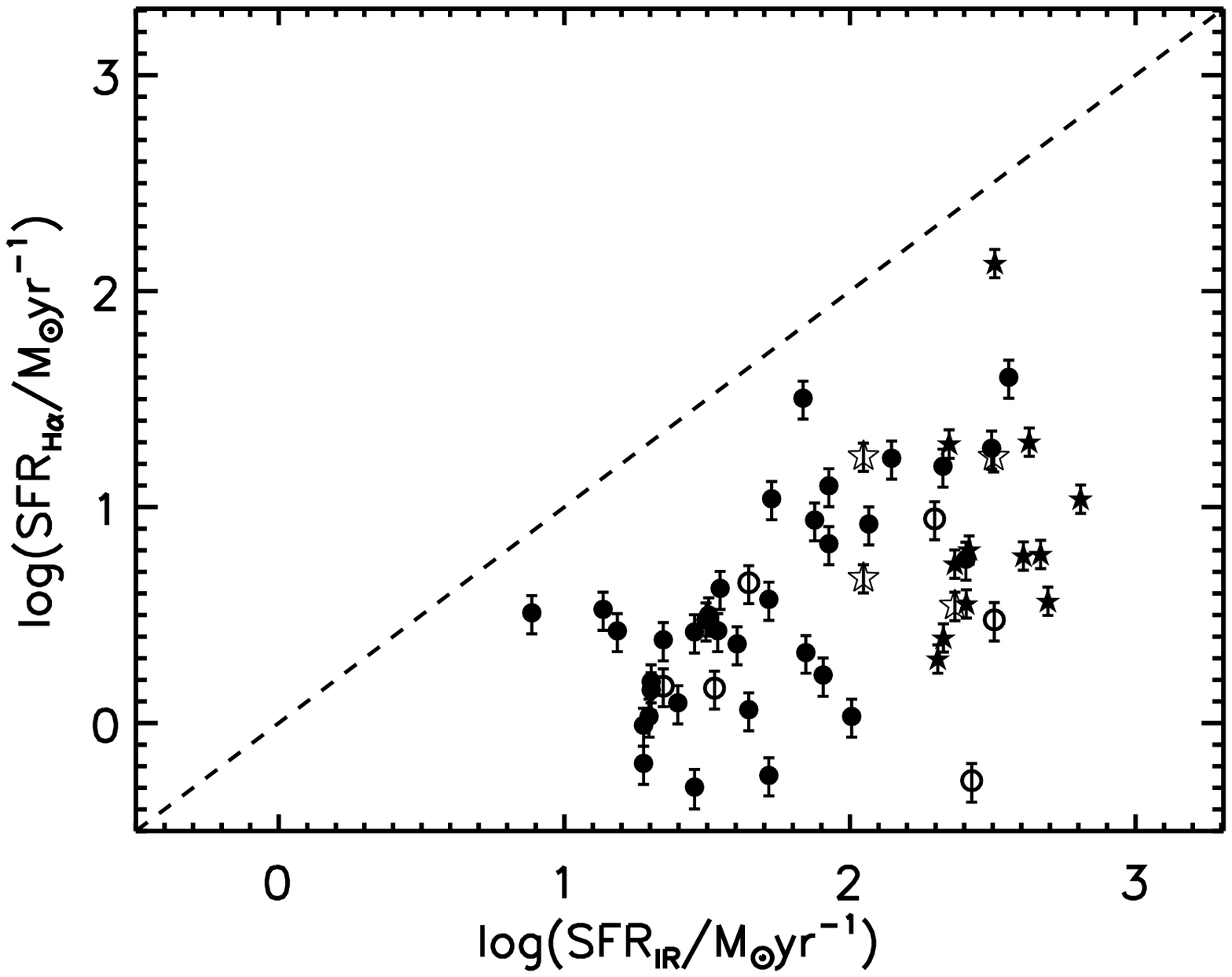}
\caption{Log of the star-formation rates (SFR) in M$_{\odot}$yr$^{-1}$
derived using the the H${\alpha}$ luminosity (SFR$_{\rm H\alpha}$),
without reddening correction, plotted against the log of the SFR
obtained using the IR luminosities. We used the results of
\cite{Garcia-Marin09b} for the figure from their sample of ULIRGs. The
circles correspond to the objects in our sample, while stars correspond
to objects in \cite{Garcia-Marin09b} sample. Those objects classified
as Sy galaxies are indicated in the figure with open symbols.}
\label{SFRs}
\end{figure}

As a first approach we compare the SFRs obtained using the
H${\alpha}$ luminosities (SFR$_{\rm H\alpha}$), without reddening
correction, and the IR luminosities (SFR$_{\rm IR}$) of the
sources. To calculate these values we used the calibration by
\cite{Kennicutt98}

\begin{equation}
{\rm SFR}_{\rm H\alpha} (M_{\odot}~yr^{-1}) = 7.9 \times 10^{-42}~L_{\rm H\alpha}~(erg~s^{-1}),
\end{equation}

\begin{equation}
{\rm SFR}_{\rm IR} (M_{\odot}~yr^{-1}) = 4.5 \times 10^{-44}~L_{\rm IR}~(erg~s^{-1}),
\end{equation}

where L$_{\rm IR}$ is the IR luminosity integrated over the range 8 -
1000 $\mu$m. Figure \ref{SFRs} shows the log of SFR$_{\rm H\alpha}$
plotted against the log of SFR$_{\rm IR}$. For the figure we also used
the results of \cite{Garcia-Marin09b}, based on optical integral field
spectroscopic observations of a sample of 22 ULIRGs. The circles in
the figure correspond to the objects in our sample, while stars
correspond to objects in the \cite{Garcia-Marin09b} sample. Consistent
with the results of \cite{Dopita02}, the figure shows that the SFRs
derived using the reddened L$_{\rm H\alpha}$ substantially
underpredict those obtained using L$_{\rm IR}$. We find mean and
median values for the ratio SFR$_{\rm H\alpha}$/SFR$_{\rm IR}$ of 0.08
and 0.06 respectively. In addition, objects classified as Sy galaxies
are indicated in the figure with open symbols. The figure shows that
with the exception of IRAS F05189-2524, the few objects classified as
Sy galaxies do not stand out from the others in the plot.
 
The results shown in Fig. \ref{SFRs} were, a priori, expected since
objects such as LIRGs and ULIRGs are known to suffer severe extinction
effects. In order to correct for these effects, we first estimated the
reddening factor $E(B - V)$. For this purpose, we used he
H${\alpha}$/H${\beta}$ values available in the literature, along with
the interstellar extinction law based on \cite{Savage79}. We assumed
an intrinsic H${\alpha}$/H${\beta}$ ratio of 2.85 for HII galaxies
(typical for Case B recombination decrement for T$_{\rm e} \sim$10$^
{4}$ K and N$_{\rm e} \sim$10$^ {4}$ cm$^{-3}$) and 3.10 for AGN
\citep{Ferland83}. Col. 4 in Table 3 shows the measured $E(B - V)$
values.

Note that the $E(B - V)$ values shown in Table 3 have been calculated
using the H${\alpha}$/H${\beta}$ ratios from the literature, which are
based on long-slit spectroscopic observations. These observations
usually concentrate on the nuclear regions of the objects. However,
reddening effects are more important in the nuclear regions, and
decrease significantly towards the extended regions of the objects
\citep{Garcia-Marin09b}. Therefore, to correct for reddening effects,
we have used the ``nuclear'' $E(B - V)$ values in Table 3 to deredden
only the fraction of the H${\alpha}$ emission within a slit of 2
arcsec width (typical slit width), which was estimated in the previous
section and shown in Table 3. As described in that section, a slit PAs
0$^{\circ}$ and 90$^{\circ}$ was used for each galaxy. When correcting
from reddening, the slit PAs that include the larger fraction of the
observed H${\alpha}$ emission were used. To deredden the H${\alpha}$
fluxes we used the standard expression

\begin{equation}
F_{i}(\lambda) = F_{o}(\lambda)10^{0.4E (B - V) f(\lambda)},
\end{equation}

where $F_{i}$ and $F_{o}$ are the intrinsic and the observed flux
respectively, and $f({\lambda})$ is the reddening law. The total
observed H${\alpha}$ emission for the galaxies is calculated as the
sum of the reddening-corrected H${\alpha}$ emission within the slit
plus the H${\alpha}$ emission outside the slit, not corrected for
reddening. Although less important than in the nuclear regions,
reddening effects are also present in the extended regions of the
objects \citep{Alonso-Herrero09}. Therefore, this approach will tend
to underestimate the values of the dereddened SFR$_{\rm
H\alpha}$. However, since the reddening in the extended regions is
usually substantially smaller than in the nuclear regions
\citep{Garcia-Marin09b,Alonso-Herrero09}, this underprediction is not
expected to significantly affect the conclusions reached in this
section.

\begin{figure}
\centering
\includegraphics[width=8.5cm]{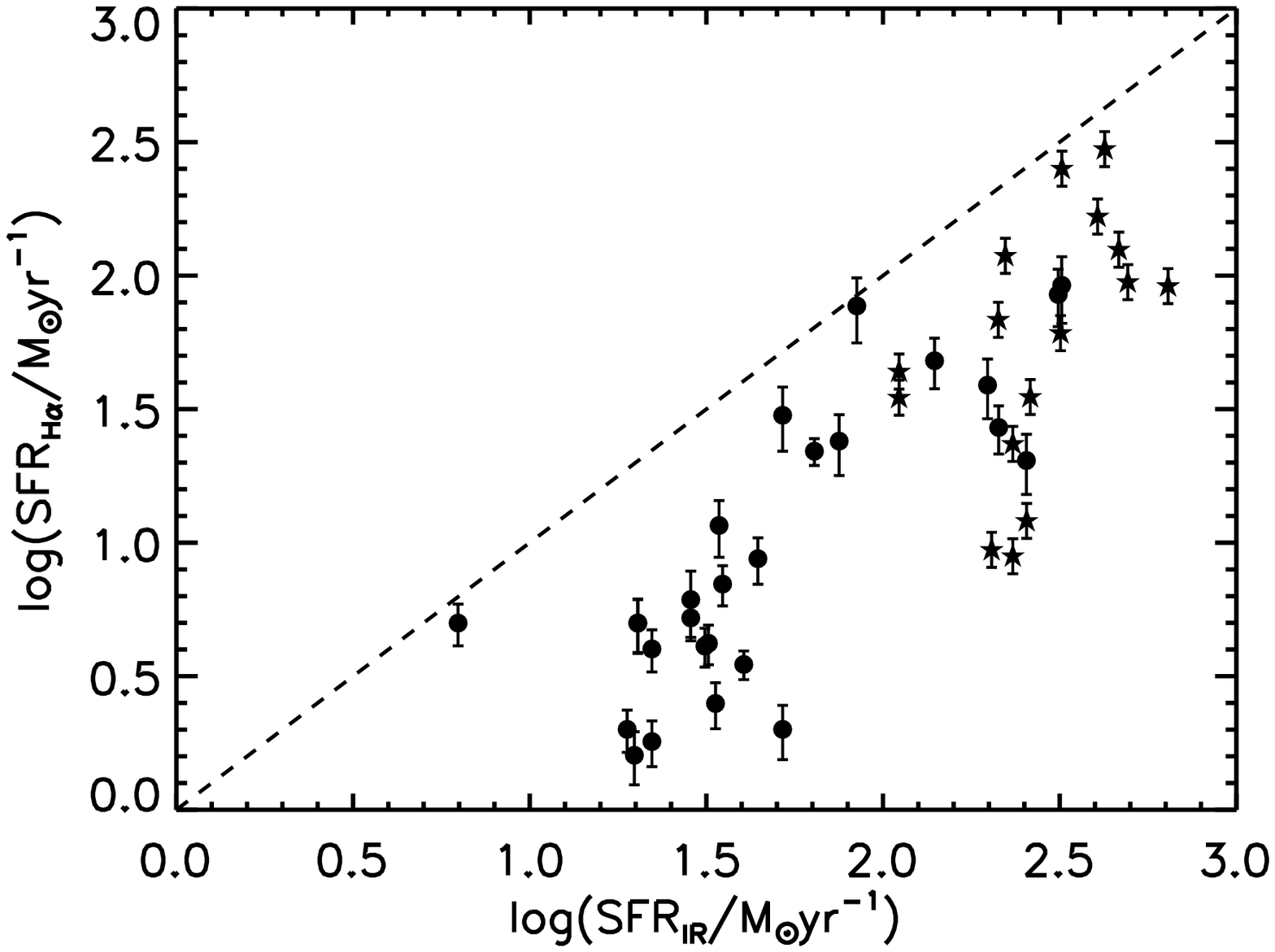}
\caption{Same as Fig. \ref{SFRs}, but the H${\alpha}$ luminosity has
been corrected from reddening effects. The nuclear $E(B - V)$ values
shown in Table 3 have been used for the objects in our sample (see
text for details). On the other hand, a 2D reddening correction was
applied for the ULIRGS in the \cite{Garcia-Marin09b} sample. The AGN
dominated source IRAS F05189-2524 and all the objects in our sample
with unavailable H${\alpha}$/H$_{\beta}$ values are not used for the
plot. }
\label{SFRs_de-red}
\end{figure}

The log of the reddening-corrected SFR$_{\rm H\alpha}$ plotted against
the log of the SFR$_{\rm IR}$ is presented in Fig.  \ref{SFRs_de-red},
together with the results of \cite{Garcia-Marin09b} for their sample
of ULIRGs. The AGN-dominated source IRAS F05189-2524 and those objects
with unavailable H${\alpha}$/H$_{\beta}$ values have been left out of
the plot. The \cite{Garcia-Marin09b} IFS observations cover the
H$\alpha$ -- H$\beta$ spectral range, and therefore, they were able to
perform a detailed 2D reddening correction of the $H{\alpha}$ emission
from the galaxies in their sample. As is obvious from the figure, the
correlation certainly improves. However, all the points in the figure
still fall below the dashed line. After accounting for reddening
effects, the mean and median values of the ratio SFR$_{\rm
H\alpha}$/SFR$_{\rm IR}$ are 0.27 and 0.22. Overall, we conclude that
the SFRs derived using the dereddened L$_{\rm H\alpha}$ generally
underpredict those obtained using L$_{\rm IR}$ by a factor of 4-5,
indepently of the L$_{\rm IR}$ of the sources. At this stage it is
important to mention that the reddening correction described in this
section accounts only for foreground screen extinction and regions
that are not completely optically thick. ``Internal'' extinction
(i.e. dust in the photoionized nebula) or heavily obscured regions
that are only visible in the far-IR are not taken into account when
applying a reddening correction based on the H${\alpha}$/H$_{\beta}$
ratio. This would explain the lower values of SFR$_{\rm H\alpha}$
compared to those of SFR$_{\rm IR}$ even after correcting for
reddening effects.

\section{Summary and Conclusions}

We presented a catalogue of the VIMOS continuum, H${\alpha}$ and
H${\alpha}$ equivalent widths (EW) images of a sample of 31 LIRGs and
7 ULIRGs (a total of 47 individual galaxies). The continuum images
trace the stellar emission, while the H${\alpha}$ images trace the
ongoing star-formation activity and/or ionizing shocks. We performed a
morphological study of the continuum and H${\alpha}$ images and
compared the H${\alpha}$ and the IR luminosities as tracers of the
star-formation activity. The main results are summarized as follows:

\begin{itemize}

\item{\it The morphologies of the continuum, H${\alpha}$ and EW
images:} the morphologies of the H${\alpha}$ images are substantially
different from those of the continuum images. The H${\alpha}$ images
frequently reveal clumpy structures, such as HII regions in spiral
arms, tidal tails, rings, bridges, extended up to few kpc from the
nuclear regions that are not visible in the continuum. The different
morphologies are explained in terms of bright, extranuclear
star-formation activity along the tidal tails or the spiral arms
and/or ionizing shocks in the extra-nuclear extended regions (Colina
et al. 2005, Monreal-Ibero et al. 2010, and references therein).

We also compared the location of the surface brightness peaks in the
continuum and the H${\alpha}$ images. For the majority of the objects
in our sample ($\sim$80\%) the peaks of the stellar and the ionized
gas emissions are separated by less than 1 kpc, with a median value of
0.2 kpc for the whole sample.

\item{\it The concentration of the continuum and the ionized gas
emissions and their connection with the dynamical status of the
objects:} we investigated the morphologies of the images on the basis
of the concentration of the emission {\it C$_{2kpc}$}, defined as the
ratio of the flux contained within an aperture of 2 kpc of diameter
centered on the nucleus of the object, and the observed flux from the
galaxy within the same physical scale for all galaxies in our
sample. In the first place, we find that the fraction of the observed
H${\alpha}$ emission from the central 2 kpc is higher than that of the
continuum for the majority (85\%) of the sources used for the study
presented here.

If we concentrate on the H${\alpha}$ images, it is remarkable that
62\% of the objects in our sample have more than the half of their
observed H${\alpha}$ emission (not corrected from extinction) outside
the central 2 kpc. This result further emphasizes the importance of
the extended star-formation activity in (U)LIRGs.

On the other hand, we do not find clear evidence for a correlation
between {\it C$_{2kpc}$} and the IR luminosity of the sources. A
possible explanation for this apparent lack of correlations are
reddening effects. It is possible that objects with higher L($_{\rm
IR}$) have the star formation more concentrated towards the nuclear
region. However, it is likely that they also have higher dust
concentrations towards these regions and therefore, the H${\alpha}$
emission measured in the central 2 kpc is not significantly higher
than in objects with lower luminosities. Finally, our results suggests
that the star-formation activity is more concentrated in objects
classified as 1 or 2 than in those classified as 0. However, samples
with a greater number of objects of each morhological class are
required to confirm these result.

\item{\it The H${\alpha}$ emission as a tracer of the star formation:}
we find that the SFRs derived using the H${\alpha}$ luminosities
generally underpredict those obtained using the MFIR luminosities,
even after correcting from reddening effects.

\end{itemize}

This study shows the utility of the IFS technique for studying the
structure of the stellar light, ionized gas and ongoing star-formation
activity in LIRGs and ULIRGs. In future publications we will probe the
full potential of our dataset by investigating other aspects such as
ionization or kinematics.

\begin{acknowledgements}

We thank Macarena Garc\'ia Mar\'in for her help during the early
stages of this project. We also thank the anonimous referee for useful
comments that helped to greatly improve the paper. Based on
observations carried out at the European Southern observatory, Paranal
(Chile), Programs 076.B-0479(A), 078.B-0072(A) and 081.B-0108(A). We
thank Edward L. Wright for the \cite{Wright06} cosmology
calculator. This paper uses the plotting package jmaplot developed by
Jes\'us Ma\'iz Apell\'aniz,
\texttt{http://dae45.iaa.csic.es:8080/$\sim$jmaiz/software}. This
research made use of the NASA/IPAC Extragalactic Database (NED), which
is operated by the Jet Propulsion Laboratory, California Institute of
Technology, under contract with the National Aeronautics and Space
Administration. The Digitized Sky Surveys were produced at the Space
Telescope Science Institute under U.S. Government grant NAG
W-2166. The images of these surveys are based on photographic data
obtained using the Oschin Schmidt Telescope on Palomar Mountain and
the UK Schmidt Telescope. The plates were processed into the present
compressed digital form with the permission of these institutions.

This work has been supported by the Spanish Ministry of Science and
Innovation (MICINN) under grant ESP2007-65475-C02-01. AM-I is
supported by the Spanish Ministry of Science and Innovation (MICINN)
under program ``Specialization in International Organisms'', ref
ES2006-0003.

\end{acknowledgements}

\clearpage

\bibliographystyle{aa} \bibliography{JRZRefs}

   \addtocounter{figure}{-8}
 \begin{figure*}
   \centering
   \vskip -0.5cm
   \includegraphics[width=13.0cm]{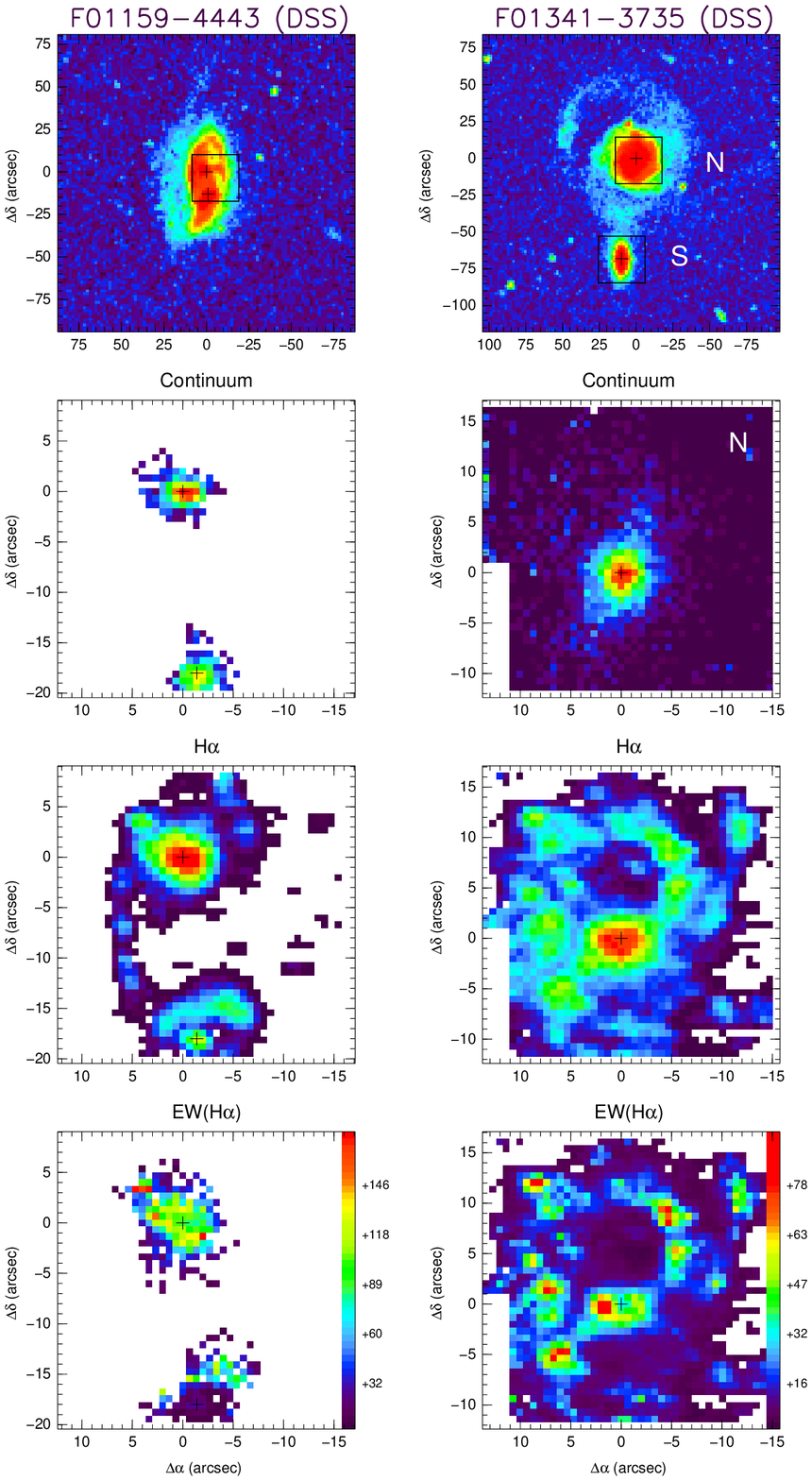}
   \caption{{\it Top panel}: DSS image of the galaxy. For those
   objects for which HST images are available in the literature these
   are used for the figure. The exceptions are IRAS F13229-2934 and
   IRAS F21453-3511. For these galaxies the HST image does not cover
   the entire emission from the galaxies and we prefered to use the
   DSS images. The box on the DSS or HST images indicates the VIMOS
   field of view, while the horizontal line in the bottom left of the
   same images corresponds to a scale of 10 kpc. {\it Second panel}:
   the continuum within the wavelength range 6390 -- 6490~\AA~(rest
   frame). {\it Third panel}: the H${\alpha}$ emission from the
   galaxy. Both the continuum and the H${\alpha}$ maps are represented
   in logarithmic scale, and in arbitrary flux units. {\it Fourth
   panel}: H${\alpha}$ equivalent width (H${\alpha}$-EW) in \AA. With
   the exception of IRAS F10567-4310 and IRAS F17138-1017, the peak of
   the continuum emission, identified as the nucleus of the galaxy, is
   indicated in all the maps with a ``plus'' sign. For these two
   galaxies the morphology of the continuum is less constrained
   because of the presence of vertical patterns (see text for
   details). In these cases the `plus' sign corresponds to the
   location of the peak of the H${\alpha}$ emission. Note that two
   pointings were used to cover the full emission from the galaxy IRAS
   F09437+0317N (IC564), which are referred to in the figure as N(P1)
   and N(P2).}
   \label{panel}
   \end{figure*}

   \addtocounter{figure}{-1}
   \begin{figure*}
   \centering 
%   \vskip -0.5cm
%   \vskip -0.5cm
   \includegraphics[width=14.0cm]{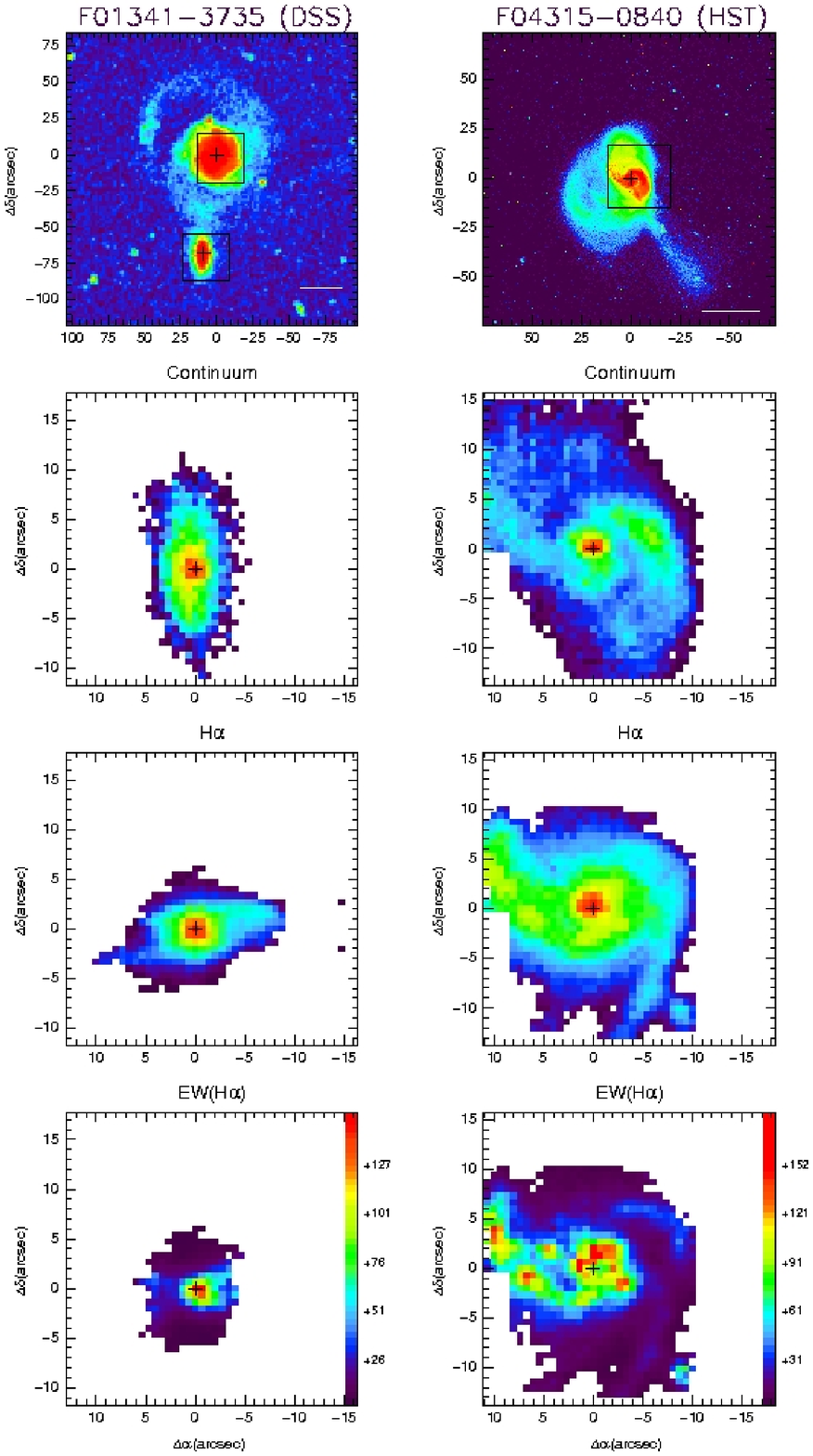}
   \caption{{\it Continued}}   
   \end{figure*}
 
   \addtocounter{figure}{-1} 
   \begin{figure*}
   \centering
%   \vskip -0.5cm
   \includegraphics[width=14.0cm]{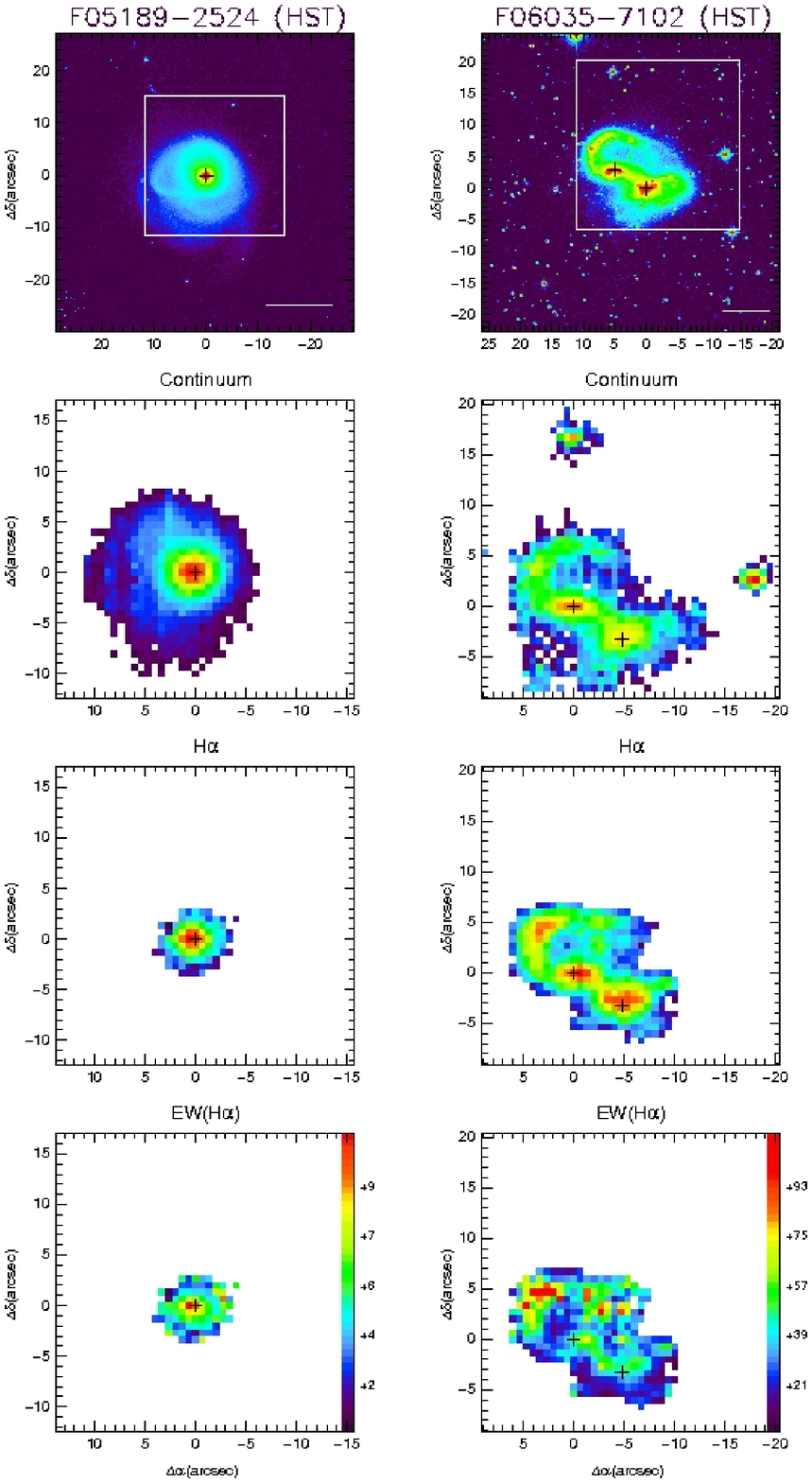}
   \caption{{\it Continued}}   
   \end{figure*}

   \addtocounter{figure}{-1}
   \begin{figure*}
   \centering
%   \vskip -0.5cm
   \includegraphics[width=14.0cm]{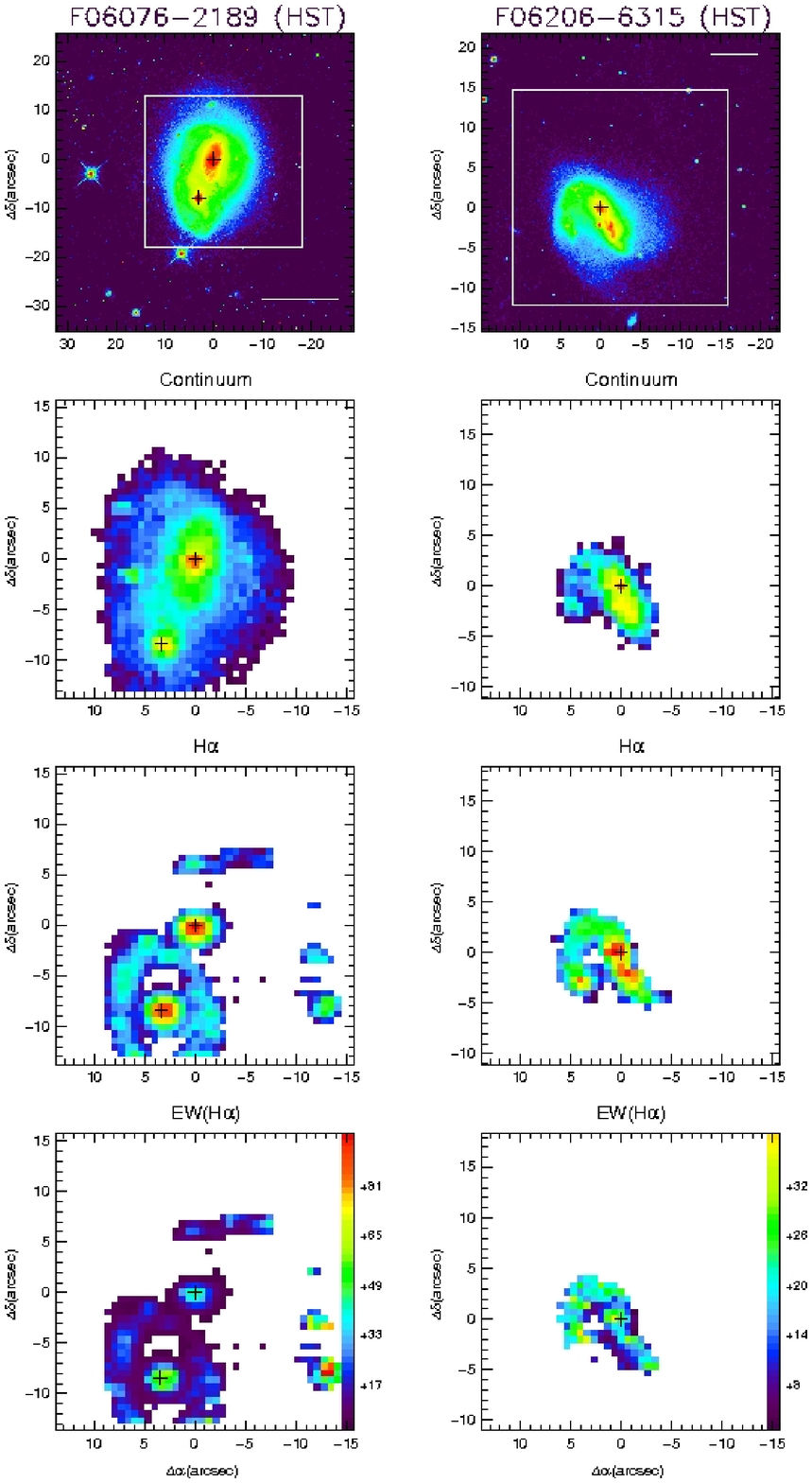}
   \caption{{\it Continued}}   
   \end{figure*}

   \addtocounter{figure}{-1}
   \begin{figure*}
   \centering
   \includegraphics[width=7.cm]{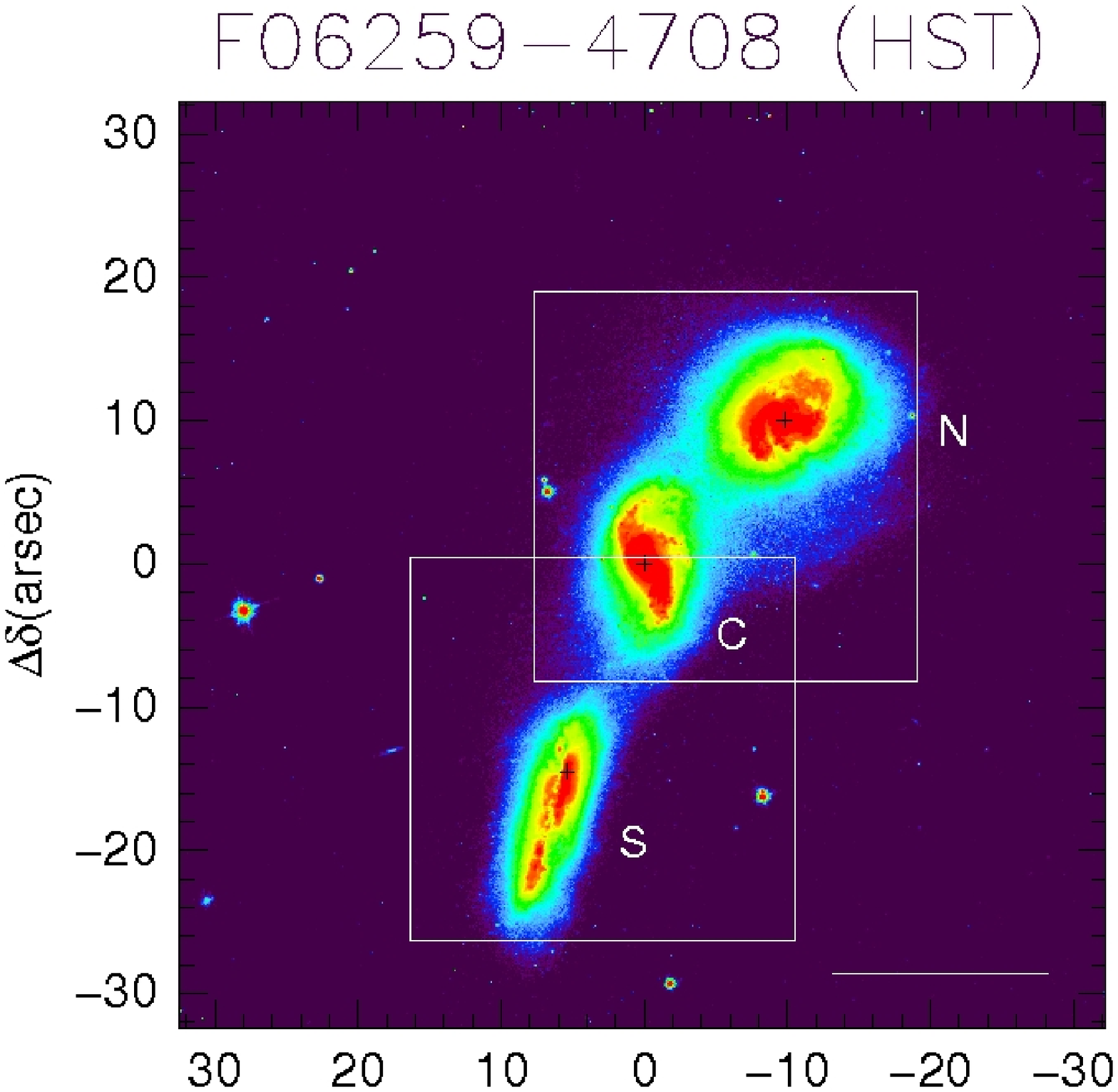}\\
   \vspace{-0.65cm}\includegraphics[width=14.cm]{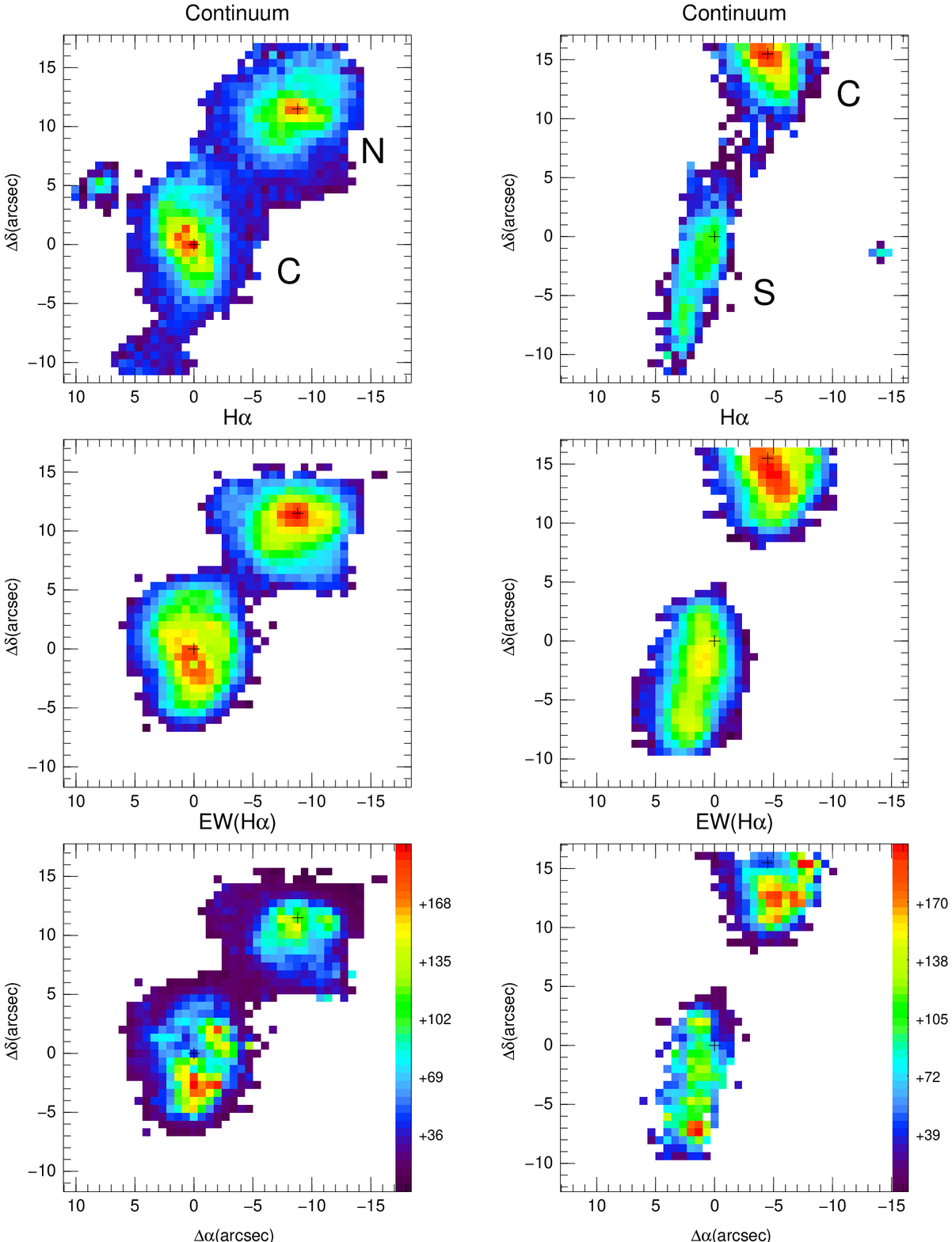} 
   \caption{{\it Continued}}   
   \end{figure*}

   \addtocounter{figure}{-1}
   \begin{figure*}
   \centering
%   \vskip -0.5cm
   \includegraphics[width=14.0cm]{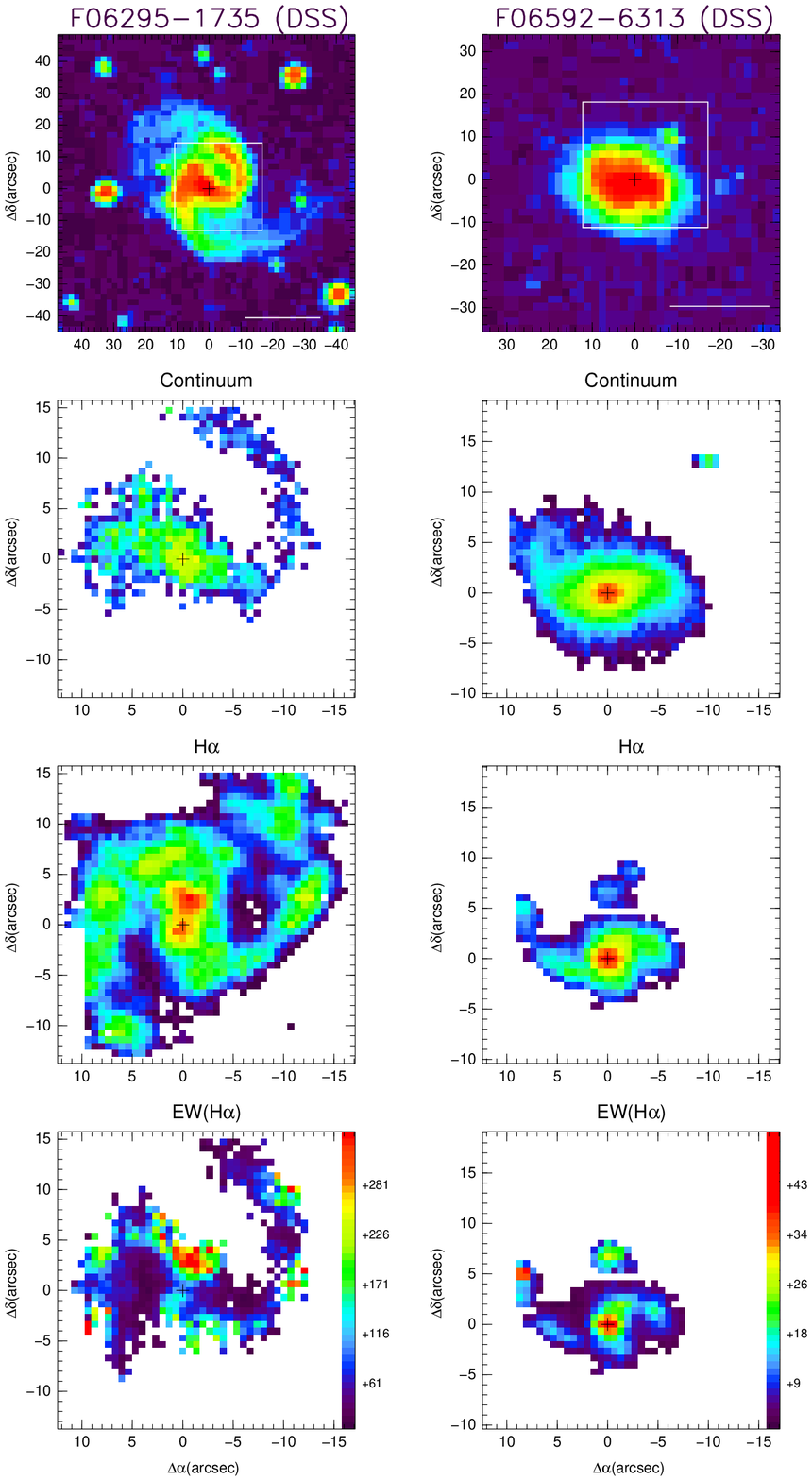}
   \caption{{\it Continued}}   
   \end{figure*}

   \addtocounter{figure}{-1}
   \begin{figure*}
   \centering
   \includegraphics[width=7.cm]{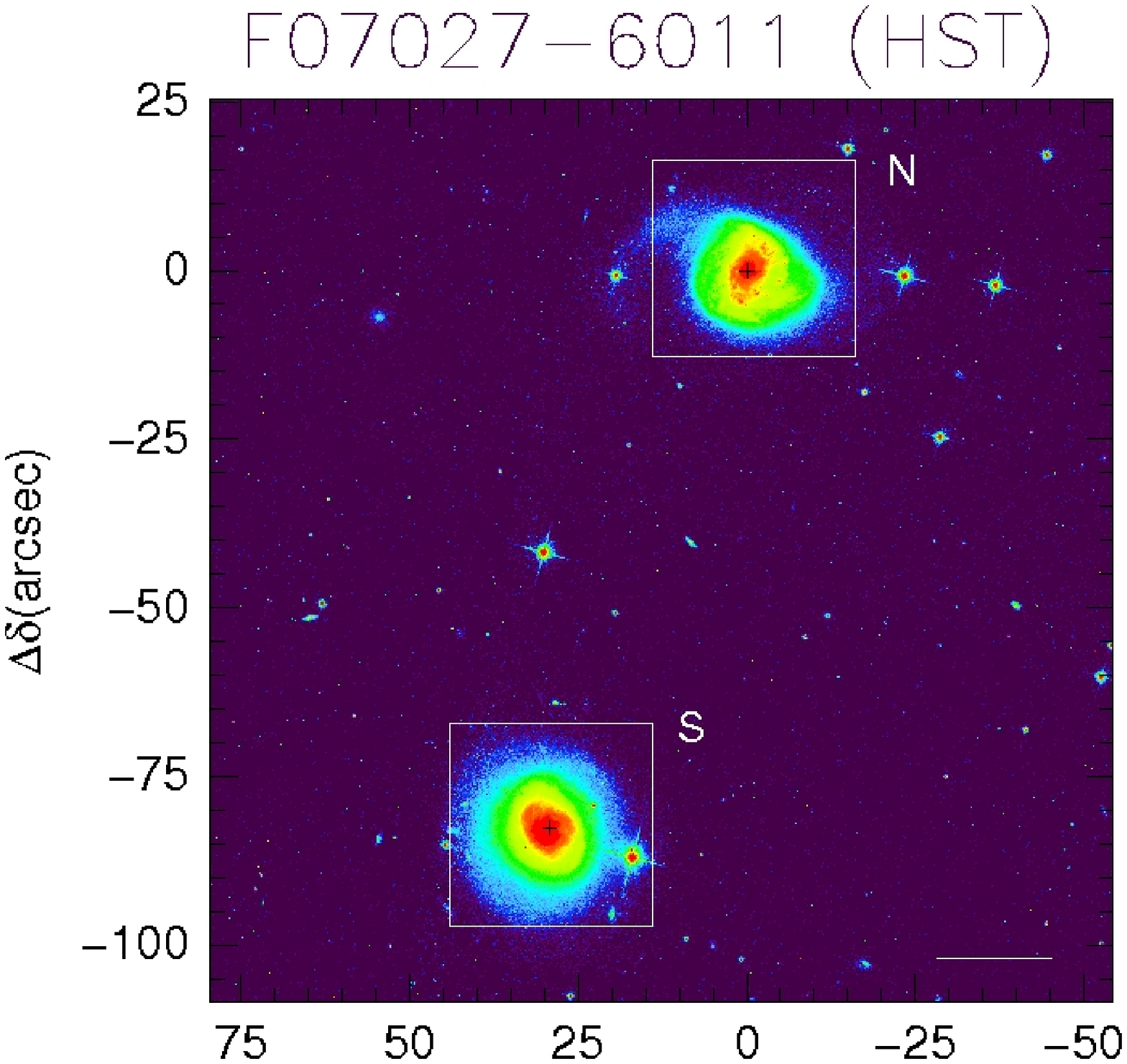}\\
   \vspace{-0.65cm}\includegraphics[width=14.cm]{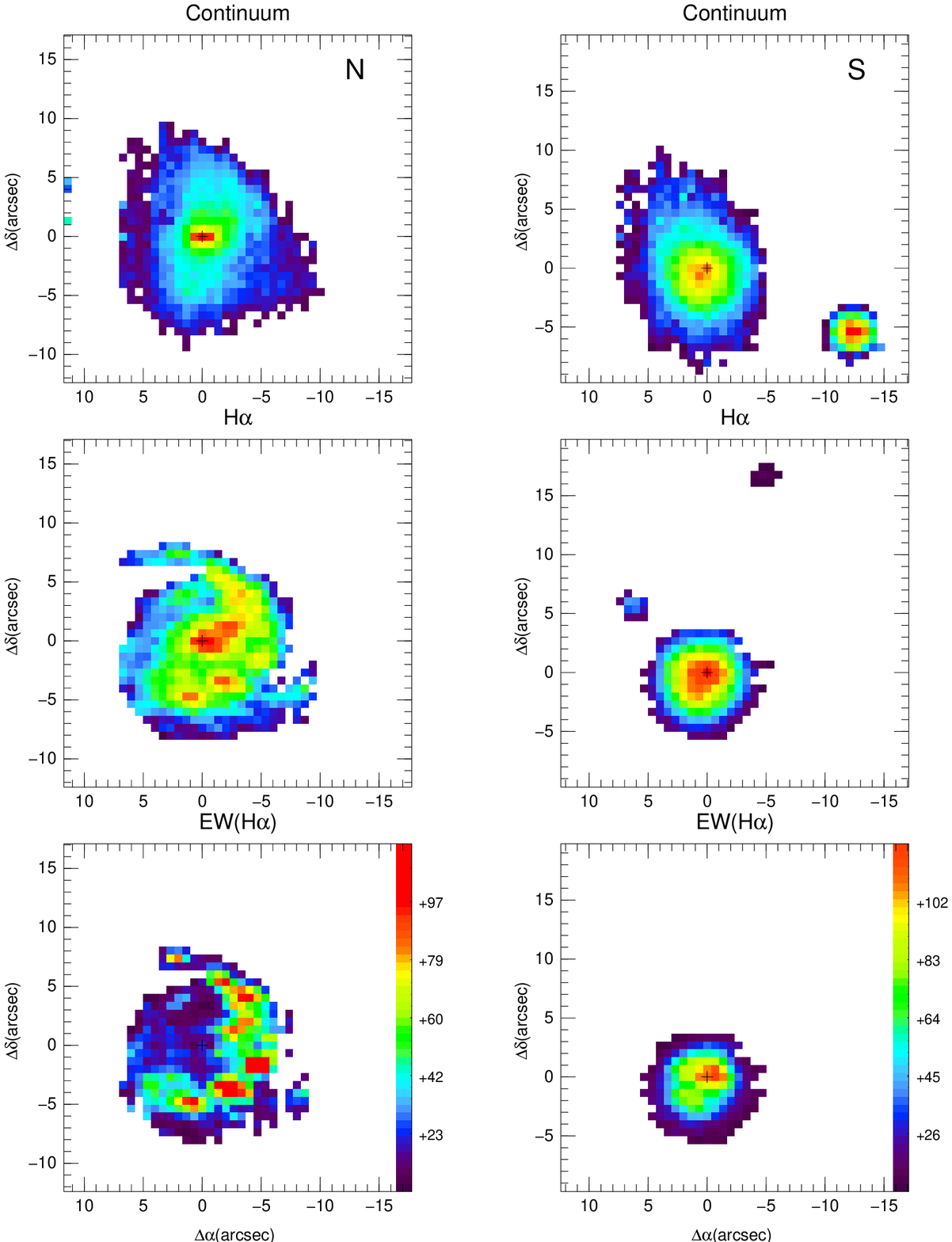}
   \caption{{\it Continued}}   
   \end{figure*}

   \addtocounter{figure}{-1}
   \begin{figure*}
   \centering
%   \vskip -0.5cm
   \includegraphics[width=15.cm]{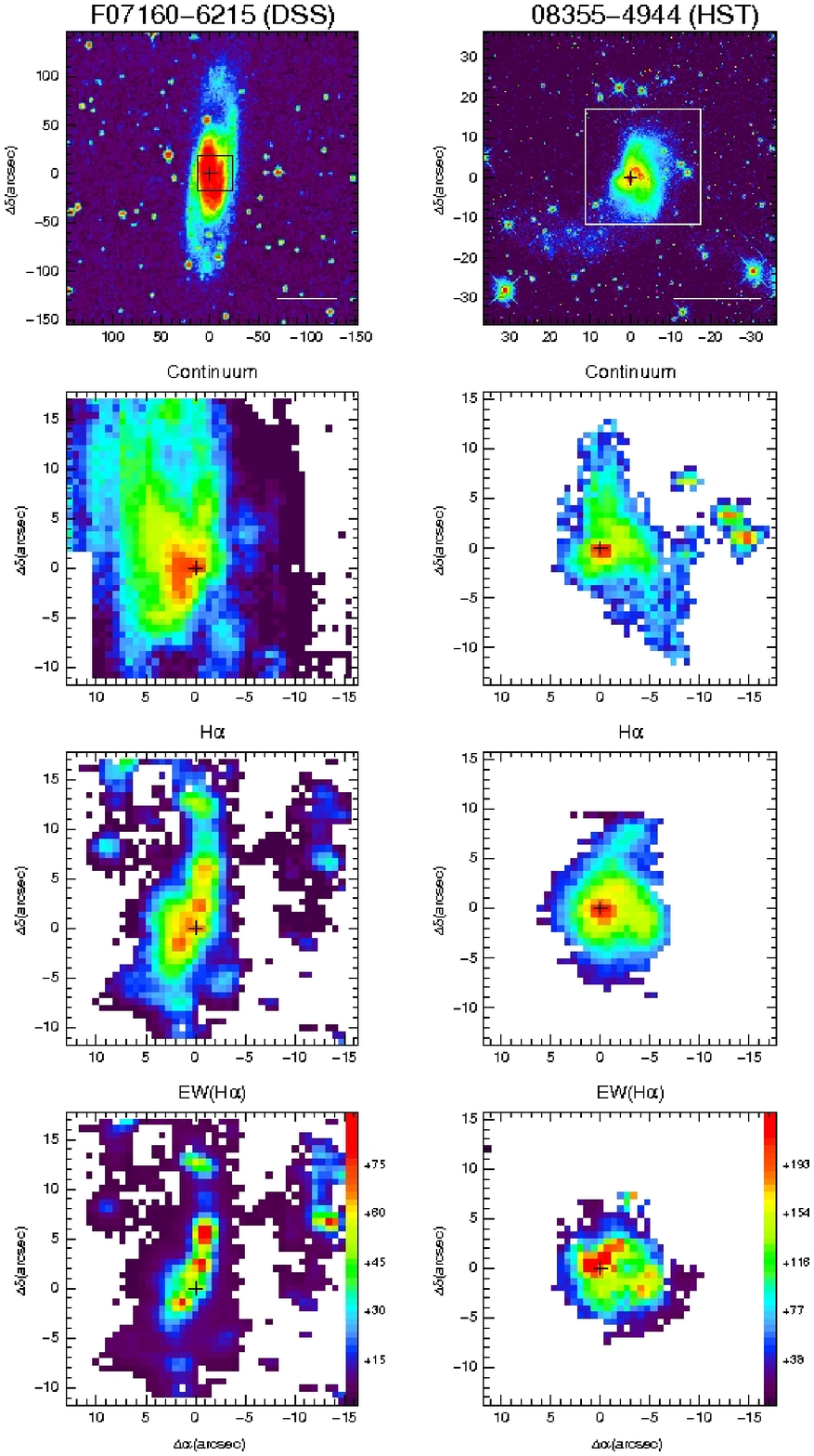}
   \caption{{\it Continued}}
   \end{figure*}

   \addtocounter{figure}{-1}
   \begin{figure*}
   \centering
%   \vskip -0.5cm
   \includegraphics[width=15.cm]{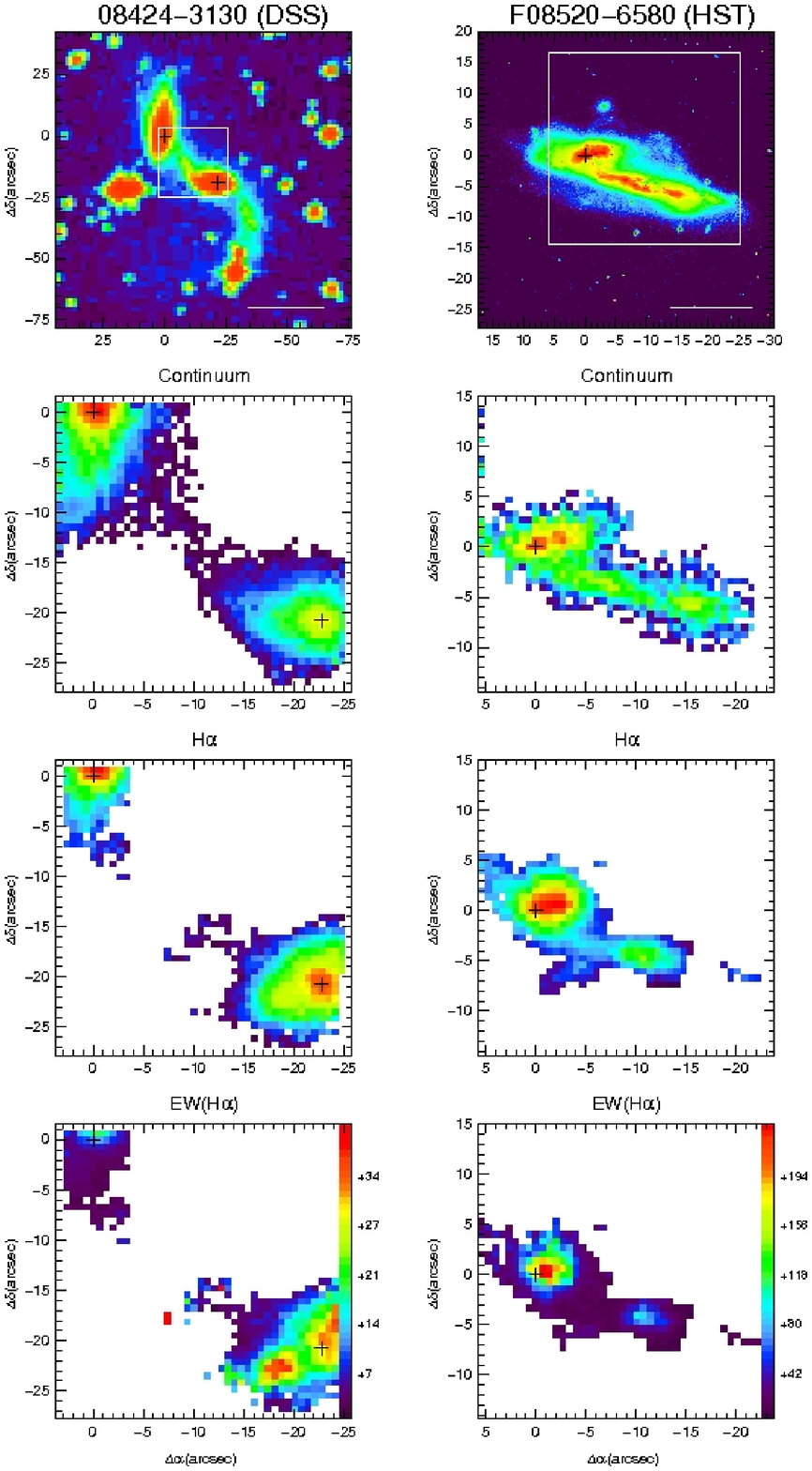}
   \caption{{\it Continued}}
   \end{figure*}

   \addtocounter{figure}{-1}
   \begin{figure*}
   \centering
%   \vskip -0.5cm
   \includegraphics[width=14.cm]{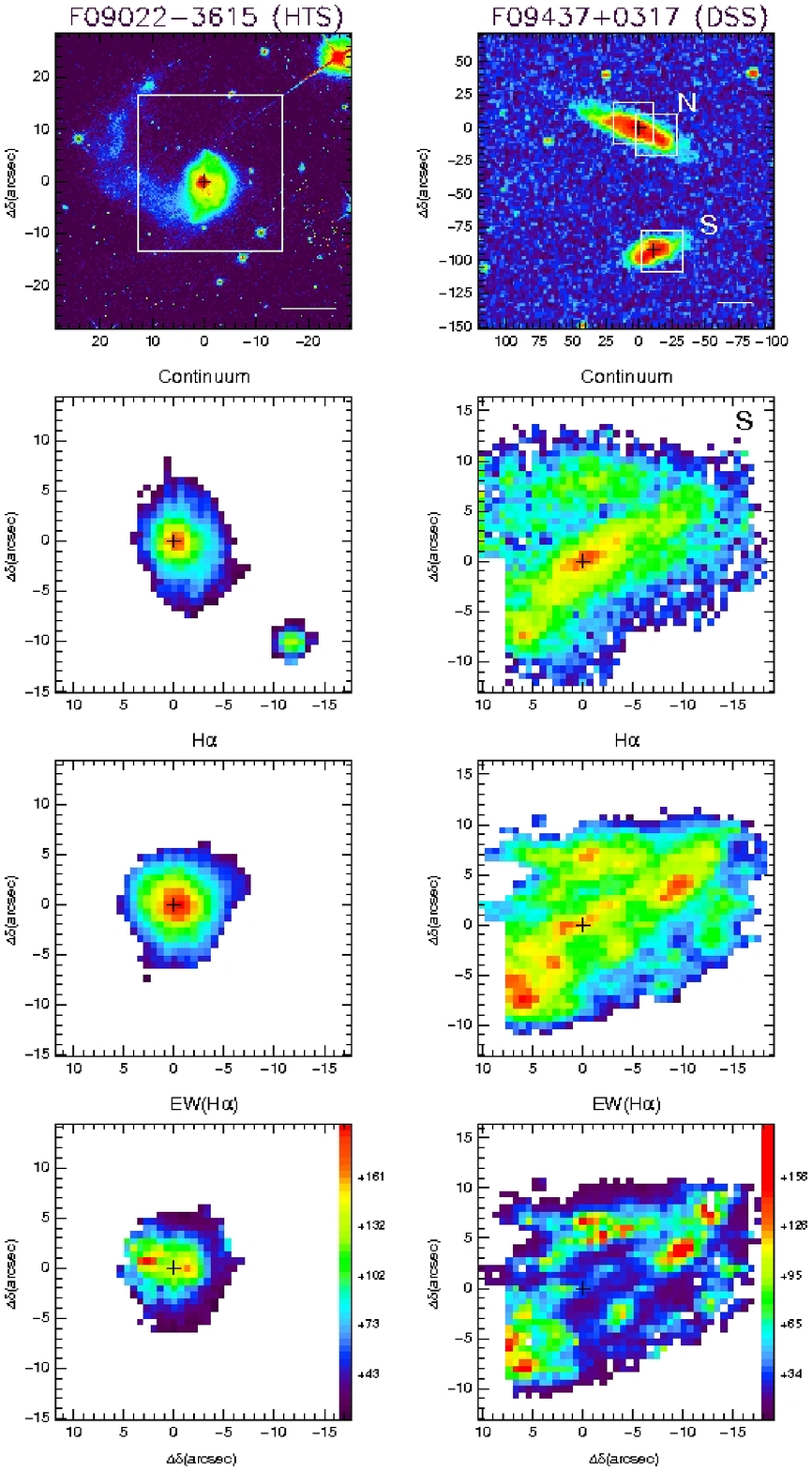}
   \caption{{\it Continued}}
   \end{figure*}ç
%\newpage
   \addtocounter{figure}{-1}
   \begin{figure*}
   \centering
   \includegraphics[width=7.cm]{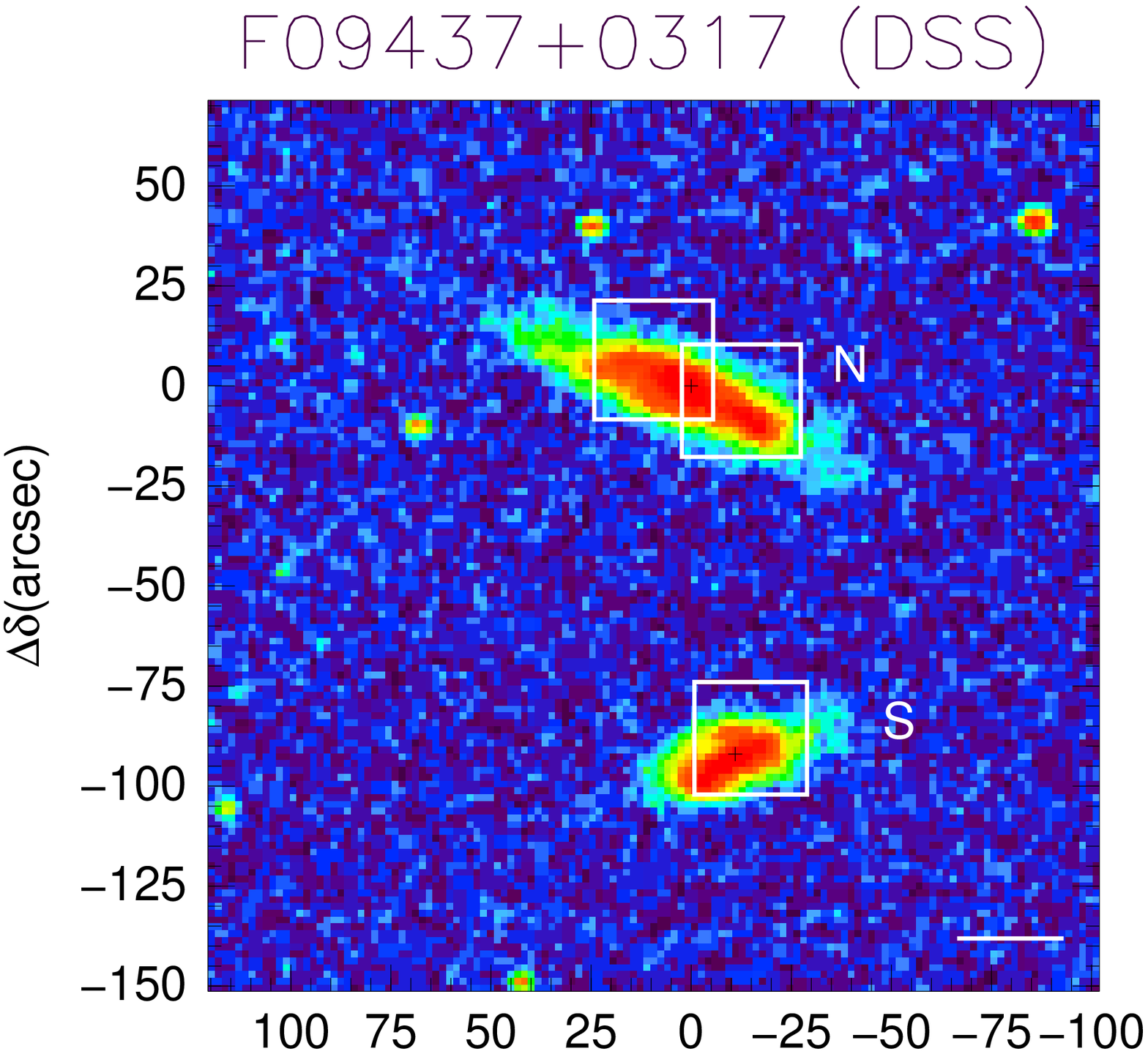}\\
   \vspace{-0.65cm}\includegraphics[width=14.cm]{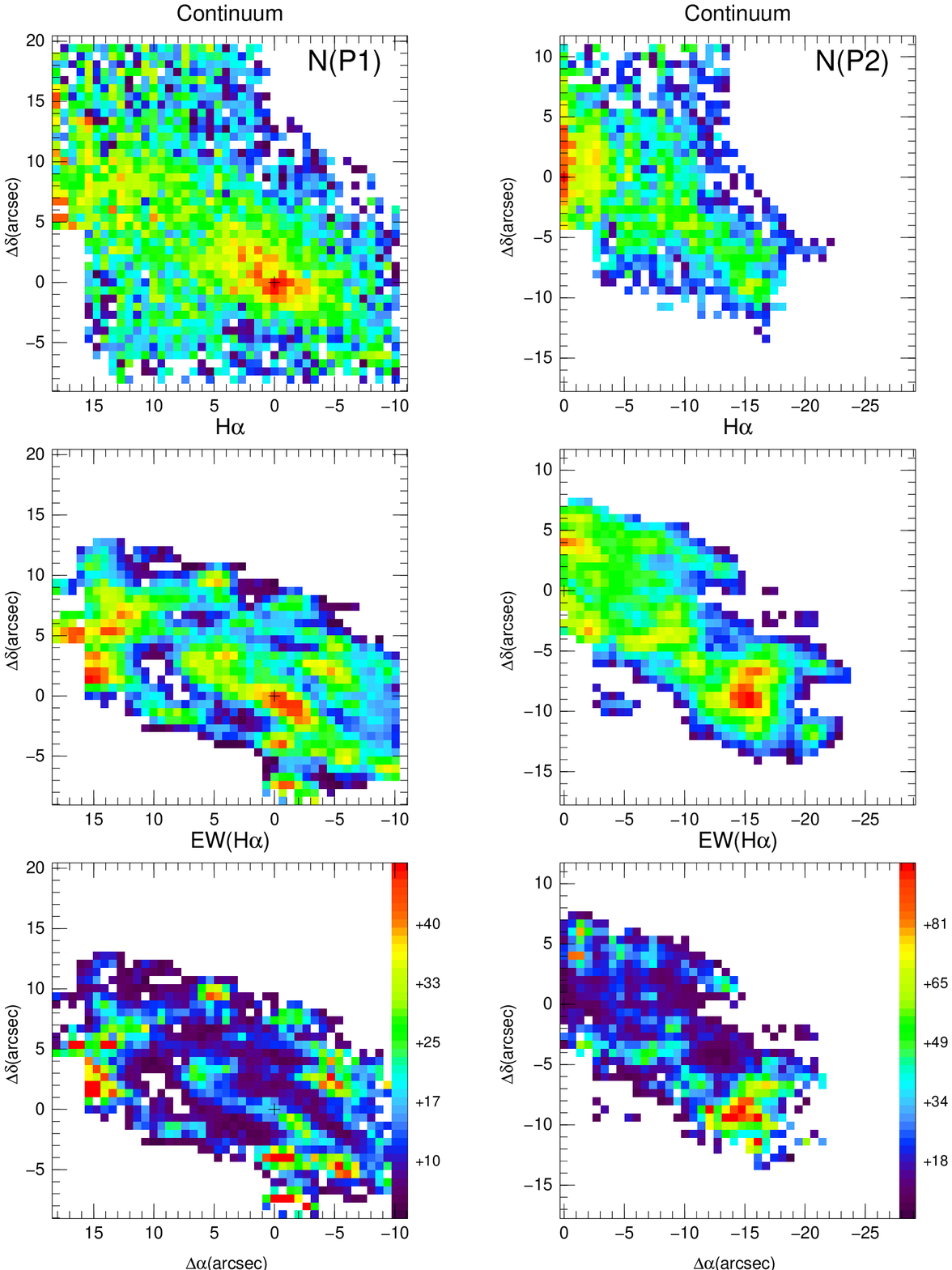}
   \caption{{\it Continued}}
   \end{figure*}

   \addtocounter{figure}{-1}
   \begin{figure*}
   \centering
%   \vskip -0.5cm
   \includegraphics[width=14.cm]{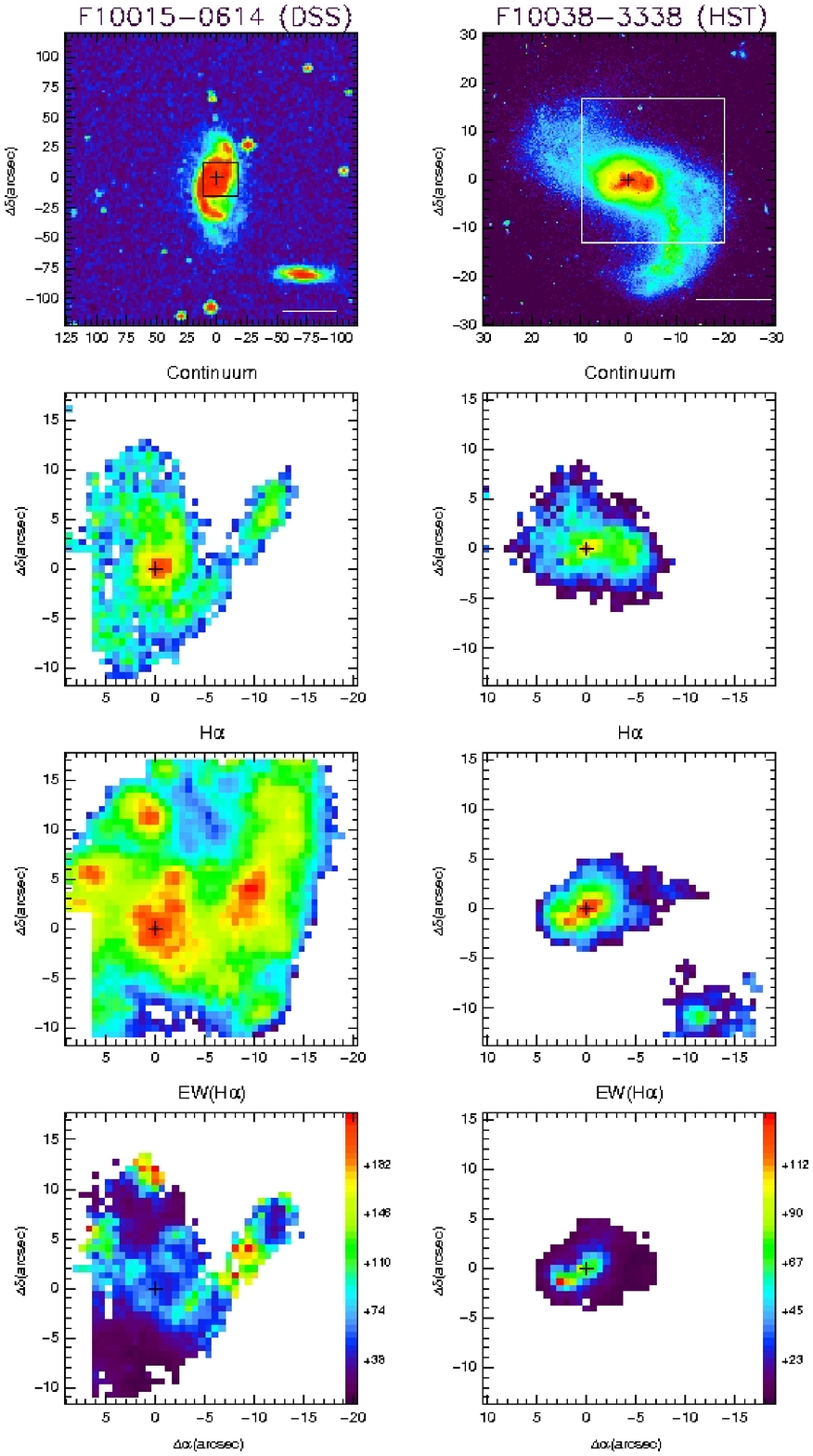}
   \caption{{\it Continued}}
   \end{figure*}

   \addtocounter{figure}{-1}
   \begin{figure*}
   \centering
%   \vskip -0.5cm
   \includegraphics[width=14.cm]{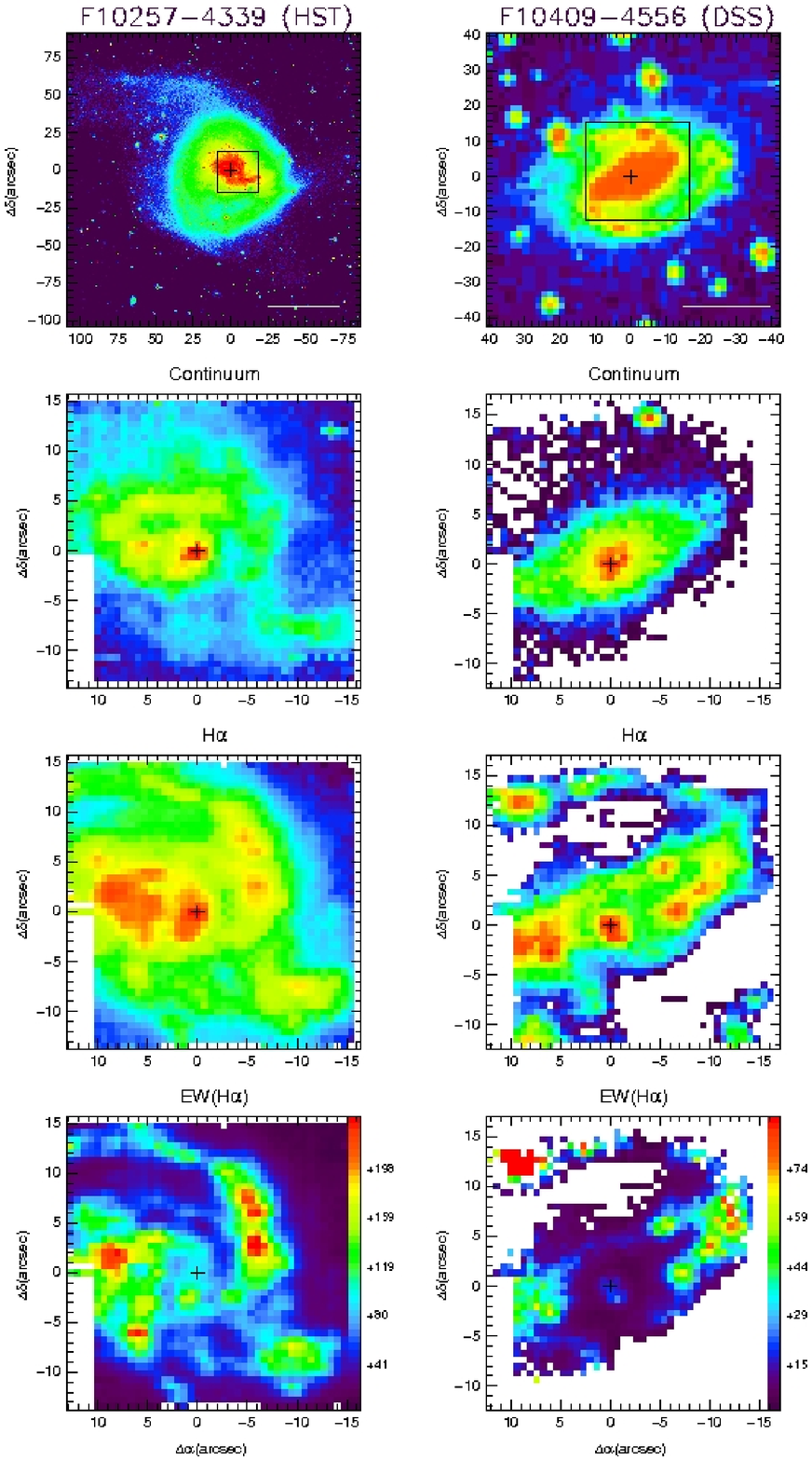}
   \caption{{\it Continued}}
   \end{figure*}

   \addtocounter{figure}{-1}
   \begin{figure*}
   \centering
%   \vskip -0.5cm
   \includegraphics[width=14.cm]{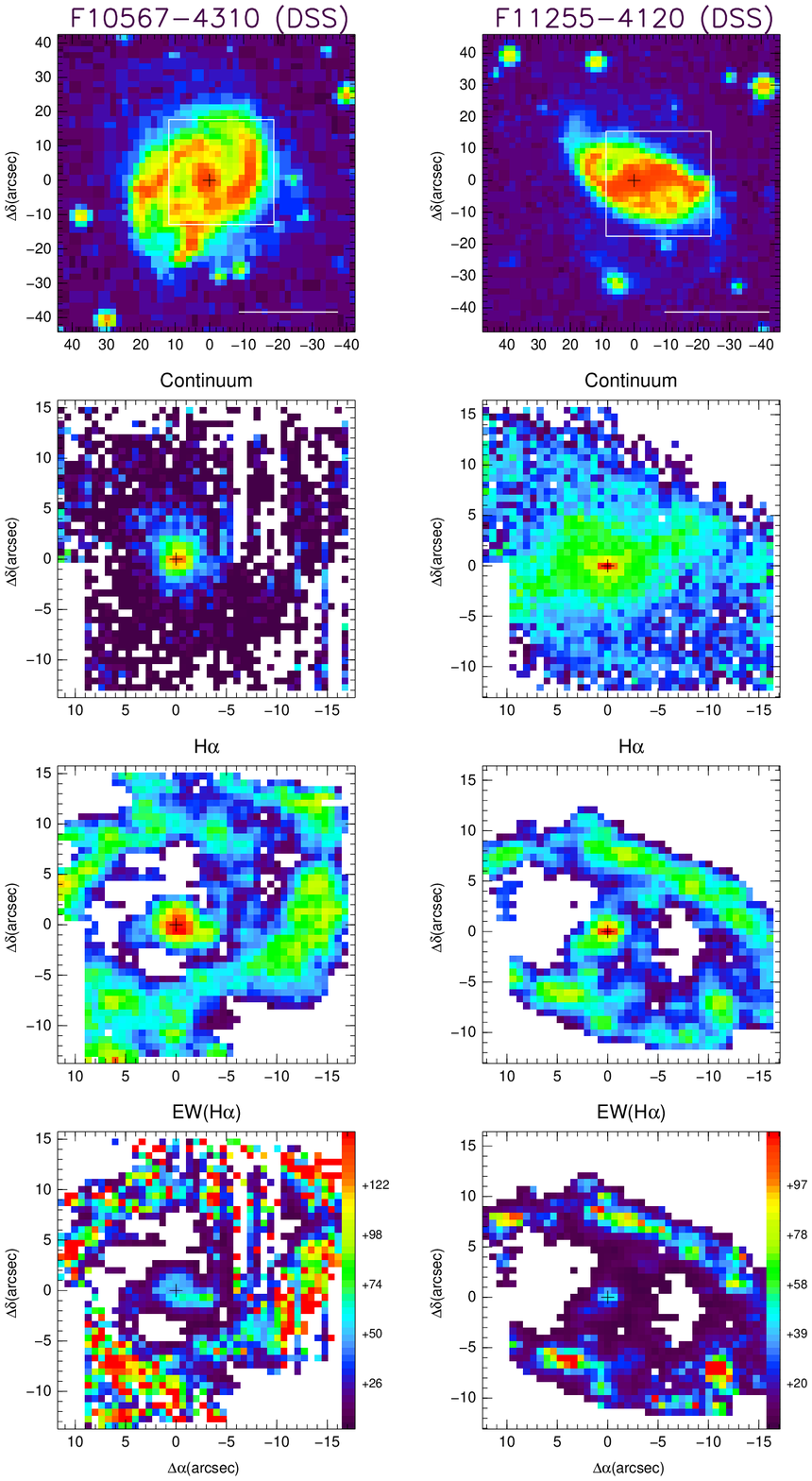}
   \caption{{\it Continued}}
   \end{figure*}

\pagebreak
   \addtocounter{figure}{-1}
   \begin{figure*}
   \centering
%   \vskip -0.5cm
   \includegraphics[width=14.cm]{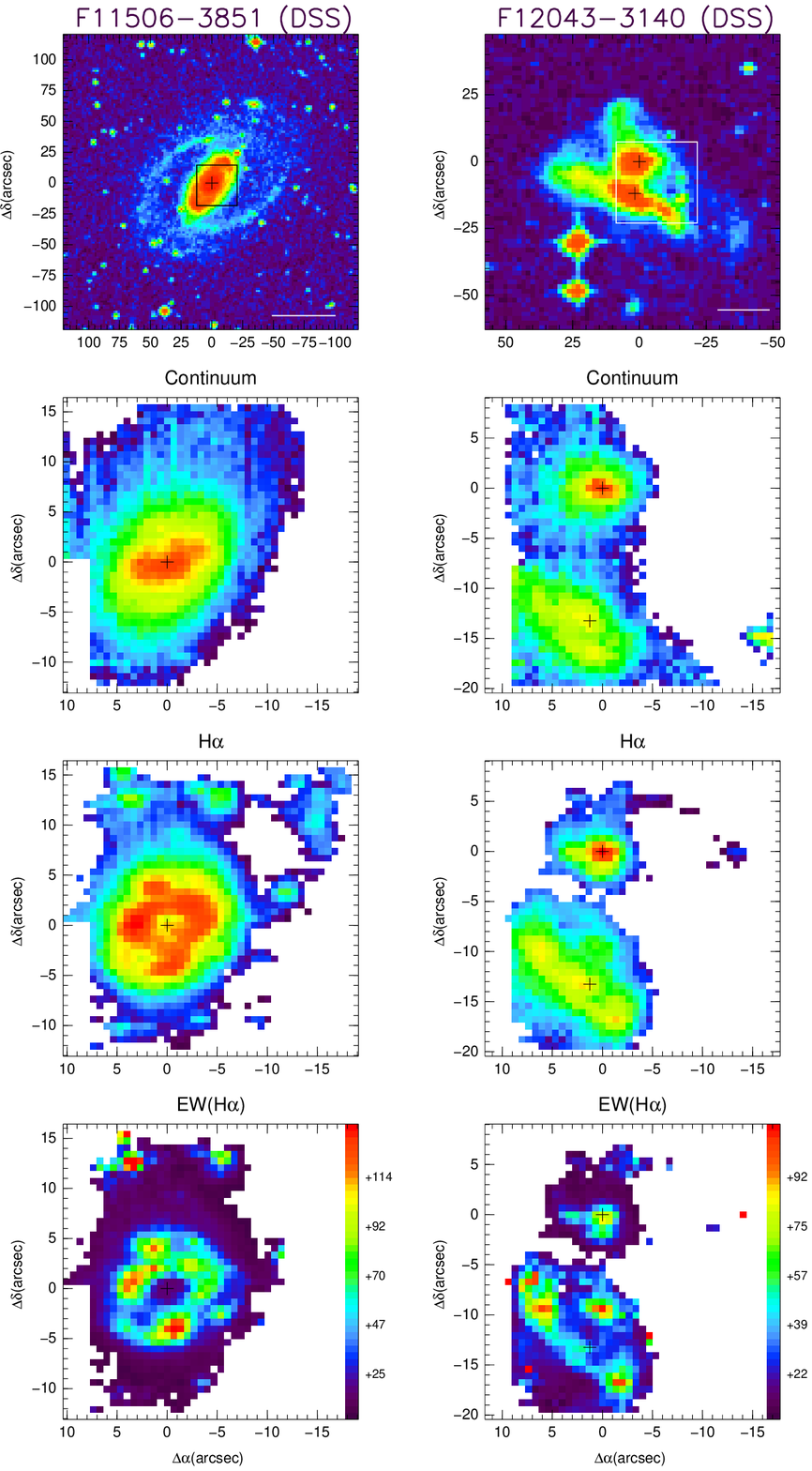}
   \caption{{\it Continued}}
   \end{figure*}

   \addtocounter{figure}{-1}
   \begin{figure*}
   \centering
%   \vskip -0.5cm
   \includegraphics[width=16.cm]{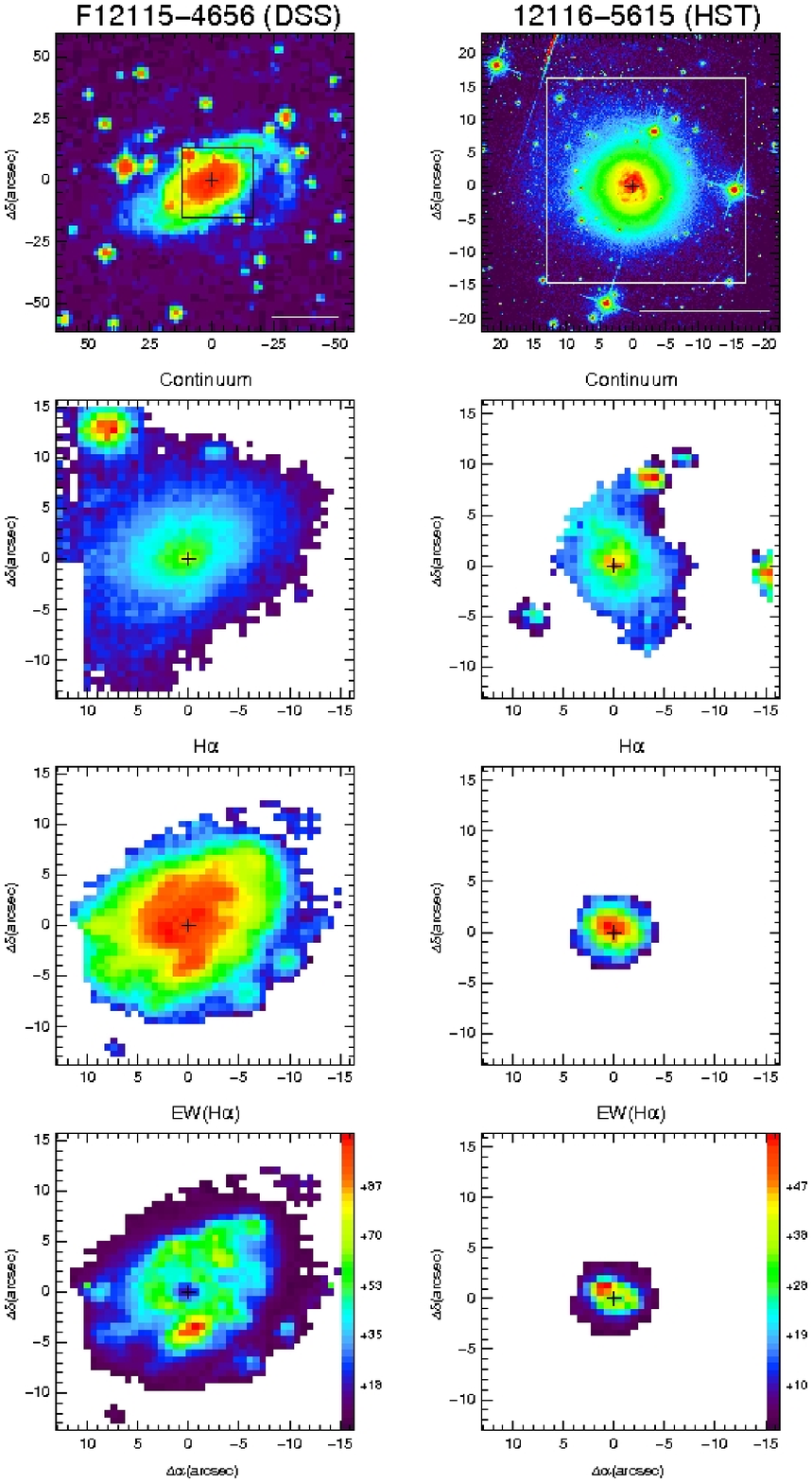}
   \caption{{\it Continued}}
   \end{figure*}

   \addtocounter{figure}{-1}
   \begin{figure*}
   \centering
%   \vskip -0.5cm
   \includegraphics[width=14.cm]{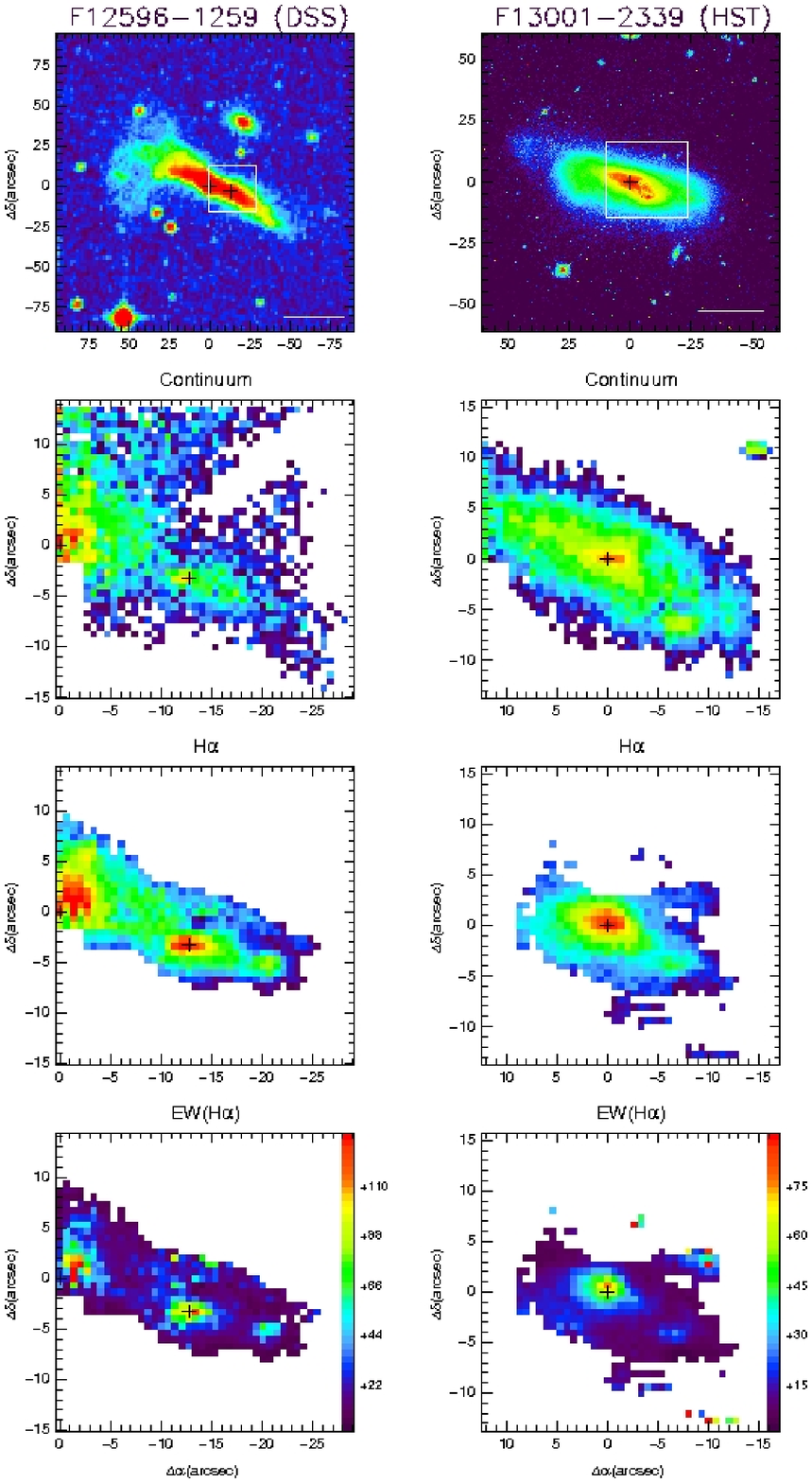}
   \caption{{\it Continued}}
   \end{figure*}

   \addtocounter{figure}{-1}
   \begin{figure*}
   \centering
%   \vskip -0.5cm
   \includegraphics[width=14.cm]{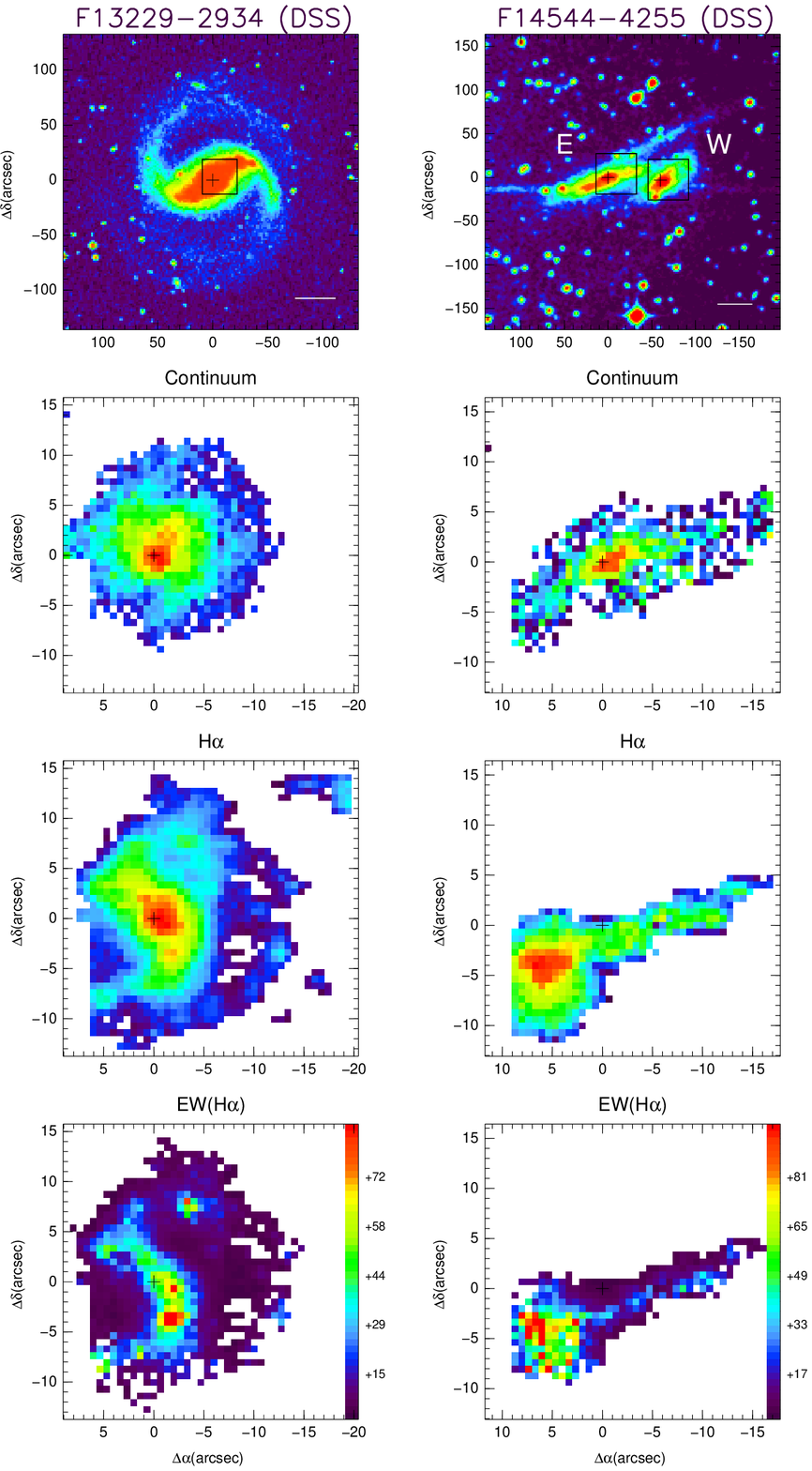}
   \caption{{\it Continued}}
   \end{figure*}

   \addtocounter{figure}{-1}
   \begin{figure*}
   \centering
%   \vskip -0.5cm
   \includegraphics[width=14.cm]{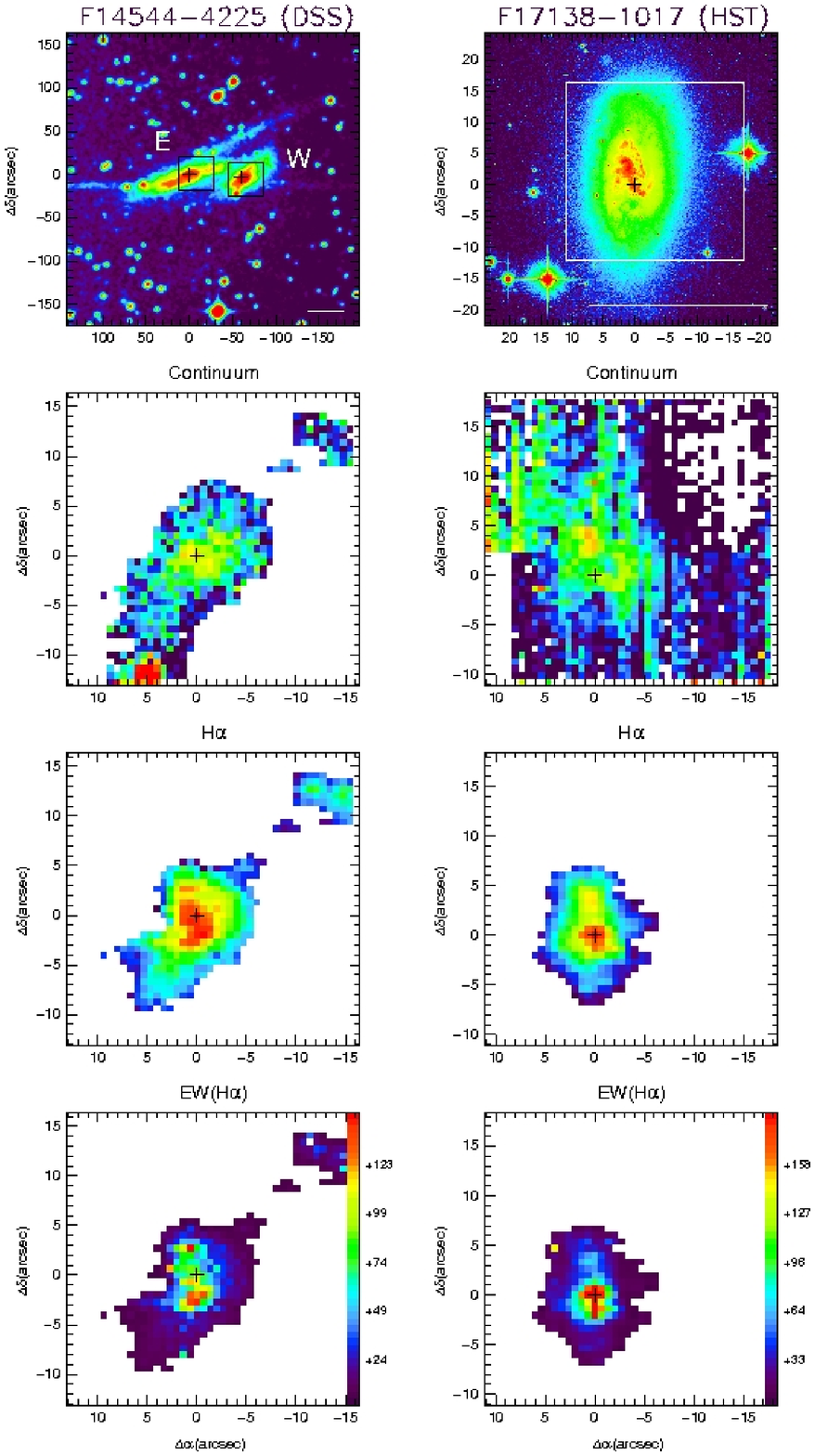}
   \caption{{\it Continued}}
   \end{figure*}

   \addtocounter{figure}{-1}
   \begin{figure*}
   \centering
   \includegraphics[width=7.cm]{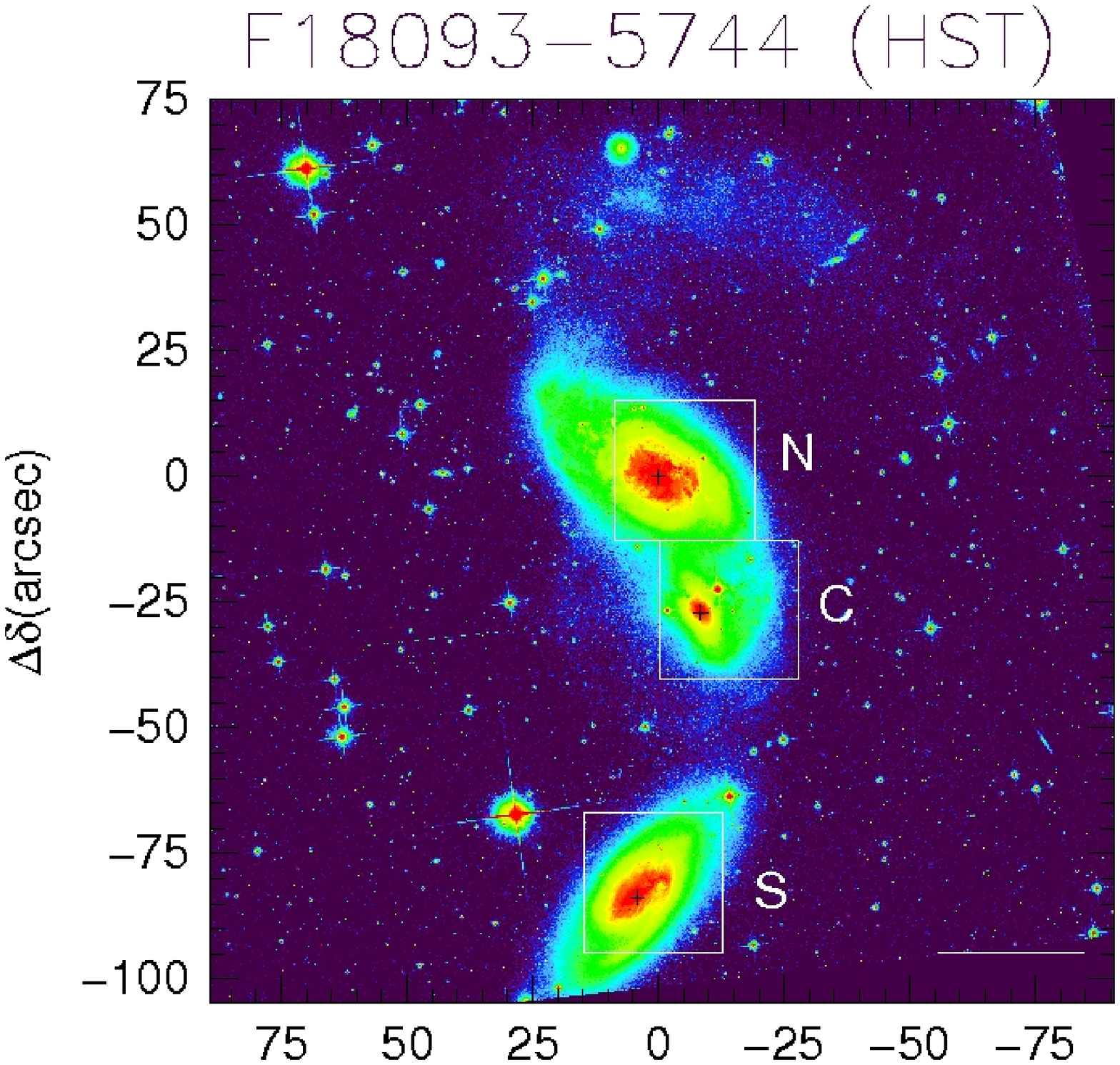}\\
   \vspace{-0.65cm}\includegraphics[width=14.cm]{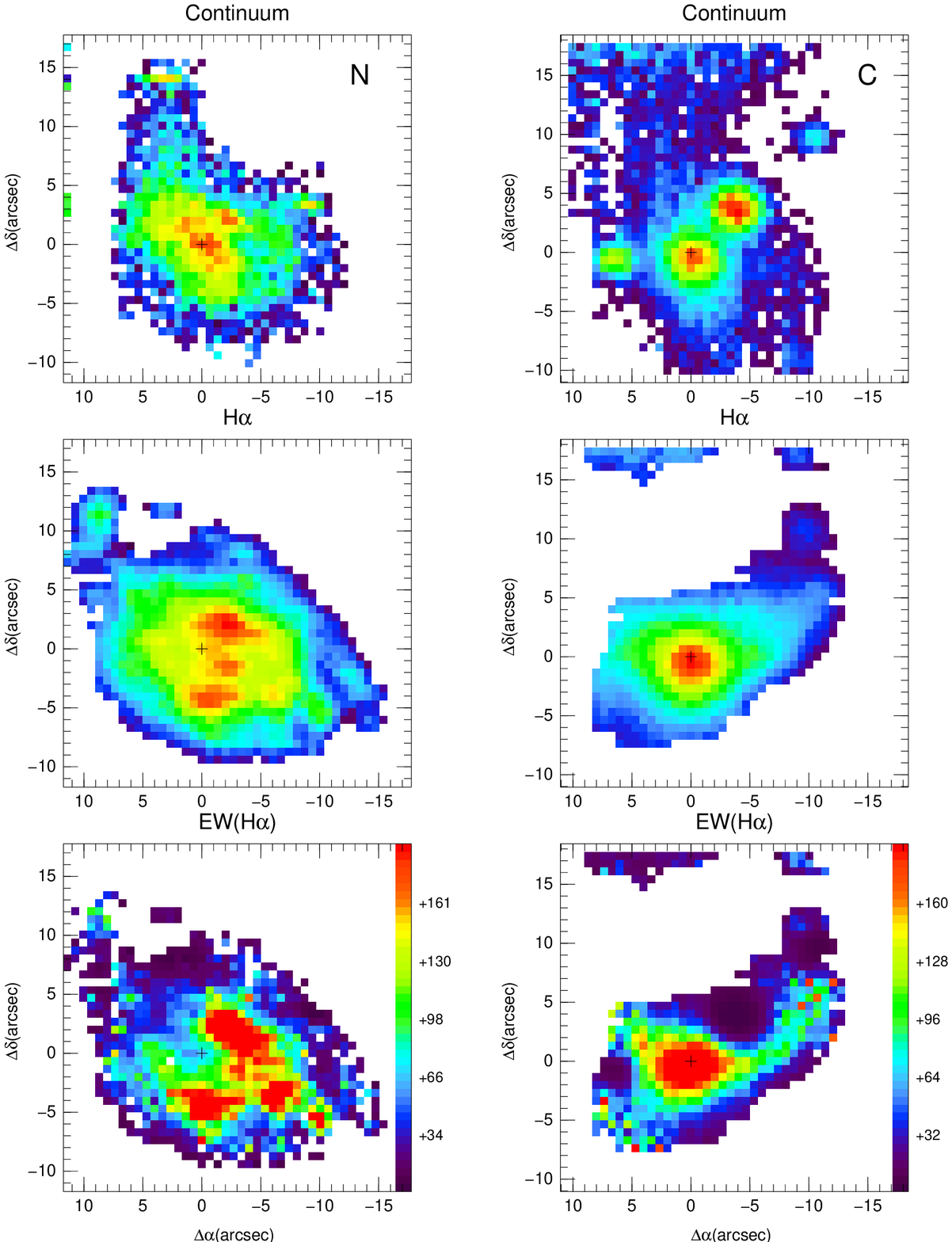}
   \caption{{\it Continued}}
   \end{figure*}

   \addtocounter{figure}{-1}
   \begin{figure*}
   \centering
%   \vskip -0.5cm
   \includegraphics[width=14.cm]{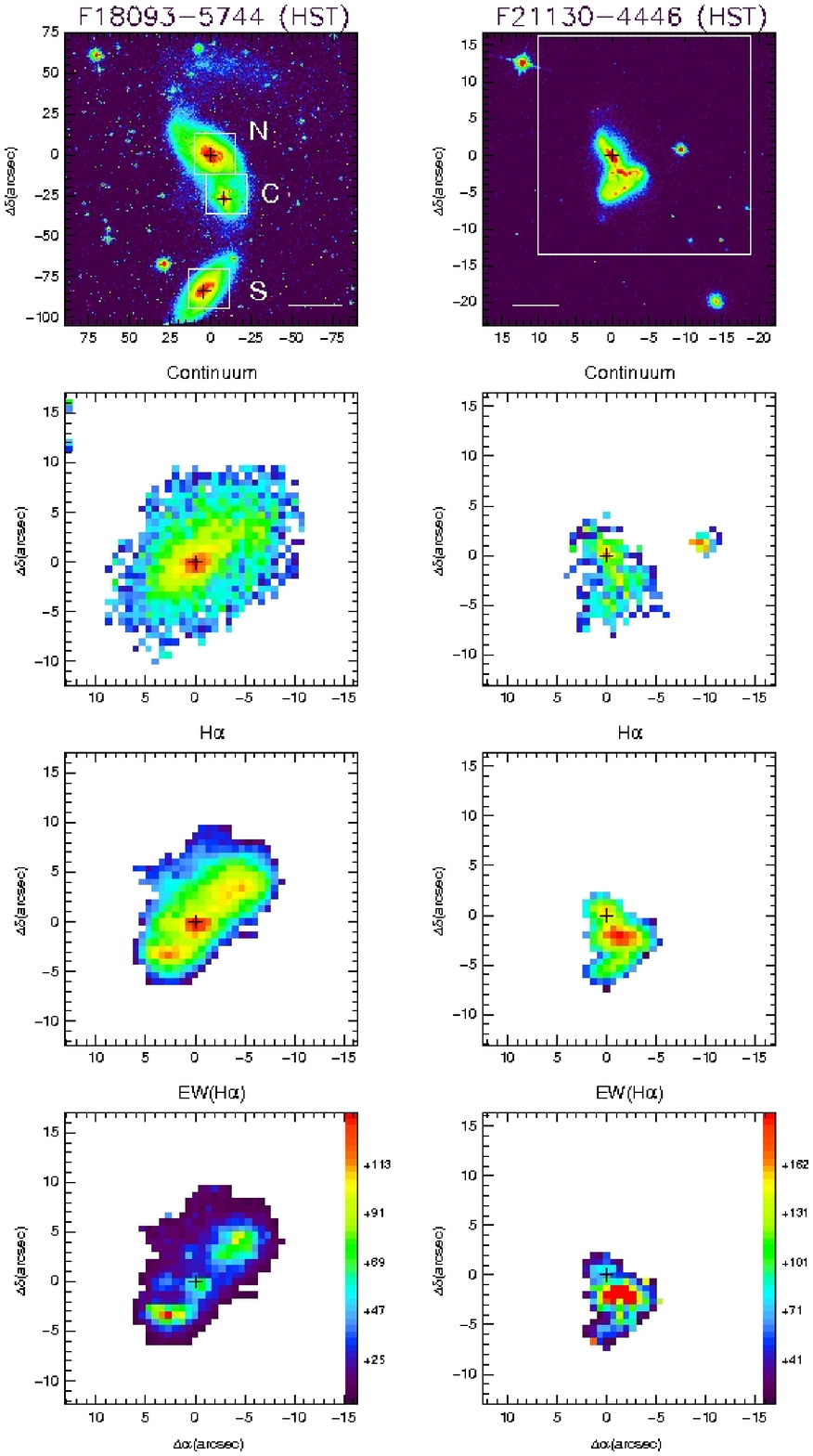}
   \caption{{\it Continued}}
   \end{figure*}

   \addtocounter{figure}{-1}
   \begin{figure*}
   \centering
%   \vskip -0.5cm
   \includegraphics[width=14.cm]{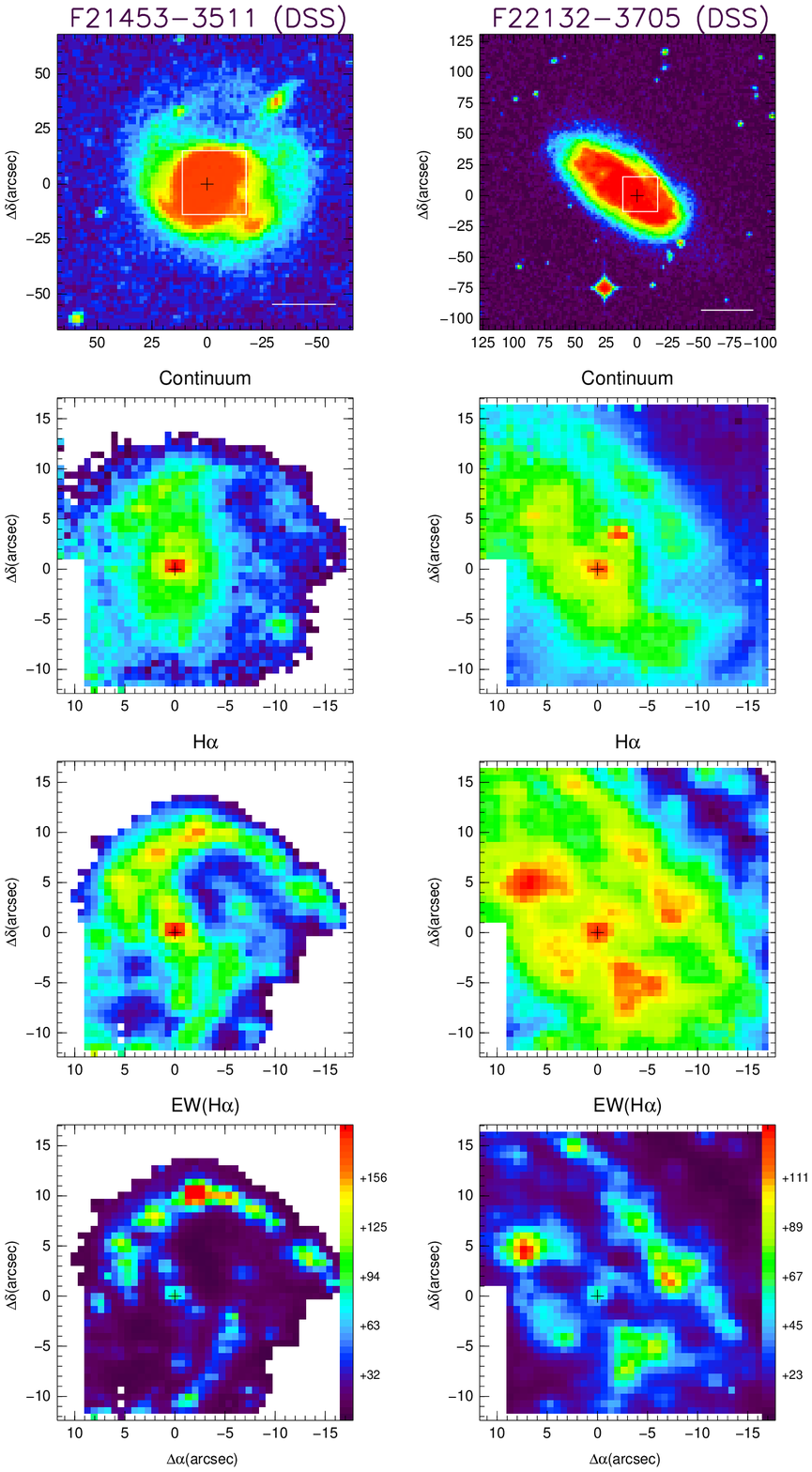}
   \caption{{\it Continued}}
   \end{figure*}

   \addtocounter{figure}{-1}
   \begin{figure*}
   \centering
%   \vskip -0.5cm
   \includegraphics[width=14.cm]{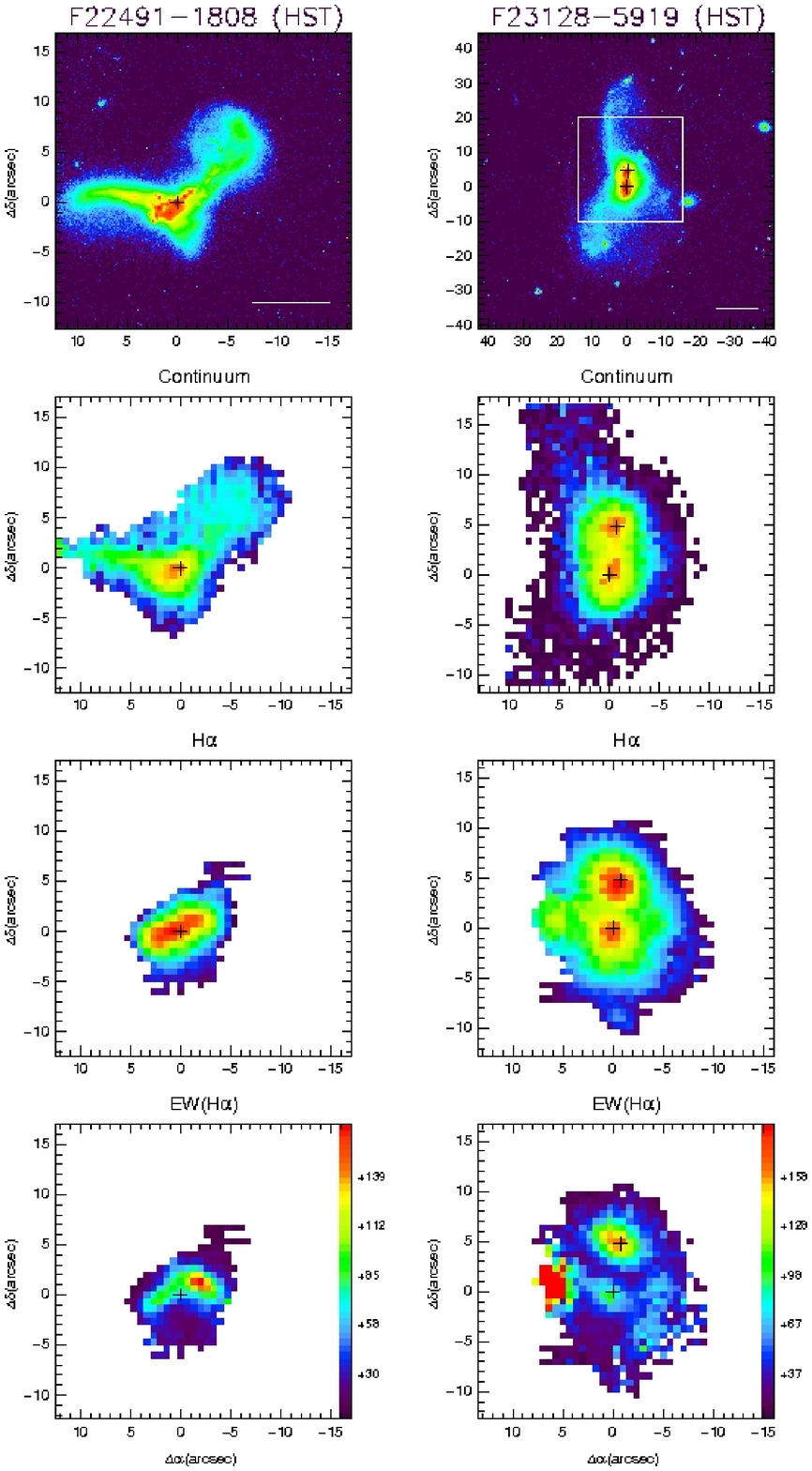}
   \caption{{\it Continued}}
   \end{figure*}

\clearpage
\appendix
\section{Notes on individual sources}

The comments about the morphology of the object refer mainly to Fig.
1 in this paper. All the comments referring to the line ratios in the
extended regions of the systems are taken from Paper II. The derived
ages for the ionizing stellar populations presented in this section
are based on the \cite{Leitherer99} (hereafter LH99) models, for solar
metalicity, instantaneous starburst and \cite{Salpeter55} IMF. Note
that because the spectra were not corrected for stellar continuum
emission, these ages represent upper limits.

\hspace{0.5cm}{\bf IRAS~F01159$-$4443 (ESO 244-G012)}: this is an
 interacting pair with a nuclear separation (NS) of NS
 $\sim$8.5~kpc. The northern galaxy has an optical spectrum of an
 HII-galaxy, while the southern source is classified as ambiguous at
 optical wavelengths \citep{Kewley01,Corbett03}. Owing to limited S/N,
 none of the extended emission seen in the DSS image is visible in our
 continuum image, where we only detect the brightest, nuclear
 emission. On the other hand, the H$\alpha$ emission line map reveals
 several knots (probably associated with star formation) in the
 northern galaxy, while half a ring of a radius of $1.7-2.5$~kpc is
 visible in the southern source. This image also reveals a prominent
 bridge joining the galaxies towards the east of the system. In
 addition, several local peaks of emission are observed to the west,
 with the brightest ones approximately 6 and 5 kpc from the northern
 and southern galaxy, respectively. This is one of the few galaxies in
 the sample for which a 2D study is already available in the
 literature \citep{Rampazzo05}, which makes it a good test-case for
 the VIMOS data. Generally. our results are consistent with those
 presented there, although they less deep.

{\bf IRAS 01341$-$3734 (ESO-297-G011/G012)}: these two galaxies are
 separated by $\sim$25~kpc, which implies the need of two VIMOS
 pointings to cover the system. According to the IR luminosity
 distribution provided in \citet{Surace04} and re-scaling to our
 adopted distance, the northern galaxy would be outside the LIRG
 luminosity range ($\log$(L$_{\rm IR}$/L$_{\odot}$) = 10.65) while the
 southern one would remain as a LIRG ($\log$(L$_{\rm IR}$/L$_{\odot}$)
 = 11.06). The northern galaxy has a substantially extended continuum
 emission that entirely covers our field of view. The H$\alpha$ map of
 the source shows a ``tightly wound spiral arm'' of condensations and
 knots that is not visible in the continuum image and extends
 $\sim$4~kpc from the nuclear region. The line ratios at the locations
 of the knots are typical of an HII region, which is also the case of
 the nuclear spectrum \citep{Kewley01,Corbett03}. Using the values of
 the H$\alpha$ equivalent widths (60 $\lsim$ H${\alpha}$-EW $\lsim$
 140~\AA) and the LH99 models we derived ages for the stellar
 populations at these locations in the galaxy of t $\lsim$ 6.5 Myr.

In the southern galaxy, the H$\alpha$ emission is oriented
 perpendicular to its major axis forming two plumes. As suggested by
 \cite{Dopita02}, this can be interpreted as if the gas were being
 blown out in the polar direction like in M~82. This galaxy is also
 classified as an HII-galaxy at optical wavelengths. The H$\alpha$
 equivalent width values in the nuclear region, and therefore the
 derived ages for the stellar populations, are similar ages to those
 found in the northern source.

{\bf IRAS 04315$-$0840 (NGC 1614):} this is a well studied, late
merger, with bright, spiral structures at scales of few kpc (1-3
kpc). In addition, the ACS {\it HST} image of the galaxy shows
relatively faint extended emission, with a loop-like feature to the
southeast of the system and a tidal tail that is extended $\sim$ 1
arcmin (20 kpc) to the southwest of the nuclear region. The H$\alpha$
image of the source, sampling the central $\sim$9.5 $\times$ 9.5
kpc$^{2}$, shows several knots and condensations extended over the
spiral arm to the east of the system, which is the faintest in
continuum emission. These knots are clearly visible in the
H${\alpha}$-EW image, where additionally a ring-like structure emerges
surrounding the nuclear region (diameter, d $\sim$ 0.7 kpc). The
presence of this ring of star formation was already reported by
\cite{Alonso-Herrero01} in their {\it HST} NICMOS detailed study of a
sample of this LIRG. This galaxy has been classified as an HII-galaxy
at all wavelengths studied (Veilleux et al. 1995; Alonso Herrero et
al. 2001; Corbet et al. 2003), and the line ratios at the location of
the ring are consistent with photoionization by stars. Using the
H$\alpha$ equivalent width values at these locations (100 $\lsim$
H${\alpha}$-EW $\lsim$ 190~\AA) and the LH99 models, we derived ages
of t$ \lsim$ 6 Myr for the stellar populations at these locations in
the galaxy.

{\bf IRAS 05189$-$2524}: according to our VIMOS images, this ULIRG is
 a compact object, especially as seen in the H$\alpha$ emission line
 map. Low surface brightness tidal structures extending up to $\sim$12
 kpc from the nuclear region, are seen in the {\it HST} ACS images of
 the galaxy. The nuclear optical spectrum of the galaxy is that of a
 Sy2 galaxy.

{\bf IRAS 06035$-$7102}: this is a double system with a nuclear
 separation of $\sim$9~kpc. The peaks of the continuum and the
 H$\alpha$ emission have an offset of 1 and 1.4 kpc for the eastern
 and the western sources respectively. Both the continuum and the
 H$\alpha$ images show a prominent tidal tail to the NE of the system
 extended $\sim$21~kpc, which coincides with the location of several
 knots and condensations seen in the {\it HST} WFPC2 image. This
 galaxy is classified as a ULIRG and therefore excluded in the
 study of the ionization mechanisms in the extended regions of LIRGs
 presented in Paper II. However, \cite{Duc97} classified this galaxy
 as an HII-galaxy in their long-slit spectroscopic study of a large
 sample of 24 ULIRGs. If we assume that the knots observed along the
 already mentioned tidal tails are regions of enhanced star formation,
 which is a reasonable assumption, we can use the H$\alpha$ equivalent
 width values at these locations to estimate an age of the stellar
 populations. We find H$\alpha$ equivalent width values of 75 $<$
 H${\alpha}$-EW $<$ 120~\AA, which corresponds to ages of t$ \lsim$ 6.5
 Myr.

{\bf IRAS 06076$-$2139}: this system consists of two galaxies in
interaction with a rather complex morphology. The {\it HST} image of
the source shows a ring of a diameter of $\sim$ 8 kpc surrounding the
nuclear region of the sourthern galaxy, while relative faint, extended
emission is observed on both sides of the northern galaxy. The ring
feature is clearly visible in the H$\alpha$ image. A detailed study of
the galaxy using the current VIMOS dataset was already presented in
\cite{Arribas08}. The authors found that although thet interact, it is
unlikely that these two galaxies finally merge. The southern nucleus,
as well as the ring and the external clumps, has HII-like line
ratios. The derived ages for the stellar populations at these
locations are t$ \lsim$ 10 Myr. The two clumps of ionized gas emission
observed to the west of the system similar properties to that of TDGs
candidates detected in ULIRGs \citep{Monreal-Ibero07,Arribas08}.

{\bf IRAS 06206$-$6315}: the HTS WFPC2 image of the system shows a
double nuclei structure (NS $\sim4.3$~kpc) that is also visible in the
VIMOS continuum, and even more clearly in the H$\alpha$ image. The
latter shows a tidal tail starting in the north and bending towards
the southeast, which contains a local peak of emission. IRAS
06206$-$6315 is classified as a ULIRG and has an optical spectrum of a
Sy2 galaxy.

{\bf IRAS F06259$-$4708 (ESO 255-IG007)}: two VIMOS pointing were used
 during the observation of this triple system. The two brightest
 galaxies, are separated by a distance of $\sim$11~kpc, while the
 third one is located at $\sim$15~kpc towards the southeast of the
 main pair. The central galaxy presents two prominent spiral
 arms/tidal tails in the {\it HST} ACS images, which can be also
 delineated in both our continuum and H$\alpha$ VIMOS images,
 specially the one for the ionized gas. The {\it HST} ACS image of the
 southern galaxy, which is the faintest in both the continuum and the
 H$\alpha$ images, shows dust features crossing the galaxy body. Our
 VIMOS images show an offset between the continuum and the ionized gas
 emission of $\sim$1.4 kpc. High H$\alpha$ equivalent width values
 (100 $\lsim$ H${\alpha}$-EW $\lsim$ 290~\AA) are found in the
 circumnuclear regions of the northern galaxy, towards the west and
 the south of of the central galaxy and crossing the body of the
 southern source from north to south. The line ratios at these
 locations of the galaxies are consistent with photoionization by
 stars. We derive ages of t$ \lsim$ 6 Myr for the stellar populations
 located in these regions.

{\bf IRAS F06295$-$1735 (ESO 557-G002)}: the DSS image of this barred
 spiral shows a companion galaxy $\sim$42~kpc towards the
 south. Interestingly, neither the arms nor the bar in the H$\alpha$
 image coincide with those in the continuum image. \cite{Corbett03}
 classified this galaxy as an HII-galaxy in their long-slit
 spectroscopic study of LIRGs. Furthermore the line ratios at almost
 all locations in the galaxy are consistent with photoionization by
 stars. However, owing to the poor sensitivity for the continuum image
 the values of the H$\alpha$ equivalent width are relatively
 unconstrained. Therefore, no attempt was made to estimate the ages of
 the young stellar populations for this galaxy.

{\bf IRAS 06592$-$6313}: This spiral galaxy presents a condensation in
 the H$\alpha$ image outside its main body at $\sim$3~kpc towards the
 north, which is not present in the continuum image. The optical,
 nuclear spectrum of this source is that of an HII-galaxy
 \citep{Corbett03}. Moderate H$\alpha$ equivalent width values (20 $<$
 H${\alpha}$-EW $<$ 50~\AA) are found in the nuclear region, in the
 extreme of the eastern spiral arm and in the condensation to the
 north of the system. The derived stellar ages are t $<$ 8 Myr.

{\bf IRAS F07027$-$6011 (AM 0702-601)}: this system consists of two
 galaxies separated by $\sim$54~kpc. On the basis of its nuclear
 spectrum, the northern galaxy is classified as a Sy2 at optical
 wavelengths \citep{Kewley01}. The H$\alpha$ image of this galaxy
 shows two spiral arms towards the north and south, as well as a chain
 of knots embedded within the main body of the galaxy, which are
 extended $\sim$3~kpc towards the southwest. This chain structure is
 even clearer in the corresponding H$\alpha$ equivalent width image,
 where it also extends towards the northwest of the galaxy. The
 location of these knots corresponds to regions of line ratios
 consistent with photoionization by stars. Using the H${\alpha}$
 equivalent width values (100 $\lsim$ H${\alpha}$-EW $\lsim$ 175~\AA)
 and the LH99 models we derive ages of t$ \lsim$ 6 Myr for the stellar
 populations located in these knots.

The H$\alpha$ emission from the southern galaxy is more concentrated
 than that of the continuum, and is associated with several
 circumnuclear condensations seen in the {\it HST} ACS images of the
 galaxy. There is no nuclear spectroscopic information available in
 the literature for this galaxy. The line ratios in the circumnuclear
 region of the galaxy are typical of HII-like regions. The
 H${\alpha}$ equivalent width values and the stellar ages at this
 location are similar to those found in the northern galaxy.

{\bf IRAS F07160$-$6215 (NGC 2369)}: the VIMOS field of view covers
 the central $7\times7$~kpc$^2$ of this spiral galaxy. Both the
 continuum and the ionized gas images show irregular
 structures. Particularly, the H$\alpha$ image, which is in itself
 substantially different from the continuum image, is full of knots
 and condensations. The clumpy structures in the central 5 kpc visible
 in our H$\alpha$ image are also observed in the high-resolution
 HST-NICMOS Pa${\alpha}$ image of the galaxy
 \citep{Alonso-Herrero06}. The regions with relatively high
 H${\alpha}$ equivalent width values (50 $\lsim$ H${\alpha}$-EW
 $\lsim$ 100~\AA) observed across the nuclear region and to the west
 of the system have line ratios consistent with photoionization by
 stars. The derived stellar ages are t$ \lsim$ 6.5 Myr.

{\bf IRAS 08355$-$4944}: the DSS and our VIMOS continuum image show a
single nucleus galaxy with two tidal tails towards the north and the
southwest of the system. However, a double nucleus structure (NS =
0.34 kpc) emerges in the high-resolution {\it HST} ACS images of this
object. The morphology of the H$\alpha$ image is very different from
that of the continuum. A tidal structure, with a slightly different
orientation than the tidal tails observed in the continuum, is visible
to the north of the system. The ACS images present two condensations
to the west, which are associated with a relatively bright area in the
H$\alpha$ map. Substantially high H$\alpha$ equivalent width values
(140 $\lsim$ H${\alpha}$-EW $\lsim$ 250~\AA) are observed in the
circumnuclear region of the galaxy. The line ratios at this location
are typical of HII-like regions. Using the LH99 models, we derived
stellar ages of t$ <$ 6 Myr.

{\bf IRAS 08424$-$3130 (ESO 432-IG006)}: the Digital Sky Survey (DSS)
image of the galaxy shows a pair of spiral galaxies (NS $\sim$ 9 kpc)
in interaction, with tidal structures such as a bridge of emission
between the two systems and a prominent tidal tail towards the
southwest of the southern galaxy. Only part of the nuclear region of
both galaxies is covered by our VIMOS field of view. The regions with
high H$\alpha$ equivalent width values (H${\alpha}$-EW $\sim$ 35~\AA)
observed in the nucleus and to the east of the nuclear region in the
southwestern source have line ratios typical of HII-like regions. The
ages derived for the young stellar populations at these locations are
t$ \lsim$ 7 Myr.

{\bf IRAS F08520$-$6850 (ESO 60-IG016)}: the {\it HST} ACS image of
this object shows two disk galaxies in interaction, which is
consistent with the morphology observed in both our continuum and the
H$\alpha$ images. Owing to the several dust features crossing the main
body of the western source in the ACS image it is hard to decide the
location of the nucleus in this galaxy and therefore to estimate a
nuclear separation. This dust obscuration would explain the relatively
faint continuum and H$\alpha$ emissions from this galaxy. In the
eastern source, there is an offset between the peaks of the continuum
and the ionized gas emission of $\sim$1.9~kpc. Note that the H$\alpha$
image shows a small tidal structure towards the south of the system
that is not visible in continuum emission. The higher H$\alpha$
equivalent width values are concentrated in the nuclear region of the
eastern galaxy, where they reach values as high as 410~\AA. However,
this region is not included in the study presented in Paper II and we
did not find spectroscopic information of the source in the
literature. Therefore, no attempt to estimate stellar ages was made
for IRAS F08520$-$6850.

{\bf IRAS F09022$-$3615}: no tidal features are seen in the VIMOS
 images of this source, which is classified as a ULIRG. However, both
 the DSS and the {\it HST} ACS images show a prominent tidal tail that
 emerges from the south and bends towards the east forming a
 semicircular structure of about $\sim$45~kpc. In addition, the ACS
 image of the source reveals a complex nuclear structure, with several
 knots and condensations. Our H$\alpha$ image shows a rather simple
 structure, with the bulk of the emission concentrated in the nuclear
 region. Two regions with high H$\alpha$ equivalent width values
 ($\sim$ 170~\AA) are observed immediately to the east and west of the
 nucleus, indicated with a cross in the VIMOS images. Classified as a
 ULIRG, this source was not included in the work presented in Paper
 II. In addition, no nuclear spectral classification was found in the
 literature for this object. Therefore, no attempt to estimate stellar
 ages was made for IRAS F09022$-$3615.

{\bf IRAS F09437+0317 (IC563/IC564)}: this is a system of two galaxies
 (north:IC564/south:IC563) with a nuclear separation of NS $\sim$ 39
 kpc. Three VIMOS pointings were required to cover most of the
 emission from the system. IRAS F09437+0317 is a LIRG but each
 individual galaxy falls outside the LIRG luminosity range. For the
 adopted distance, the logarithms of the infrared luminosities in
 solar units are 10.90 and 10.95 for IC563 and IC564 respectively
 \citep{Surace04}. The continuum and the H$\alpha$ images show marked
 differences for both galaxies. For IC563, a bar is detected in our
 continuum image at P.A$\sim$45$^{o}$, although it is not as well
 traced in the H$\alpha$ image. The H$\alpha$ emission from this
 galaxy is concentrated in the extremes of the bar, to the northwest
 and the southeast of the nuclear region. Indeed, the peak of the
 H$\alpha$ emission, which is located to the southeast of the system,
 has an offset of $\sim$4~kpc which respect to the maximum of the
 continuum emission. High H$\alpha$ equivalent width values (150
 $\lsim$ H${\alpha}$-EW $\lsim$ 215~\AA) are measured in the extremes
 of the bar, and towards the north of the galaxy. The line ratios at
 these locations are typical of HII-like regions. Using the LH99
 models, we derived stellar ages of t$ <$ 6 Myr.

 In the northern galaxy (IC564), two pointings sample the northeast
 and the southwest of the galaxy, and are referred to as northern
 pointing 1 and 2 (NP(1) and NP(2)) in Fig. 1. The peak of the
 H$\alpha$ emission is located $\sim$6.5 kpc to the east of the
 nuclear region (both the peak of the continuum and the H$\alpha$
 emission fall in the northeastern pointing). The H$\alpha$ image
 shows several concentrations of emission extended throughout the
 entire body in the galaxy. Equivalent width values in the range 60
 $\lsim$ H${\alpha}$-EW $\lsim$ 140~\AA~are found towards the east and
 southwest of the galaxy. The line ratios at these locations in the
 galaxy are consistent with photoionization by stars. Using the LH99
 models, we derived stellar ages of t$ \lsim$ 6.5 Myr.

{\bf IRAS F10015$-$0614 (NGC 3110)}: this galaxy shows two well
defined spiral arms in the DSS image. The spiral arm extended from the
east to the south of the galaxy falls outside our field of view. The
H$\alpha$ emission from the galaxy shows clumpy structures to the west
and the northeast of the nuclear region extended through the spiral
arms, which is also visible in Pa${\alpha}$ emission
\citep{Alonso-Herrero06}. H$\alpha$ equivalent widths ranging from
100~\AA~up to values as high as 266~\AA~are found coinciding with the
location of such clumpy structures. The derived ages for the stellar
populations in these regions are t$ \lsim$6 Myr. However, note that
because of the low S/N of the continuum image in this particular case
the values of the H$\alpha$ equivalent width are less
constrained. Finally, the DSS image shows a companion galaxy observed
$\sim$37~kpc towards the southwest, whose with might be interacting.

{\bf IRAS F10038$-$3338 (IC2545)}: the ACS image of the galaxy shows
 two close (NS $\sim$ 0.6 kpc) nuclei as well as two prominent tidal
 tails bending from the east towards the north and from the west
 towards the south respectively. This last tail seems to be associated
 with a relatively bright region in the H$\alpha$ emission line
 map. However, because of S/N limitations, these tidal structures are
 not visible in our continuum map. High H$\alpha$ equivalent width
 values (110 $\lsim$ H${\alpha}$-EW $\lsim$ 140~\AA) are found
 confined to a small region of few spaxels immediately towards the
 southeast of the nucleus in our VIMOS images (marked with a
 cross). The line ratios at this location are that of an HII-like
 region, and the derived ages for the stellar populations are t$
 \lsim$ 6 Myr.

{\bf IRAS F10257$-$4338 (NGC 3256)}: the {\it HST} ACS image of the
galaxy shows extended emission up to $\sim$30 kpc from the nuclear
region. In addition, a rather complex structure emerges in the center
of the galaxy, with several knots, condensations, and other tidal
structures. This complex structure is also clearly visible in our
VIMOS images, which samples the central $\sim$6 $\times$ 6~kpc$^2$ of
the galaxy. \cite{Lipari04} and \cite{Alonso-Herrero06} presented
ESO-NTT and HST WFPC2-WF2 H$\alpha$, and Pa$\alpha$ images of the
galaxy, which show clumps located to the east and west of the nuclear
region. These knots are also visible in our H$\alpha$ images and
moreover, in the corresponding H$\alpha$ equivalent width image of the
galaxy. The line ratios at the location of these knots are consistent
with photoionization by stars. Using the H$\alpha$ equivalent width
values, which are indeed remarkably high (200 $\lsim$ H${\alpha}$-EW
$\lsim$ 280~\AA) and the LH99 models, we estimate an age of t$ \lsim$
5 Myr for the stellar populations in these knots.

{\bf IRAS F10409$-$4556 (ESO 264-G036)}: this is an isolated barred
 spiral galaxy as seen in the DSS image. The bar is visible both in
 the continuum and H$\alpha$ images at P.A.$\sim$70$^{o}$. On the
 other hand, the spiral arm towards the north of the galaxy is only
 detected in the ionized gas map. The southern spiral arm falls mostly
 outside the VIMOS field of view. There are two relatively symmetric
 regions with enhanced H$\alpha$ emission associated with the bar
 structure. In addition, a notable increase of the ionized gas
 emission is also observed in the extreme of the spiral arm extended
 to the north of the galaxy. These regions have line ratios consistent
 with that of HII-like regions, and therefore trace the location of
 ongoing star-formation activity. The H$\alpha$ equivalent width
 values are 30 $\lsim$ H${\alpha}$-EW $\lsim$ 90~\AA~and
 H${\alpha}$-EW $\sim$ 160~\AA~for the regions associated to the bar
 and the spiral arm respectively. Using the LH99 models, we obtain
 ages of t$ \lsim$ 6 and $\lsim$ 5 Myr for the young stars at these
 locations in the galaxy. Owing to the low signal of the continuum
 image in the extreme of the northern spiral arm the H$\alpha$
 equivalent width values at that location are less constrained.

{\bf IRAS F10567$-$4310 (ESO 264-G057)}: the DSS image of the galaxy
shows a spiral structure with two spiral arms to the east and west of
the nuclear region. The image also shows what might be a bar structure
crossing the nuclear region at P.A.$\sim 10^{o}$ and an additional
extended emission to the southeast of the galaxy. Unfortunately, the
the data-cube of IRAS F10567-4310 showed vertical patterns over the
entire field of view. Possibly due to a non-linear effect, it was not
possible to remove these patterns during the reduction
process. However, these patterns are only important if the S/N is low,
and therefore, although they affect the morphology of the continuum
(and the corresponding H${\alpha}$-EW) image, they have no effect on
the H${\alpha}$ emission map. The H${\alpha}$ image shows a rather
complex morphology, with an impressive ``tightly wound spiral arm''
extended over the entire VIMOS field of view ($\sim$10 $\times$ 10
kpc$^2$).

{\bf IRAS F11255$-$4120 (ESO 319-G022)}: this is a barred spiral with
 a ring extended up to $\sim$4 kpc from the nuclear region, clearly
 detected in our H$\alpha$ image. Interestingly, the orientation of
 the bar seen in continuum emission (P.A.$\sim 110^{o}$) is different
 from that of the ionized gas emission (P.A. $\sim 150^{o}$). Several
 knots, along the ring structure, are observed in the H${\alpha}$ and
 the H${\alpha}$-EW images. The line ratios measured over the
 entire ring are consistent with photoionization by stars. The
 H${\alpha}$-EW values at the location of the knots are $\gsim$
 100~\AA~, being as high as 348~\AA.  Using the LH99 models we obtain
 ages t$ \lsim$ 6 Myr for the stellar population located in this ring
 of star formation.

{\bf IRAS F11506$-$3851 (ESO 320-G030)}: the DSS image of the galaxy
 shows an overall spiral structure with extended emission over $\sim$
 30 kpc. Our VIMOS images covers the central $\sim6 \times
 6$~kpc$^2$. The H${\alpha}$ image shows four concentration of ionized
 gas emission towards the north and the northwest of the nuclear
 region. In addition, a ring structure, not detected in the continuum
 image, emerges in the H${\alpha}$, and moreover, the H${\alpha}$-EW
 image. The presence of this ring was already reported by
 \cite{Alonso-Herrero06} in their study of NICMOS-Pa$\alpha$ images of
 LIRGs. The line ratios over such a ring structure are typical for
 HII-like regions. High H${\alpha}$-EW values (100 $\lsim$
 H${\alpha}$-EW $\lsim$ 150~\AA) are found for the knots observed
 within the ring, revealing the location of a stellar population of an
 age of t$ \lsim 6$ Myr.

{\bf IRAS 12043$-$3140 (ESO 440-IG058)}: this system consists of two
 merging galaxies with a separation of $\sim$6 kpc. The northern
 galaxy is very compact and has an optical spectrum that is a mix
 between LINER and HII \citep{Corbett03}. The southern source presents
 several knots in the continuum and H$\alpha$ maps as well as two
 tidal plumes that are visible in the corresponding DSS image. The
 knots are aligned forming what seems to be a ring-like structure
 as seen in the H${\alpha}$-EW image. This galaxy is spectroscopically
 classified as an HII-galaxy at optical wavelengths, and the radio
 observations of \citep{Condon96} suggest that it dominates the far-IR
 emission. The line ratios observed throughout the ``ring structure''
 are consistent with photoionizations by stars. H${\alpha}$-EW values
 (50 $\lsim$H${\alpha}$-EW $\lsim$ 110~\AA) are found through the
 ``ring structure'' in the southern source. Using the LH99 models, we
 derive ages for the stellar populations at these locations of t$
 \lsim$ 6.5 Myr.

{\bf IRAS 12115$-$4656 (ESO 267-G030)}: This galaxy, which is included
in the catalog of interacting galaxies of \cite{Arp87}, might be
interacting with IRAS 12112-4659, located $\sim$ 260 arcsec (97.5 kpc)
to the southwest of the system. A prominent spiral arm/tidal tail is
observed to the northwest of the galaxy in the DSS image of the
object, but not detected in the VIMOS continuum or H${\alpha}$
images. The H${\alpha}$ image shows a concentration of the emission
towards the nuclear region, while a ring structure emerges in the
corresponding H${\alpha}$-EW image. The line ratios over the whole
extension of the ring are consistent with those of HII-like
regions. Using the values of the H${\alpha}$-EW, ranging from
30~\AA~to values as high as 107~\AA~to the south of the nucleus, we
derived ages of t$ \lsim$ 6.5 Myr for the stellar populations within
this star-forming ring.

{\bf IRAS 12116$-$5615}: this is one of the few objects for which the
morphological classification is controversial. The overall structure
of the galaxy is substantially symmetric, which would be in favor of
morphological class 0. However, the {\it HST} ACS image of the galaxy
reveals a more complex nuclear morphology, with a bright structure
emerging from the east of the nuclear region and bending towards the
northwest of the system. It is possible that this complex, nuclear
structure is related to a past/recent interaction, in which case
this galaxy would be classified as type 2, our preferred morphological
classification.

The H${\alpha}$ image of the source shows a compact morphology, with
most of the emission concentrated in the central 3 $\times$ 3
kpc$^{2}$ region. Since no nuclear optical diagnostic was found for
this objects and the nuclear regions of the system were excluded
from the work presented in Paper II, no attempt was made to estimate
the ages of the young stellar populations for this galaxy.

{\bf IRAS 12596$-$1529 (MGC-02-33-098)}: the DSS image of this source
shows a distorted, elongated morphology along P.A.$\sim$65$^{\circ}$
with diffuse emission to the northwest of the galaxy, which falls
outside our VIMOS field of view. This galaxy is interacting with
MCG-02-33-099, $\sim$115 arsec ($\sim$37kpc) to the southeast of the
system. IRAS 12596$-$1529 was misclassified as 2 in Paper I. A double
nucleus structure is not clearly visible in the DSS image, the
presence of two nuclei separated by $\sim$12 arcsec (NS $\sim$ 3.8
kpc) has already been reported in the past by other authors
\citep{Veilleux95,Kewley01,Alonso-Herrero06}, and therefore, the
galaxy is morphologically classified as 1.

Most of the H$\alpha$ emission from the system is concentrated around
the two nuclei, coinciding with the location of high H${\alpha}$-EW
values (100 $\lsim$H${\alpha}$-EW $\lsim$ 130~\AA). Both nuclei are
also spectroscopically classified as HII-like regions at optical
wavelengths.  \citep{Veilleux95,Corbett03}. Using the LH66 models we
derived ages of t$ \lsim$ 6 Myr at these location in the system.

Finally, it is important to mention that the data cube of IRAS
F12596-1529 showed vertical patterns within a region to the east of
our VIMOS FOV. In this case these vertical patterns are due to an
incorrect fiber tracing during the reduction and affects both the
continuum and the H${\alpha}$ maps. After substantial experimentation,
it was not possible to entirely correct for this effect during the
reduction process.

{\bf IRAS~F13001$-$2339 (ESO 507-G070)}: the DSS image of the source
shows a relatively small (tidal?) structure towards the northeast of
the main body of the galaxy, which is in itself rather symmetric. In
addition, the high resolution {\it HST} ACS image of the galaxy also
shows a second high surface brightness region $\sim$ 4 kpc towards the
southwest of the nuclear region. Although this region could be
identified as a secondary nucleus, it has no associated extended
structure, and it seems more like a massive star-forming region. At
this stage, it is straightforward to understand that it is not trivial
to classify this object as 0, 1, or 2. Our preferred classification for
IRAS~F13001$-$2339 is as type 2.

A hint of this already mentioned secondary bright concentration is
visible in our continuum image, but it disappears in the image tracing
the ionized gas emission. In addition, the H$\alpha$ image of the
galaxy shows a small structure extending $\sim$4.5 kpc to the
northwest of the nuclear region. The optical, nuclear spectrum of the
source is that of a LINER \citep{Corbett03}, which is also consistent
with the line ratios found in the extended regions.

{\bf IRAS~F13229$-$2934 (NGC 5135)}: the {\it HST} WFPC2 image the
galaxy shows a spiral structure, with several clumps in the nuclear
region and relatively faint emission extending up to $\sim$40 arcsec
($\sim$ 11 kpc) to the southeast of the galaxy. The spiral structure
is still distinguishable in our VIMOS continuum images. The morphology
of the ionized gas emission is substantially different than that of
the continuum, with some filaments extending towards the west of the
galaxy and a region of enhanced emission coinciding with the nuclear
clumps observed in the {\it HST} WFPC2 and NICMOS-Pa$\alpha$ images of
the source \citep{Alonso-Herrero06}. This region is better delineated
in the H${\alpha}$-EW image, where some clumps with high
H${\alpha}$-EW values (90 $\lsim$ H${\alpha}$-EW $\lsim$ 120~\AA) are
also visible. Although the nuclear spectrum of the galaxy is that of a
Sy2-galaxy, the line ratios found throughout this region are
consistent with photoionization by stars. We derived ages of t$ \lsim$
6 Myr for the stellar populations located in the already mentioned
clumpy structures.

{\bf IRAS~F14544$-$4255 (IC 4518)}: this system is a merger between
two galaxies with a nuclear separation of NS $\sim$ 12 kpc, which were
observed separately using two VIMOS pointings. The eastern source is
an elongated galaxy with a faint tail extended towards the northwest
of the system, as seen in the DSS image. The western source is more
compact, although some extended emission is also observed towards the
northwest of the galaxy. In the western source, there is a knot of
emission to the south of the nuclear region, also observed in our
VIMOS continuum image. This is spectroscopically confirmed as a late
type (G) star in our Galaxy. The morphology of the ionized gas
emission observed for the western galaxy is consistent with that of
the continuum emission. However, this is not the case for the eastern
source. For this galaxy the region of the major enhancement of
H$\alpha$ emission to the southeast of the galaxy corresponds to a
region with substantially faint continuum emission. Indeed, the peak
of the ionized gas is located $\sim$ 2.5 kpc to the southeast of the
peak of the continuum emission.

The optical spectrum of the western source is that of a Sy2. On the
other hand, no nuclear spectroscopic classification was found for the
eastern galaxy. However, the line ratios found towards the southeast
of the galaxy are consistent with photoionization by stars. Using the
H${\alpha}$-EW values (90 $\lsim$ H${\alpha}$-EW $\lsim$ 140~\AA) at
this location we estimate stellar ages of t$ \lsim$ 6 Myr.

%A similar type of structure is observed in the DSS image of the system
%to the east of the eastern source. Unfortunately, it falls outside the
%VIMOS fielf of view and we cannot confirm whether this knot certainly
%belongs to the galaxy

{\bf IRAS~F17138$-$1017}: the {\it HST} ACS image of the source shows
an overall symmetric, spiral morphology, based on which the galaxy
would be morphologically classified as 0. However, the image also
shows a rather complex structure in the central regions, where some
condensations and dust lanes are observed. Furtthermore, a double
nucleus structure emerges in the 3.4$\mu$m image of \cite{Zhou93},
with a nuclear separation of NS $\sim$ 1 kpc. The adopted
morphological class for this galaxy is type 2.

Unfortunately, the data-cube of IRAS~F17138$-$1017, as in the case of
IRAS F10567-4310, showed vertical patterns over the entire field of
view. As mentioned before, these residuals are only important if the
S/N is low, and therefore, although they affect the morphology of the
continuum (and the corresponding H${\alpha}$-EW) image, they have no
effect on the H${\alpha}$ emission map. Interestingly, the H${\alpha}$
image of the galaxy shows a very different morphology than the one
observed in the {\it HST} ACS image. The structure of the ionized gas
is substantially compact, with no signs of spiral structure and is
concentrated towards the central region. In general, this is
consistent with the structure observed in the Pa$\alpha$ image of the
galaxy \citep{Alonso-Herrero06}. The latter also reveals the presence
of knots and condensations in the nuclear region.

{\bf IRAS~F18093$-$5744}: this system consists of three galaxies in
interaction \citep{West76}, of which each individual galaxy was
observed separately. The nuclear separations between the northern (IC
4687) and the central galaxy (IC 4686) and between the central and the
southern galaxy (IC 4689) are NS $\sim$10 kpc and $\sim$20 kpc
respectively. The IRAS pointing in centered on the northern pair
(IC4687/IC4686). Using the results of \cite{Surace04} and re-scaling
to our adopted distance we derived an infrared luminosity of
$\log$(L$_{\rm IR}$/L$_{\odot}$) = 11.49 from these two
galaxies. Bearing in mind that the total $\log$(L$_{\rm
IR}$/L$_{\odot}$) is 11.57, the contribution to this quantity from the
southern source is $\log$(L$_{\rm IR}$/L$_{\odot}$) = 10.79. The {\it
HST} ACS image of the system shows a spiral-like morphology for IC
4687, with several knots and concentrations in the nuclear region. On
the other hand, IC 4686 appears to be a very compact object, whose
continuum emission is contaminated by a late type star (G or K) in our
galaxy. In addition, the {\it HST} ACS image shows extended emission
clearly linking the galaxies IC 4687 and IC 4686. In the case of IC
4689, a spiral morphology is observed in the {\it HST} ACS image,
without clear evidence of strong interaction.

The H${\alpha}$ map of IC 4687 shows three main regions of enhanced
ionized gas emission corresponding to the center of the galaxy and
other two regions immediately to the north and south of the nuclear
region. The high-resolution HST-NICMOS Pa$\alpha$ image of the galaxy
reveals that these regions correspond to several knots and
condesations \citep{Alonso-Herrero06}. In addition, a fourth region is
observed to the southwest of the nucleus, although with weaker
H${\alpha}$ emission. In IC 4686 the H${\alpha}$ emission from the
source is highly concentrated in the nuclear region of the
galaxy. Ionized gas emission is also observed extending towards the
north-west of the nuclear region. This emission seems to delineate the
continuum, extended emission seen in the {\it HST} image, linking this
galaxy with IC 4687. Both IC 4687 and IC 4686 have been classified as
HII galaxies on the basis of their nuclear spectra
\citep{Kewley01,Corbett03}. Additioanlly all extranuclear regions with
enhanced emission have line ratios consistent with an HII-like
region. Therefore, using the LH99 models and the extremely high
H${\alpha}$-EW values (200 $\lsim$ H${\alpha}$-EW $\gsim$ 420~\AA) we
derive ages of t$ \lsim$ 5 Myr for the stellar populations at these
locations. Finally, the H$\alpha$ image of IC4689 shows a region of
enhanced emission crossing the galaxy from the southeast to the
northwest of the galaxy, with some clumps in the extremes, and towards
the center of this region. Although no nuclear spectral classification
is available for this source, the clumps are located in regions with
line ratios typical for HII regions. The H${\alpha}$-EW values found
at these location are $\sim$100~\AA and the corresponding derived
stellar ages are t$ \lsim$ 6Myr.

{\bf IRAS~F21130$-$4446}: this ULIRG has been already identified in
the past as a double nucleus system with a nuclear separation of NS =
5.4 kpc \citep{Dasyra06a}. The {\it HST} WFPC2 of this source shows a
complex morphology, with several knots and condensations distributed
over the entire body of the galaxy. The two brightest condensations
seen in the {\it HST} image are located in the central and northern
regions and are likely associated with the nuclei of the two merging
galaxies. However, owing to the peculiar morphology of the galaxy, it
is hard to locate the exact position of the two nuclei. The peak of
the continuum emission in our VIMOS image coincides with the location
of the northern condensation. Interestingly, the most intense
H$\alpha$ emission is observed in the central region of the system,
coinciding with the location of the central condensation seen in the
{\it HST} image. \cite{Farrah03} found that star-formation activity is
responsible for most of the L$_{\rm IR}$ of this source. Using the
high H${\alpha}$-EW values (200 $\gsim$ H${\alpha}$-EW $\gsim$
400~\AA) found at this location, and the LH99 models we derive ages of
t$ \lsim$ 5Myr for the stellar populations.

{\bf IRAS~F21453$-$3511 (NGC 7130)}: the DSS image of the galaxy shows
a relatively asymmetric morphology with extended emission over
$\sim$35 kpc. The central 10 $\times$ 10 kpc$^2$ are covered by the
VIMOS field of view. The high-resolution {\it HST} WFPC2 image reveals
a nuclear spiral structure, with two spiral arms to the north and
south of the nuclear region. Such a spiral structure is also visible
in our continuum, and moreover, our H$\alpha$ image. However, note
that the extreme of the southern spiral arm seen in the {\it HST} is
not covered by our field of view.

Both our H$\alpha$ and the Pa$\alpha$ image of the galaxy
\citep{Alonso-Herrero06} show that the ionized gas emission is
concentrated in the nuclear region and the northern spiral arm. The
nuclear, optical spectrum of the source shows a mix between LINER- and
Sy-like features \citep{Veilleux95,Corbett03}. On the other hand, the
spiral arms have line ratios typical of HII-like galaxies, and
therefore it is possible to use the H${\alpha}$-EW values to estimate
the ages of the stellar populations at these locations. Although
values of $\sim$60~\AA~are found over the entire spiral structure, the
higher values are concentrated in the northern spiral arm. The
H${\alpha}$-EW values at that location are 100 $\lsim$ H${\alpha}$-EW
$\lsim$ 270~\AA, and the derive ages are t$ \lsim$ 6Myr. Finally, it
is worth mentioning that this remarkable concentration of H$\alpha$
emission in the northern spiral arm is perhaps indicative of a past
(minor?)  interaction that has enhanced the star-formation activity.

{\bf IRAS~F22132$-$3705 (IC5179)}: our VIMOS images sample the central
$\sim$7 $\times$ 7 kpc$^{2}$ of this very extended ($\sim$30 kpc)
spiral galaxy. The small concentration of continuum emission observed
$\sim$1~kpc to the north of the nucleus is a star in the field
\citep{Alonso-Herrero06}. The morphology of the ionized gas emission
is substantially clumpier than that of the continuum, with several
knots and concentrations spread over the entire extension of the
galaxy covered by the VIMOS field of view. Such knots and
condensations also observed in the high-resolution HST-NICMOS
Pa$\alpha$ image of the galaxy \citep{Alonso-Herrero06}. The optical,
nuclear spectrum of the galaxy is that of an HII-galaxy. In addition,
HII-like line ratios are found at almost all locations in the VIMOS
H$\alpha$ image. Using the H${\alpha}$-EW values (60 $\gsim$
H${\alpha}$-EW $\gsim$ 140~\AA) and the LH99 models we derive ages of
t$ \lsim$ 6.5 Myr for the stellar populations at these locations.

{\bf IRAS~F22491$-$1808}: this is a double nucleus system with several
knots and condensations located in both the nuclear region and the
tidal tails observed to the east and northwest of the system. Although
the two nuclei are not clearly distinguishable at optical wavelengths,
a double nucleus structure emerges at near-IR wavelengths, with a
nuclear separation of NS $\sim$3~kpc
\citep{Surace00a,Dasyra06a}. Interestingly, our H${\alpha}$ image of
the source shows a more compact morphology than that of the continuum,
with no evidence for the already mentioned tidal tails. A small region
with high H${\alpha}$-EW values is observed immediately to the
north-west of the nuclear region. This region coincides with the
location of some of the knots observed in the {\it HST} ACS
image. Bearing in mind that this source is classified as an HII-galaxy
at all wavelengths studied
\citep[e.g.][]{Genzel98,Veilleux99,Farrah03,Imanishi07}, we use the
H${\alpha}$-EW values (H${\alpha}$-EW $\sim$ 150~\AA) and the LH99
models to determine the age of the stellar populations found at this
location. We obtain ages t$<$ 6 Myr, which are consistent with the
estimated ages obtained in the imaging study of \cite{Surace00a} for
the knots at this location in the galaxy .

{\bf IRAS~F23128$-$5919 (AM 2312-591)}: this is a double nucleus
system (NS $\sim$4~kpc) with two prominent tidal tails to the north
and southeast of the nuclear region that extends over $\sim$ 45
kpc. Our VIMOS image covers the central $\sim$25 kpc of the
galaxy. The double nucleus structure is clearly visible in both the
continuum and the H$\alpha$ image, although the tidal tails are not
observed in ionized gas emission. In addition there is a third
concentration of H$\alpha$ emission to the east of the double nucleus
structure. The optical spectrum of this galaxy shows a mix between
LINER, Sy2 and HII-like features \citep{Duc97,Kewley01}. This galaxy
is classified as a ULIRG and therefore, was not included in the study
presented in Paper II. Hence, no attempt was made to estimate the ages
of the young stellar populations in IRAS~F23128$-$5919. However, note
that the concentration of H$\alpha$ emission seen to the east of the
system coincides with the location of the highest H${\alpha}$-EW
values obtained for all the galaxies in our sample, with values as
high as 850~\AA.

\end{document}